\documentclass[12pt,a4paper,fleqn]{book}
\usepackage[tbtags]{amsmath}
\usepackage{amssymb}
\usepackage{mathrsfs}

\usepackage{fancyhdr}
\makeatletter
\def\cleardoublepage{\clearpage\if@twoside \ifodd\c@page\else%
    \hbox{}%
    \thispagestyle{empty}
    \newpage%
    \if@twocolumn\hbox{}\newpage\fi\fi\fi}
\makeatother

\renewcommand{\vec}[1]{\mathbf{#1}}
\newcommand{\fillblank}{\textsf}

\begin{document}


\setlength{\parskip}{1.0ex plus 0.5ex minus 0.5ex}
\frontmatter

\begin{titlepage}
\begin{center}

\Huge
\textbf{Theories on noncommutative spaces\\ and deformed symmetries}

\vfill

\normalsize
{\Large Thesis submitted for the degree of}\\[2.2ex]
\textbf{\Large Doctor of Philosophy (Science)}\\[2ex]
{\Large of}\\[2ex]
\textbf{\Large Jadavpur University, Kolkata}

\vfill

{\Large July 2008}

\vfill

\textbf{{\Large Saurav Samanta}}\\[2ex]
{\large \mbox{Satyendra Nath Bose National Centre for Basic Sciences}}\\
{\large JD Block, Sector 3, Salt Lake, Kolkata 700098, India}

\end{center}
\end{titlepage}



\chapter*{Certificate from the supervisor}
\thispagestyle{empty}


This is to certify that the thesis entitled \fillblank{`Theories on noncommutative spaces and deformed symmetries'} submitted by \fillblank{Saurav Samanta}, who got his name registered on \fillblank{May 26, 2006} for the award of \fillblank{Ph.D.~(Science)} degree of \fillblank{Jadavpur University}, is absolutely based upon his own work under the supervision of \fillblank{Professor Rabin Banerjee} at \fillblank{S.N.~Bose National Centre for Basic Sciences, Kolkata, India}, and that neither this thesis nor any part of it has been submitted for any degree/diploma or any other academic award anywhere before.

\vspace{3.0cm}

\hfill \begin{tabular}{@{}l@{}}
\fillblank{Rabin Banerjee}\\
Professor\\
S.N.~Bose National Centre for Basic Sciences\\
JD Block, Sector 3, Salt Lake\\
Kolkata 700098, India
\end{tabular}


\chapter*{}
\thispagestyle{empty}


{\Large  Dedicated to my family}



\chapter*{Acknowledgements}
\thispagestyle{empty}

At the end of my tenure at Satyendra Nath Bose National Centre
for Basic Sciences (SNBNCBS), I would like to express my gratitude to the people
who have made this amazing journey possible and recollect some of their contributions.

My thesis supervisor Prof. Rabin Banerjee has been inspirational in this entire period and
without his expert guidance I won't have been where I am today. It has been a wonderful learning experience and a great privilege to work under Prof. Rabin Banerjee. His valuable suggestions and clear insights have lent me a fascinating point of view into the working of physics and by extension, nature. To him goes my heartfelt respect and gratitude.

Prof. Binayak Dutta Roy has always been available with his able guidance and considerable repertoire of knowledge. I sincerely thank him for all those intense academic discussions. I would like to thank Dr. Pradip Mukherjee for sharing with me his insightful and innovative ideas which led to some very fruitful discussions. I am grateful to Dr. Biswajit Chakraborty for his enthusiastic discussions on physics. 

    I am indebted to Prof. S. Dattagupta, ex. Director of SNBNCBS, for allowing me to research here. I thank
Prof. A. K. Raychaudhuri , present Director, for many facilities I enjoyed in this centre.

I thank all the academic and administrative staff of Bose centre for helping me in many ways. Particularly I would like to mention the Library staff for their excellent support.

I have been fortunate enough to count some very bright and brilliant persons as my `group-mates' namely Kuldeep, Shailesh, Bibhas, Sujoy. The named four and others like Sunandan, Arindam, Chandrasekhar and Saikat have contributed much to my stay at SNBNCBS,  both in  an academic sense and otherwise. It has been a rewarding experience working with them.

To my seniors Shanta-da, Abhishek-da, and Mani-da I would like to express my gratitude for the brotherly care and support provided by them during my initial days at the center. To Ankush, who was more a friend than a senior I convey my warmest thanks and regards.

 My batch mates -- Kartick, Anjan, Kunal, Manoranjan and Manoj provided pleasant companionship for which I shall be ever grateful.

The varied group comprising of hyper enthusiastic Sagar at one extreme to sedate Ashish at another with impish Santosh and the unlikely trio of Arya, Tamoghna and Bipul somewhere around the middle, constitute my junior batch at SNBNCBS! With a list like that, it does not take a physicist to gather that my experiences with them has been dramatic, funny and in general out of the ordinary! I highly admire my young friends Amartya, Anirban and Shantanu for their weird imagination. I shall never forget our long discussions on fiction and fantasy.

In my home village I have many friends and well-wishers. Gautam and Rinku deserve special mention for their company throughout the years. I am indebted to my seniors Tarun, Atanu, Aniruddha for their helpful attitude on many occasions. I express my love and gratitude to my cousin Anirban and nephews Arya, Adrish, Chayan, Ayan, Sayan, Alapan.

 My thesis will be incomplete if I don't acknowledge the role my family has played. Without their unflinching support this journey would have been impossible. My sister has been a source of joy in my life and without her my existence would have been dry indeed. To my parents who have borne great demands on my part with a smiling face I convey my deep felt respect and love. Having stayed away from home for a large part of my life due to educational constraints, their encouragement and belief in me has provided the motivation to withstand the hardships and difficult situations for all these years.


\chapter*{List of publications}
\thispagestyle{empty}


\begin{enumerate} \raggedright
\item \textbf{Noncommutativity from embedding techniques.}\\ Saurav Samanta\\
    {\it Mod. Phys. Lett.} {\bf A 21} 675 (2006) [arXiv: hep-th/0510138] 
\item \textbf{Deformed symmetry in Snyder space and relativistic particle dynamics}\\
    Rabin Banerjee, Shailesh Kulkarni and Saurav Samanta\\
    {\it JHEP} {\bf 0605} 077 (2006) [arXiv: hep-th/0602151]
\item \textbf{Remarks on the noncommutative gravitational quantum well}\\
    Rabin Banerjee, Binayak Dutta Roy and {Saurav Samanta}\\
    {\it Phys. Rev.} {\bf D 74} 045015 (2006) [arXiv: hep-th/0605277]
\item \textbf{Gauge Symmetries on theta-Deformed Spaces}\\ Rabin Banerjee and Saurav Samanta\\
   {\it JHEP} {\bf 0702} 046 (2007) [arXiv:hep-th/0611249] 
\item \textbf{Gauge generators, transformations and identities on a noncommutative space}\\ Rabin Banerjee and Saurav Samanta\\
  {\it Eur. Phys. J.} {\bf C 51} 207 (2007) [arXiv: hep-th/0608214]
\item \textbf{Lie algebraic noncommutative gravity.}\\ Rabin Banerjee, Pradip Mukherjee and Saurav Samanta\\
 {\it Phys. Rev.} {\bf D 75} 125020 (2007) [arXiv: hep-th/0703128]
\item \textbf{Diffeomorphism symmetry in the Lagrangian formulation of gravity}\\Saurav Samanta\\{[arXiv:0708.3300]}
\item \textbf{String non(anti)commutativity for Neveu-Schwarz boundary conditions}\\ Chandrasekhar Chatterjee, Sunandan Gangopadhyay, Arindam Ghosh Hazra and Saurav Samanta\\{To appear in {\it IJTP}} [arXiv:0801.4189]
\item \textbf{Noncommutative Black Hole Thermodynamics}\\ Rabin Banerjee, Bibhas Ranjan Majhi and Saurav Samanta\\{\it Phys. Rev.} {\bf D 77} 124035 (2008) [arXiv:0801.3583]
\end{enumerate}

This thesis is based on the papers numbered by [1,2,3,4,5,6,7].



\chapter*{}
\pagenumbering{roman}
\thispagestyle{empty}

\begin{center}
\uppercase{Theories on noncommutative spaces\\ and deformed symmetries}
\end{center}



\tableofcontents



\mainmatter

\chapter{\label{chap:intro}Introduction}


\section{Historical background}
The idea of noncommutativity first comes in physics via quantum mechanics where the position and momentum variables $\hat{x},\hat{p}$ satisfy the Heisenberg commutation relation $[\hat{x},\hat{p}]=i\hbar$. A consequence is the uncertainty relation $\Delta x\Delta p\ge\frac{\hbar}{2}$. So the concept of point does not exist in quantum phase space, instead it is replaced by Planck cell. It was von Neumann who first attempted to describe such a space by ``pointless geometry". In this way Neumann algebra was formulated which leads to noncommutative geometry for studying topological spaces whose commutative $C^*$-algebras of functions are replaced by noncommutative algebras\cite{connes}. Here the notion of point is discarded and space is understood in purely algebraic terms. The generalization of differential calculus in a noncommutative setting was also done by the mathematicians \cite{mat1,mat2}. Various mathematical tools were further developed by Connes and his collaborators to define a noncommutative Yang--Mills theory\cite{y1} and particle models\cite{y2}.

In physical problems the concept of phase space noncommutativity was extended to noncommutativity among coordinates by Heisenberg himself. In a letter to Peierls\cite{pauli1}, he proposed coordinate uncertainty relation as a solution to avoid the singularities of the electron self energy. Later these ideas were used by Peierls in Landau level related problem. Pauli also came to know this idea from Heisenberg and informed Oppenheimer about it\cite{pauli2}. In $1947$, Hartland Snyder, a student of Oppenheimer published the first paper\cite{sny} on this subject. By a dimensional descent from five dimensions he obtained a Lorentz invariant discrete space-time in which a natural unit of length exists. His hope was that such a length scale may remove the infinities of field theory. In the same year C. N. Yang generalized Snyder's work in curved de Sitter space\cite{yang}. Although the main stream interest was shifting from noncommutative physics due to the contemporary success of renormalization theory, mathematicians were still working in this field. The main source of inspiration came from the noncommutativity of quantum mechanics and many ideas were borrowed from that.

In order to quantize a theory, Dirac gave a simple prescription of replacing the classical variables by operators and Poisson bracket by commutator bracket (divided by $i\hbar$). The mapping between classical kernels which are $c$-number phase space function $f(x,p)$ and the corresponding operator $\hat{f}(\hat{x},\hat{p})$ with proper ordering was first given by Weyl\cite{Weyl}. As an example, a polynomial function of $\hat{x},\hat{p}$ is ordered in a completely symmetrized way which is usually called the Weyl ordering prescription. The importance of the composition rule of classical kernels in an operator product was first realized by von Neumann\cite{neumann} in a study of the uniqueness of the Schr\"odinger representation. Then Groenewold did the necessary calculation\cite{groenewold} to obtain the composition rule of two operators $\hat{f}$ and $\hat{g}$ corresponding to the classical kernels $f$ and $g$ and how it is related to $fg$. This was used by Moyal to formulate the quantum mechanics in phase space\cite{moyal}.

In 1964, Mead in his paper\cite{mead} carefully analyzed many thought experiments to study the effect of gravitation on measuring the position of a particle. He concluded that there is always an error $\Delta x\gtrsim \sqrt{G}$ where $G$ is the gravitational constant in natural units. A similar restriction also exists for the measurement of time interval. The result can be understood heuristically by realizing that to get high resolution one needs high energy photons but higher energy means high gravitational interaction and this will seriously affect the spacetime.

Uncertainty relation, in quantum mechanics, comes as a consequence of the non vanishing commutator algebra which is an inbuilt kinematic structure of the theory. Hence there were attempts to give a kinematical meaning to the position uncertainty described by Mead. In Townsend's paper\cite{townsend} Planck length appears in gravity due to the noncommutativity of the generators of translations. Maggiore extended the ideas\cite{mag1,mag2} to get a general noncommutative algebra where position-position bracket was nonzero. Doplicher, Fredenhagen and Roberts developed a model of quantum spacetime\cite{dop1,dop2} which imply uncertainty relations among different coordinates. They also gave some general ideas on the definition of fields and their interactions over this space.

String theory which is supposed to describe quantum theory of gravity contains an intrinsic length, the length of the string $l_s$. Naturally it is not possible to probe a distance smaller than $l_s$. Detailed analysis based on high energy scattering amplitudes\cite{ven,gross,amati} showed the existence of minimum length. Not only that, the techniques of noncommutative geometry have been applied to rigorously study the duality symmetries of string theory\cite{lizzi,landi}. Later it was shown that noncommutative geometry arises due to the toroidal compactification of the matrix model\cite{aconnes}. In fact these matrix models lead to noncommutative Yang-Mills theory as their effective field theory. Seiberg and Witten in their seminal paper \cite{SW} extended the ideas of noncommutativity in string theory with a nonzero $B$ field. They showed an equivalence between ordinary gauge fields and noncommutative gauge fields which is usually called the Seiberg--Witten map. Till now, this map has been applied in various research areas from Hall effect, fluid dynamics, field theory to gravity. Different types of noncommutative structures have also been studied without referring to their commutative counterpart. A general overview of this broad subject may be found in \cite{dou,boz,NIK}.

The simplest form of noncommutativity is the canonical noncommutativity where position position bracket is a constant. The algebra itself and field theories defined on such a space are known to violate the Lorentz invariance. However, a deformed Poincar$\acute{\textrm e}$ symmetry can be developed where the usual Poincar$\acute{\textrm e}$ algebra is satisfied but the generators have a deformed coproduct rule\cite{chainew,wessju}. As a result, noncommutative field theory possesses the symmetry under the deformed (Hopf) Poincar$\acute{\textrm e}$ algebra. Interesting consequences of this deformed symmetry have been studied in \cite{113,114}. The issues regarding the parity and time reversal symmetry have also gained attention in the literature\cite{M.M,Hayakawa,bala}.

Apart from the space-time symmetry, the gauge symmetries of a noncommutative field theory is also an important topic. The study, without the use of Seiberg-Witten map, was initiated by \cite{Hayakawa}. A noncommutative version of the standard model was also developed\cite{t1,t2}. Noncommutative gauge theory has also been studied using the Seiberg--Witten map\cite{schn,W1,W2,W3}. 

Recently an interesting study was done in the literature\cite{vas,wess} by twisting the coproduct rule of both the Poincar$\acute{\textrm e}$ generators and gauge generators. The result obtained was quite remarkable: the noncommutative theory turned out to be invariant under the commutative space gauge transformations. The issue, whether this twisted gauge invariance is a real physical symmetry or not has been discussed later\cite{chai,zet}. Gravity, which is also a gauge theory, has been formulated over noncommutative spacetime. There are different approaches, in\cite{chams} the theory was constructed by gauging the noncommutative $ISO(3,1)$ group using the Seiberg--Witten map, but there are also constructions based on a deformed Poincar$\acute{\textrm e}$ algebra\cite{Aschierie}. A different approach was followed in \cite{gaume} where noncommutative gravity was obtained from string theory in the Seiberg--Witten limit.

\section{Structure of the thesis}
This thesis, based on the work \cite{sa,ssamanta,ssam,saurav,sau,samanta,sausa} , is devoted to study different aspects of quantum mechanics, field theory and gravity on noncommutative spaces. We give particular emphasis on the symmetry properties (both spacetime and gauge) of noncommutative theories. The outline of our thesis is as follows

In chapter 2, we discuss the noncommutative algebra of the generalized Landau problem where a charged particle is subjected to a quadratic potential with a perpendicular constant magnetic field. This is done by two approaches. In the first method we use the Batalin--Tyutin embedding technique to reveal the noncommutative structures. Different types of noncommutativity follow from different gauge choices. In the other approach, the model is mapped to a chiral oscillator problem. Both methods establish a duality among the noncommutativity of coordinates and the noncommutativity of momenta. 

The gravitational well problem on a noncommutative phase space has been discussed in chapter 3. We use the WKB approximation to study the effect of noncommutativity analytically. Comparison with recent experimental data with cold neutrons at Grenoble imposes an upper bound on the noncommutative parameter.


One feature of noncommutative algebra is that it violates the usual spacetime symmetries. For example, in the canonical noncommutativity translatioal symmetry is obeyed but the Lorentz symmetry is violated. In the Snyder algebra situation is quite opposite--Lorentz symmetry is restored but the translational symmetry is lost. These symmetry related issues are addressed in chapter 4. Schr\"odinger generators which consist of the Galilean generators and the nonrelativistic version of two conformal generators do not satisfy the standard algebra on the canonical noncommutative space. Following an algebraic approach, we construct the deformed generators which satisfy the usual commutative space algebra. Then a dynamical model is constructed whose symplectic analysis gives the canonical noncommutative algebra. Using the N\"other's theorem we obtain the same deformed generators from this model. This shows the consistency of two approaches. Similar analysis is also performed for the Snyder type noncommutativity to obtain the deformed Poincar$\acute{\textrm e}$-conformal generators.

In the next chapter, we discuss the gauge symmetries of the noncommutative Yang--Mills theory in a Lagrangian framework. By abstracting a connection between gauge symmetry and gauge identity, we analyze star (deformed) gauge transformations with usual Leibniz rule as well as undeformed gauge transformations with a twisted Leibniz rule. Explicit structures of the Lagrangian gauge generators in either case are computed. We show that, in the former case, the relation mapping the generator with the gauge identity is a star deformation of the commutative space result. In the latter case, on the other hand, this result gets twisted to yield the desired map.

Hamiltonian analysis of the same noncommutative gauge theory is studied in chapter 6. Both types of gauge transformations are considered. Using the constraint analysis we show that the structure of the Hamiltonian gauge generator is identical in either case. The difference comes in the computation of the graded Poisson brackets to get the gauge transformations of the fields. The analysis for the undeformed gauge transformations provides a novel interpretation of the twisted coproduct rule. We find that it is same as the normal coproduct with the stipulation that the gauge parameter is taken outside the star operation at the end of the computations.

In chapter 7, we discuss general relativity over a Lie algebra valued noncommutative spacetime. We follow the minimal (unimodular) formulation where the physical symmetries are manifest. Gauging the Poincar$\acute{\textrm e}$ group, we exploit the Seiberg -- Witten map technique to formulate the theory as a perturbative Lagrangian theory. Detailed expressions of the Seiberg -- Witten maps for the gauge parameters, gauge potentials and the field strengths have been worked out. We find a remarkable result that the first order correction of the noncommutative gravity action is zero, exactly as is the case for canonical noncommutativity. Finally, we end up with conclusions in chapter 8.


\chapter{\label{chap:landau}Generalized Landau problem and phase space noncommutativity}

In order to study noncommutativity in quantum mechanics, Landau problem is a standard model where a charged particle moving on a plane is subjected to a constant magnetic field in the perpendicular direction. Usually people study it by introducing noncommuting coordinates by hand or by introducing noncommuting momenta due to the peculiar structure of canonical momenta which depends on the magnetic field. Both theoretical\cite{LUKE,x,x1,x2,x3,x4} and phenomenological\cite{y,y-1,y-2,y-3} studies have been done to reveal various interesting points of this problem. However in this chapter we take a generalized version of the Landau problem where an additional quadratic potential is present together with the usual constant magnetic field. Here noncommutativity is not introduced by hand; rather it is a consequence of the modified symplectic structure. We follow two methods, first the Batalin--Tyutin\cite{m} embedding approach and next, the doublet splitting \cite{c} method to analyze the noncommutative structure of this problem.

 We treat the generalized problem as a constrained Hamiltonian system for which a first order formulation is most natural. In this formulation the number of variables is doubled; moreover second class constraints occur. The Poisson brackets therefore get replaced by the Dirac brackets\cite{n} which are finally elevated to the level of commutators. The Dirac brackets among both sets of dynamical variables lead to noncommuting structures.

Next, we embed this second class system in an extended space by introducing new pairs of canonical variables such that the original system is converted into a first class one. This embedded system is therefore considered as a true gauge theory. By choosing the unitary gauge which amounts to setting the new variables to zero, the original second class system is recovered. We then discuss two particular gauge choices in some details. These choices are done such that, in either case, the two sets of dynamical variables can be regarded as coordinates and their conjugate momenta. However one gauge choice leads to commuting coordinates but noncommuting momenta while the other choice yields commuting momenta but noncommuting coordinates. Since these distinct structures follow from the same master gauge theory, a duality is established between them.



This type of duality between different noncommutative structures is further confirmed by doublet splitting where a general mapping between the variables of generalized Landau problem and the variables of chiral oscillators is established. Instead of giving a theory on quantum mechanical representation\cite{la} we stress on the dual nature of noncommuting Poisson brackets and Lagrangian framework.
\section{The model: generalized Landau problem}
The classical equations of motion for an electron\footnote{We have rationalized $e=c=1$.} moving in the $x_1-x_2$ plane under the influence of a constant perpendicular magnetic field $B$ are,
\begin{equation}
m\ddot x_i=B\epsilon_{ij}\dot x_j.
\end{equation}

The above equations of motion follow from the Lagrangian,
\begin{equation}
L=\frac{m}{2}\dot x_i^2+\frac{1}{2}B\epsilon_{ij}x_i\dot x_j.
\end{equation}
The canonical momentum 
\begin{equation}
p_i=\frac{\partial L}{\partial \dot x_i}=m\dot x_i-\frac{1}{2}B\epsilon_{ij}x_j
\end{equation}
is clearly different from the kinematic momentum $m\dot x_i$ by a term proportional to the magnetic field. This leads to the canonical Hamiltonian
\begin{equation}
H=\frac{\pi_i^2}{2m}=\frac{1}{2m}\left(p_i+\frac{1}{2}B\epsilon_{ij}x_j\right)^2.
\end{equation}
Now we generalize the Landau problem by introducing an oscillating potential with spring constant $k$ in the $x_1-x_2$ plane. The equations of motion are
\begin{equation}
m\ddot x_i-B\epsilon_{ij}\dot x_j+kx_i=0.
\label{old1}
\end{equation}
It is convenient to express the second order system in its first order form. Furthermore following 't Hooft\cite{a,aG,aGH}, let the equations of motion of a system $\dot q_i=\{q_i,H\}=f_i(q)$, be a function
of position alone. Then denoting $p_i$ as the conjugate momenta, we note that the Hamiltonian $H=\Sigma_ip_if_i(q)$ does not have a lower bound. A positive-definite function $\rho$, considered as the physical Hamiltonian, can be constructed such that $\{\rho,H\}=0$. But this change from the original (unbounded) Hamiltonian $H$ to the bounded positive (semi) definite Hamiltonian $\rho$ leads to a modified algebra that can be obtained as follows\cite{c}:
\begin{equation}
\dot q_i=\{q_i,\rho\}=\{q_i,q_j\}\partial_j\rho(q).
\end{equation}
To reproduce the original set of equations of motion, obviously one should take
\begin{equation}
\{q_i,q_j\}\partial_j\rho(q)=f_i(q),
\end{equation}
leading to a nontrivial algebra of $q_i$, eventually leading to noncommuting structures.

Now the noncommutativity of the generalized Landau problem appears by writing the second order system into a pair of first order equations by doubling the degrees of freedom\cite{c}. Consider the pair of first order equations
\begin{eqnarray}
\dot x_i=\alpha q_i+\beta \epsilon_{ij}x_j
\label{x}\\
\dot q_i=\omega x_i+\lambda \epsilon_{ij}q_j
\label{q}
\end{eqnarray}
which lead to the Landau type equations in both $x_i$ and $q_i$\cite{c},
\begin{equation}
\ddot r_i=(\beta+\lambda)\epsilon_{ij}\dot r_j+(\beta \lambda+\alpha \omega)r_i, \ r_i=x_i,q_i.
\label{old2}
\end{equation}
By identifying,
\begin{eqnarray}
&&\frac{B}{m}=(\beta+\lambda)
\label{rb1}\\
\textrm{and} &&\frac{k}{m}=-(\beta \lambda+\alpha \omega)
\label{rb2}
\end{eqnarray}
(\ref{old2}) is regarded as a generic version of (\ref{old1}). A Hamiltonian is now constructed\cite{c,a},
\begin{equation}
H=(\alpha q_i+\beta \epsilon_{ij}x_j)\pi^x_i+(\omega x_i+\lambda \epsilon_{ij}q_j)\pi^q_i
\label{ham}
\end{equation}
where $(x_i,\pi^x_i)$ and $(q_i,\pi^q_i)$ are canonical pairs. The equations of motion $\dot r_i=\{r_i,H\}$ just yield (\ref{x}) and \,(\ref{q}). As usual, this $H$ is not bounded from below. A positive definite $\rho$, commuting with $H$, has to be obtained. This $\rho$ gets identified with the physical Hamiltonian\cite{c,a,aG,aGH}. A natural choice satisfying $\{\rho,H\}=0$ is
\begin{equation}
\rho=\frac{q^2}{2m}+\frac{1}{2}kx^2, \  q^2=q_i^2 \ \textrm{and} \ x^2=x_i^2
\label{19}
\end{equation}
where 
\begin{equation}
\alpha=-\frac{\omega}{km}.
\label{hooft} 
\end{equation}
The corresponding algebra is
\begin{equation}
\{x_i,x_j\}=\frac{\beta}{k} \epsilon_{ij}, \  \{x_i,q_j\}=-\frac{\omega}{k} \delta_{ij}, \  \{q_i,q_j\}=m\lambda \epsilon_{ij}.
\label{chcm}
\end{equation}
This algebra leads to noncommuting structures for both $x_i$ and $q_i$ so that the equations of motion (\ref{x}) and (\ref{q}) can be reproduced from $\dot r_i=\{r_i,\rho\}, \ r_i=x_i,q_i$.
Now we are in a position to construct the Lagrangian for this generalized Landau problem. The physical concept behind this construction is given in \cite{fa}.

First a $\Lambda$ matrix is constructed from the basic brackets \,(\ref{chcm}) 
\begin{eqnarray}
\Lambda_{ij}&=&\left[\left\{\Gamma_i,\Gamma_j\right\}\right], \Gamma=(x,q)\nonumber\\
&=&\left(\begin{matrix}\frac{\beta}{k}\epsilon_{ij}&-\frac{\omega}{k}\delta_{ij}\cr \frac{\omega}{k}\delta_{ij}&m\lambda\epsilon_{ij}\end{matrix}\right)
\end{eqnarray}
Its inverse is
\begin{eqnarray}
\Lambda^{ij}=\frac{k}{\omega^2-mk\beta\lambda}\left(\begin{matrix}mk\lambda\epsilon_{ij}&\omega\delta_{ij}\cr-\omega\delta_{ij}&\beta\epsilon_{ij}\end{matrix}\right).
\label{mat}
\end{eqnarray}
Using (\ref{hooft}) and (\ref{rb2}) we show 
\begin{eqnarray}
\omega^2-mk\beta\lambda=k^2.
\label{tomega}
\end{eqnarray}
to write (\ref{mat}) as
\begin{eqnarray}
\Lambda^{ij}=\frac{1}{k}\left(\begin{matrix}mk\lambda\epsilon_{ij}&\omega\delta_{ij}\cr-\omega\delta_{ij}&\beta\epsilon_{ij}\end{matrix}\right).
\end{eqnarray}
The Lagrangian is therefore\cite{c},
\begin{eqnarray}
L&=&\frac{1}{2}\Gamma_i\Lambda^{ij}\dot\Gamma_j-\rho(\Gamma)\label{hooftkl}\\
&=&\frac{1}{2k}(mk\lambda\epsilon_{ij}x_i\dot x_j+\beta\epsilon_{ij}q_i\dot q_j+\omega x_i\dot q_i-\omega q_i\dot x_i)-(\frac{1}{2m}q_i^2+\frac{k}{2}x_i^2).
\label{430}
\end{eqnarray}
We get the following equations of motion from the above Lagrangian
\begin{eqnarray}
&&m\lambda\dot x_i-\frac{\omega}{k}\epsilon_{ij}\dot q_j+k\epsilon_{ij}x_j=0\\
&&\frac{\beta}{k}\dot q_i+\frac{\omega}{k}\epsilon_{ij}\dot x_j+\frac{1}{m}\epsilon_{ij}q_j=0.
\end{eqnarray}
These are compatible with \,(\ref{x}) and \,(\ref{q}) under the conditions (\ref{hooft},\ref{tomega}). Since \,(\ref{x}) and \,(\ref{q}) reproduced (\ref{old2}), the system described by the Lagrangian (\ref{430}) and that described by (\ref{old2}) are same. Of course both the variables $x_i$ and $q_i$ satisfy the same equation of motion (\ref{old2}) and hence there is a symmetry between them.
\section{Noncommutativity from Batalin--Tyutin framework}
According to Dirac's\cite{n} analysis of constrained systems, the constraints with weakly vanishing Poisson's bracket are called first class constraints (FCC), others are called second class constraints (SCC). The FCC constraints induce gauge invariance in the theory whereas Dirac brackets are consequences of the SCC. 

 In the Batalin--Tyutin formalism, new auxiliary variables are introduced in the system containing SCC in such a way that the original SCC can be modified to a set of FCC. Also, the original Hamiltonian has to be altered appropriately to make the resulting system gauge invariant.


The canonical momenta corresponding to the coordinates $x_i$ and $q_i$ for the Lagrangian (\ref{430}) are given by,
\begin{eqnarray}
&&\pi^x_i=\frac{\partial L}{\partial \dot x_i}=-\frac{1}{2k}(mk\lambda\epsilon_{ij}x_j+\omega q_i)
\\
&&\pi^q_i=\frac{\partial L}{\partial \dot q_i}=-\frac{1}{2k}(\beta\epsilon_{ij}q_j-\omega x_i).
\end{eqnarray}
Thus they form the constraints:
\begin{eqnarray}
&&\Omega^1_i=\pi^x_i+\frac{1}{2k}(mk\lambda\epsilon_{ij}x_j+\omega q_i)\approx 0
\label{con1}\\
&&\Omega^2_i=\pi^q_i+\frac{1}{2k}(\beta\epsilon_{ij}q_j-\omega x_i)\approx 0.
\label{con2}
\end{eqnarray}
The commutator matrix for the above constraints is given by
\begin{eqnarray}
\Omega^{XY}_{ij}&=&\{\Omega^X_i,\Omega^Y_j\};X,Y=1,2\\
&=&\frac{1}{k}\left(\begin{matrix}mk\lambda\epsilon_{ij}&\omega\delta_{ij}\cr-\omega\delta_{ij}&\beta\epsilon_{ij}\end{matrix}\right).
\label{inversemat}
\end{eqnarray}
Since the constraint matrix is nonsingular, inverse of (\ref{inversemat}) exists. It is given by,
\begin{eqnarray}
\Omega^{(-1)XY}_{ij}=\left(\begin{matrix}\frac{\beta}{k}\epsilon_{ij}&-\frac{\omega}{k}\delta_{ij}\cr \frac{\omega}{k}\delta_{ij}&m\lambda\epsilon_{ij}\end{matrix}\right).
\end{eqnarray}
According to Dirac's classification\cite{n}
 (\ref{con1}) and (\ref{con2}) are second class constraints. The Dirac brackets defined by,
\begin{eqnarray}
\{f,g\}_{DB}=\{f,g\}-\{f,\Omega^X_{i}\}\Omega^{(-1)XY}_{ij}\{\Omega^Y_{j},g\}
\label{diracbracket} 
\end{eqnarray}
gives the algebra
\begin{eqnarray}
\{x_i,x_j\}_{DB}=\frac{\beta}{k} \epsilon_{ij}, \  \{x_i,q_j\}_{DB}=-\frac{\omega}{k} \delta_{ij}, \  \{q_i,q_j\}_{DB}=m\lambda \epsilon_{ij}
\label{dirac}
\end{eqnarray}
which reproduces (\ref{chcm}). This algebra shows a more general type of noncommutativity than that of \cite{s} where momenta-momenta bracket is zero.

In order to convert the second class constraints (\ref{con1}) and (\ref{con2}) into first class constraints a canonical set of auxiliary variables is introduced
\begin{eqnarray}
\{\phi^X_i,\phi^Y_j\}=\epsilon_{XY}\delta_{ij}; \ X,Y=1,2.
\end{eqnarray}
Now we define the following constraints
\begin{eqnarray}
&&\Psi^1_i=\Omega^1_i+A(\phi^1_i+\epsilon_{ij}\phi^2_j)
\label{cons1}\\
&&\Psi^2_i=\Omega^2_i+C\phi^2_i+D\epsilon_{ij}\phi^1_j
\label{cons2}
\end{eqnarray}
with
\begin{eqnarray}
A=\left(\frac{m\lambda}{2}\right)^{1/2}, \ C=\left(\frac{1}{2m\lambda}\right)^{1/2}\left(1-\frac{\omega}{k}\right), \ D=\left(\frac{1}{2m\lambda}\right)^{1/2}\left(1+\frac{\omega}{k}\right)
\label{sC}
\end{eqnarray}
so that the algebra of the constraints (\ref{cons1}) and (\ref{cons2}) is strongly involutive ($\{\Psi^X_i,\Psi^Y_j\}=0$). Hence the constraints are first class. Also note that $A(C+D)=1$.

To obtain the first class Hamiltonian we begin by constructing the improved variables\cite{s,new,newsubir}. Improved variables are first class counterparts of the original variables $x_i$ and $q_i$. These are given by
\begin{eqnarray}
&&\tilde{x_1}=x_1+C\phi^2_1-D\phi^1_2, \  \ \tilde{x_2}=x_2+D\phi^1_1+C\phi^2_2\nonumber\\
&&\tilde{q_1}=q_1+A(\phi^2_2-\phi^1_1), \  \ \tilde{q_2}=q_2-A(\phi^1_2+\phi^2_1)\nonumber
\end{eqnarray}
where $A,C$ and $D$ are given by (\ref{sC}). One can easily check
\begin{eqnarray}
\{\tilde{r_i},\Psi^X_i\}=0; \ \tilde{r_i}=\tilde{x_i},\tilde{q_i}
\end{eqnarray}
so that they are first class indeed. They satisfy the algebra 
\begin{eqnarray}
\{\tilde{x_i},\tilde{x_j}\}=\frac{\beta}{k} \epsilon_{ij}, \  \{\tilde{x_i},\tilde{q_j}\}=-\frac{\omega}{k} \delta_{ij}, \  \{\tilde{q_i},\tilde{q_j}\}=m\lambda \epsilon_{ij}
\end{eqnarray}
which mimics (\ref{dirac}) and is a consequence of a general theorem\cite{m} which states that $\{\tilde{A},\tilde{B}\}= \widetilde{\{A,B\}_{DB}}$.

Any function of the phase space variables can be made first class by the following transformation
\begin{eqnarray}
F(x,q)\rightarrow \tilde{F}(\tilde{x}\tilde{q})=
\left.F(x,q)\right|_{x=\tilde{x},q=\tilde{q}}.
\end{eqnarray}
Hence the first class Hamiltonian is given by,
\begin{eqnarray}
\tilde{H}=\frac{1}{2m}\tilde{q_i}^2+\frac{k}{2}\tilde{x_i}^2.
\end{eqnarray}
It is interesting to note that the equations of motion are form invariant i. e. improved variables satisfy the same equations of motion (\ref{x}) and (\ref{q}). This is just a result of the form invariance of the Hamiltonian and the algebra among the basic variables.

In the enlarged space different gauge conditions can be chosen to show the different types of noncommutative structures. For example in the unitary gauge
\begin{eqnarray}
\Psi^3_i=\phi^1_i\approx 0, \ \Psi^4_i=\phi^2_i\approx 0
\end{eqnarray}
we get back the original physical subspace with the algebra (\ref{dirac}).

Next, we choose gauge condition such that $\{x_i,q_j\}_{DB}=\delta_{ij}$ in which case these variables may be regarded as canonical pairs. In one gauge we obtain noncommuting momenta while in the other, noncommuting coordinates are found. Let us choose the gauge conditions,
\begin{eqnarray}
\Psi^3_i&=&sx_i+q_i-A\phi^1_i+A\sqrt{D/C}\epsilon_{ij}\phi^1_j-A\sqrt{C/D}\phi^2_i+A\epsilon_{ij}\phi^2_j\approx 0\\
\Psi^4_i&=&x_i+(l/2A+\sqrt{CD})\phi^1_i-D\epsilon_{ij}\phi^1_j \nonumber\\
&+&C\phi^2_i+(l/2A-\sqrt{CD})\epsilon_{ij}\phi^2_j+l\epsilon_{ij}x_j\approx 0
\end{eqnarray}
where $A,C$ and $D$ are given by the expressions (\ref{sC}) and 
\begin{eqnarray}
&&l=\sqrt{-\frac{m\beta}{k\lambda}}\left(\frac{\omega}{k\lambda}-\sqrt{-\frac{m}{\lambda B}}\right)^{-1}\\
&&s=-\sqrt{\frac{kB}{\beta}}-\sqrt{-\frac{mk\lambda}{\beta}}.
\end{eqnarray}
For this choice we get the Dirac algebra 
\begin{equation}
\{x_i,x_j\}_{DB}=0, \ \{x_i,q_j\}_{DB}=\delta_{ij}, \ \{q_i,q_j\}_{DB}=B\epsilon_{ij}.
\label{mor1}
\end{equation}
which is the standard commutative Landau model algebra where the bracket among the momenta gives the magnetic field.

Alternatively, we choose the following gauge constraints
\begin{eqnarray}
\Psi^3_i&=&q_i-A\phi^1_i-(A\sqrt{D/C}+l/2C)\epsilon_{ij}\phi^1_j \nonumber\\
&+&(A\sqrt{C/D}-l/2D)\phi^2_i+A\epsilon_{ij}\phi^2_j+l\epsilon_{ij}q_j\approx 0\\
\Psi^4_i&=&vx_i+q_i-v\sqrt{CD}\phi^1_i-Dv\epsilon_{ij}\phi^1_j+Cv\phi^2_i+v\sqrt{CD}\epsilon_{ij}\phi^2_j\approx 0
\end{eqnarray}
with the following values of the coefficients,
\begin{eqnarray}
&&v=\left(\sqrt{-\frac{\beta}{mk\lambda}}+\sqrt{\frac{B}{m^2k\lambda}}\right)^{-1}\\
&&l=\sqrt{-\frac{m\beta\lambda}{k}}\left(\sqrt{-\frac{m\beta}{B}}-1\right)^{-1}
\end{eqnarray}
to get the algebra
\begin{equation}
\{x_i,x_j\}_{DB}=\frac{B}{km}\epsilon_{ij}, \ \{x_i,q_j\}_{DB}=\delta_{ij}, \ \{q_i,q_j\}_{DB}=0.
\label{mor2}
\end{equation}
This is the noncommutative Landau model with usual momenta algebra. We thus conclude that the standard (commutative) and noncommutative Landau models are dual aspects of the same parent model.
\section{Alternative approach based on doublet structure }
The original model (\ref{430}) has two sets of variables. It is possible to express this by a doublet of models with an appropriate separation of variables. This doublet structure is basically the soldering formalism discussed in various papers \cite{wot,wotrabin}. Consider the Lagrangians 
\begin{eqnarray}
&&L_{+}=-\frac{1}{2}\epsilon_{ij}z_{i}\dot z_{j}-\frac{\omega_{+}}{2}z_i^2
\label{L+}\\
&&L_{-}=\frac{1}{2}\epsilon_{ij}y_{i}\dot y_{j}-\frac{\omega_{-}}{2}y_i^2
\label{L-}
\end{eqnarray}
with positive $\omega_+$ and $\omega_-$. They represent the motion of one dimensional (chiral) oscillators rotating in the clockwise and anticlockwise directions. Suitable combination of these chiral oscillators leads to a two-dimensional oscillator which has been studied in \cite{ra} in the context of Zeeman effect. Here our motivation is to define two variables $x_i$ and $q_i$ from the chiral oscillator variables $y_i$ and $z_i$ in such a way that $x_i$ and $q_i$ satisfy the correct equations of motion and algebras of the generalized Landau problem.

The equations of motion following from (\ref{L+}) and (\ref{L-}) are
\begin{eqnarray}
&&\dot z_i=\omega_{+}\epsilon_{ij}z_j
\label{z}\\
&&\dot y_i=-\omega_{-}\epsilon_{ij}y_j
\label{y}
\end{eqnarray}

To illustrate 't Hooft's mechanism\cite{c,a,aG,aGH} we start from \,(\ref{z}). The Hamiltonian which gives the equation of motion \,(\ref{z}) is
\begin{eqnarray}
H=(\omega_+\epsilon_{ij}z_j)p_i.
\end{eqnarray}
This can be checked easily using the equation of motion $\dot z_i=\{z_i,H\}$. Since this Hamiltonian has no lower bound we take a physical Hamiltonian $\rho_+$ which commutes with $H$.
\begin{eqnarray}
\rho_+=\frac{\omega_+}{2}z_i^2.
\label{ro1}
\end{eqnarray}
To reproduce \,(\ref{z}) basic algebras should be taken as
\begin{eqnarray}
\{z_i,z_j\}=\epsilon_{ij}
\label{shan1}
\end{eqnarray}
 so that
\begin{eqnarray}
\dot z_i=\{z_i,\rho_+\}=\omega_+\epsilon_{ij}z_j.
\end{eqnarray}
leads to the correct equation of motion.
Same calculation for the other equation of motion \,(\ref{y}) gives the physical Hamiltonian
\begin{eqnarray}
\rho_-=\frac{\omega_-}{2}y_i^2.
\label{ro2}
\end{eqnarray}
and the algebra 
\begin{eqnarray}
\{y_i,y_j\}=-\epsilon_{ij}
\label{shan2}
\end{eqnarray}

Now in order to calculate the Lagrangian from the relation (\ref{hooftkl}), we construct the $\Lambda$ matrix for (\ref{shan1})
\begin{eqnarray}
\Lambda_{ij}=\{z_i,z_j\}=\epsilon_{ij}\nonumber
\end{eqnarray}
and its inverse is
\begin{eqnarray}
\Lambda^{ij}=-\epsilon_{ij}\nonumber.
\end{eqnarray}
So the Lagrangian is
\begin{eqnarray}
L_+&=&\frac{1}{2}z_i\Lambda^{ij}\dot z_j-\frac{\omega_+}{2}z_i^2\nonumber\\
 &=&-\frac{1}{2}\epsilon_{ij}z_i\dot z_j-\frac{\omega_+}{2}z_i^2.
\end{eqnarray}
which is our initial expression for chiral oscillator Lagrangian \,(\ref{L+}).
\subsection{Mapping between the equations of motion}
We make an ansatz
\begin{equation}
x_i=az_i+b\epsilon_{ij}z_j+cy_i+d\epsilon_{ij}y_j.
\label{xi}
\end{equation}
Now using (\ref{x}), \,(\ref{z}) and \,(\ref{y}) one can write $q_i$ in terms of $y_i$ and $z_i$
\begin{equation}
q_i=\frac{1}{\alpha}(\beta-\omega_+)(bz_i-a\epsilon_{ij}z_j)+\frac{1}{\alpha}(\beta+\omega_-)(dy_i-c\epsilon_{ij}y_j).
\label{qi}
\end{equation}
Taking the time derivative of the above equation and using (\ref{z}) and (\ref{y}) we get,
\begin{equation}
\dot q_i=\frac{\omega_+}{\alpha}(\beta-\omega_+)(az_i+b\epsilon_{ij}z_j)-\frac{\omega_-}{\alpha}(\beta+\omega_-)(cy_i+d\epsilon_{ij}y_j).
\label{qi1}
\end{equation}
Again using (\ref{xi}) and \,(\ref{qi}) in (\ref{q}) we obtain
\begin{equation}
\dot q_i=az_i\{\omega+\frac{\lambda}{\alpha}(\beta-\omega_+)\}+b\epsilon_{ij}z_j\{\omega+\frac{\lambda}{\alpha}(\beta-\omega_+)\}+cy_i\{\omega+\frac{\lambda}{\alpha}(\beta+\omega_-)\}+d\epsilon_{ij}y_j\{\omega+\frac{\lambda}{\alpha}(\beta+\omega_-)\}.
\label{qi2}
\end{equation}
So consistency between \,(\ref{qi1}) and \,(\ref{qi2}) demands
\begin{eqnarray}
&&\beta=-\lambda+(\omega_+-\omega_-)
\label{bl}\\
&&\omega=\frac{1}{\alpha}(\lambda+\omega_-)(\lambda-\omega_+).
\label{1000}
\end{eqnarray}
From the above two equations, using (\ref{rb1}), (\ref{rb2}) and (\ref{hooft}) we can show
\begin{eqnarray}
B=m(\omega_+-\omega_-), \ k=m\omega_+\omega_-.
\label{kB}
\end{eqnarray}
This important result shows that magnetic field appears as the difference whereas the spring constant is a product of the chiral frequencies.
\subsection{Mapping between the algebra}
Using the definitions of $x_i$ and $q_i$ from (\ref{xi}) and \,(\ref{qi}) we get
\begin{eqnarray}
&&\{x_i,x_j\}=(a^2+b^2-c^2-d^2)\epsilon_{ij}=\frac{\beta}{k} \epsilon_{ij}\\
&&\{q_i,q_j\}=\frac{1}{\alpha^2}\{(\beta-\omega_+)^2(a^2+b^2)-(\beta+\omega_-)^2(c^2+d^2)\}\epsilon_{ij}=m\lambda \epsilon_{ij}\\
&&\{x_i,q_j\}=\frac{1}{\alpha}\{-(\beta-\omega_+)(a^2+b^2)+(\beta+\omega_-)(c^2+d^2)\}\delta_{ij}=-\frac{\omega}{k}\delta_{ij}
\end{eqnarray}
where we have used (\ref{shan1},\ref{shan2}) and consistency with the algebra (\ref{chcm}).

The above three equations are not independent. From the last two equations, using (\ref{bl}) and (\ref{1000}) one can obtain the following relations
\begin{eqnarray}
a^2+b^2=\frac{\omega_+(\omega_+-\lambda)}{k(\omega_++\omega_-)}, \ c^2+d^2=\frac{\omega_-(\omega_-+\lambda)}{k(\omega_++\omega_-)}.
\label{c,d}
\end{eqnarray}
These pair of equations give the expressions so that variables of the generalized Landau problem can be defined in terms of the chiral variables with the help of (\ref{xi}) and \,(\ref{qi}). The interesting point is that the coefficients $a,b,c$ and $d$ are not completely determined. Different choices subject to (\ref{c,d}) can be made which exactly reproduce the results for different gauge fixings. 
\section{Special cases}
We note that (\ref{rb1}), (\ref{rb2}) and (\ref{hooft}) already give severe restrictions on the parameters $\alpha,\beta,\omega$ and $\lambda$. In order to give them specific values we set $\omega=-k$ so that $\{x_i,q_j\}=\delta_{ij}$. Now (\ref{hooft}) implies that this choice of $\omega$ fixes the value of $\alpha$ as $\alpha=\frac{1}{m}$. Using these values of $\omega$ and $\alpha$ we get $\beta\lambda=0$ from (\ref{rb2}). That means either $\beta$ or $\lambda$ is zero. In the following two subsections these situations are studied separately.
\subsection{Case 1}
We consider $\beta=0$ first. Then from (\ref{rb1}) $\lambda=\frac{B}{m}$.
Let us now collect the special values of the parameters mentioned so far
\begin{eqnarray}
\alpha=\frac{1}{m}, \ \beta=0, \ \omega=-k, \ \lambda=\frac{B}{m}
\label{ph1}
\end{eqnarray}
For the parameters (\ref{ph1}), the basic brackets following from (\ref{chcm}) are given by, 
\begin{equation}
\{x_i,x_j\}=0, \ \{x_i,q_j\}=\delta_{ij}, \ \{q_i,q_j\}=B\epsilon_{ij}.
\label{new}
\end{equation}
This structure is same as (\ref{mor1}) and corresponds to the conventional Landau algebra.

Now to find the connection with the chiral oscillator problem, we take (\ref{bl}) and (\ref{c,d}). From (\ref{bl}) we can choose either $\beta$ or $\lambda$ independently. We make the choice $\beta=0$. This implies that $\lambda$ is fixed by the relation,
\begin{equation}
\lambda=\omega_+-\omega_-.
\label{re1}
\end{equation}
Using the above equation we get from (\ref{c,d}),
\begin{equation}
a^2+b^2=c^2+d^2=\frac{\omega_+\omega_-}{k(\omega_++\omega_-)}.
\label{abcd}
\end{equation}
Again as mentioned earlier $a,b,c$ and $d$ are not uniquely determined by (\ref{abcd}). Different choices can be made. One can take the symmetrical combination where $a,b,c$ and $d$ are all equal. But to proceed further we make the following asymmetrical choice
\begin{eqnarray}
&&b=d=0 \ \textrm{and}\\
&&a=c=\left(\frac{\omega_+\omega_-}{k(\omega_++\omega_-)}\right)^{1/2}=\chi(\textrm{say})\nonumber
\end{eqnarray}
so that (\ref{xi}) and\,(\ref{qi}) imply
\begin{eqnarray}
&&x_i=\chi(z_i+y_i)
\label{43}\\
&&q_i=m\chi\epsilon_{ij}(\omega_+z_j-\omega_-y_j).
\end{eqnarray}
Now using (\ref{shan1},\ref{shan2}) the basic brackets are easy to calculate
\begin{eqnarray}
&&\{x_i,x_j\}=0
\label{sag1}\\
&&\{x_i,q_j\}=\frac{m\omega_+\omega_-}{k}\delta_{ij}
\label{sag2}\\
&&\{q_i,q_j\}=\frac{m\omega_+\omega_-}{k}m(\omega_+-\omega_-)\epsilon_{ij}.
\label{sag3}
\end{eqnarray}
This algebra is compatible with \,(\ref{new}) under the identifications (\ref{kB}). Thus we see, transforming our second order system to a first order one by introducing an additional variable, noncommutativity is naturally induced. Again this result is reproduced by superposition of two chiral oscillators, which are also first order systems. Since the difference in the chiral frequencies is proportional to the magnetic field the connection of two approaches gets established. 
\subsection{Case 2}
To show the situation where the momenta are commuting we take $\lambda=0$, then from (\ref{rb1}) $\beta=\frac{B}{m}$.
So all the values of parameters are listed below.
\begin{equation}
\alpha=\frac{1}{m}, \ \lambda=0, \ \omega=-k, \ \beta=\frac{B}{m}.
\label{49}
\end{equation}
In this case the basic brackets of the Landau problem following from the algebra \,(\ref{chcm}) are
\begin{equation}
\{x_i,x_j\}=\frac{B}{km}\epsilon_{ij}, \ \{x_i,q_j\}=\delta_{ij}, \ \{q_i,q_j\}=0.
\label{new1}
\end{equation}
Note that algebras given by (\ref{mor2}) and (\ref{new1}) are structurally equivalent.

Now we have to find the corresponding situation in the chiral oscillator problem. In the previous subsection $\beta$ was taken to be zero in (\ref{bl}). Now to generate commuting momenta $\lambda$ is set to be zero. 
Then we take the following asymmetrical choice of the coefficients from (\ref{c,d})
\begin{eqnarray}
&&b=d=0\nonumber \ \textrm{and}\\
&&a=\frac{\omega_+}{\sqrt{k(\omega_++\omega_-)}}, \ c=\frac{\omega_-}{\sqrt{k(\omega_++\omega_-)}}.\nonumber
\end{eqnarray}
Putting these values of the coefficients in (\ref{xi}) and (\ref{qi}) we 
observe that $x_i$ and $q_i$ are now defined by the relations
\begin{eqnarray}
&&x_i=az_i+cy_i\\
&&q_i=am\omega_-\epsilon_{ij}z_j-cm\omega_+\epsilon_{ij}y_j.
\label{am}
\end{eqnarray}
Using the algebra (\ref{shan1},\ref{shan2}) it is easy to show that they satisfy the following algebra
\begin{eqnarray}
&&\{x_i,x_j\}=\frac{m(\omega_+-\omega_-)}{mk}\epsilon_{ij}
\label{bl1}\\
&&\{x_i,q_j\}=\frac{m\omega_+\omega_-}{k}\delta_{ij}
\label{bl2}\\
&&\{q_i,q_j\}=0.
\label{bl3}
\end{eqnarray}
We note that above algebra and (\ref{new1}) also match under the same identifications (\ref{kB}).

In section 2.2 we saw from Batalin--Tyutin extended space framework of generalized Landau problem how the usual (commutative) and noncommutative Landau models were related by gauge transformations. Now we have discussed an alternative approach where the general problem is expressed, through certain parameters, by a doublet structure. The connection between two formulation is that different parametric choices correspond to distinct gauge fixings in the extended space approach.
\section{Construction of Lagrangian}
 The motivation for doing this calculation is to establish an equivalence between the generalized Landau problem and chiral oscillators at the Lagrangian level.
\subsection{Case 1}
For the values of parameters given in (\ref{ph1}) we get the Lagrangian from (\ref{430})
\begin{eqnarray}
L=q_i\dot x_i+\frac{B}{2}\epsilon_{ij}x_i\dot x_j-\frac{1}{2}\left(\frac{q_i^2}{m}+kx_i^2\right).
\end{eqnarray}
 Since $q_i$ is an auxiliary variable, it can be eliminated using its equation of motion $q_i=m\dot x_i$ 
to yield,

\begin{equation}
L=\frac{m}{2}\dot x_i^2+\frac{B}{2}\epsilon_{ij}x_i\dot x_j-\frac{1}{2}kx_i^2.
\label{lag1}
\end{equation}
Now we have to find the composite (soldered) Lagrangian\cite{c,wot,ra} for the chiral oscillators
\begin{eqnarray}
L=L_++L_-=-\frac{1}{2}\epsilon_{ij}z_i\dot z_j-\frac{\omega_+}{2}z_i^2+\frac{1}{2}\epsilon_{ij}y_i\dot y_j-\frac{\omega_-}{2}y_i^2.
\label{59}
\end{eqnarray}

We take $x_i$ as defined in (\ref{43}) and write $y_i$ in terms $x_i$ and $z_i$
\begin{eqnarray}
y_i=\frac{1}{\chi}(x_i-\chi z_i).
\end{eqnarray}

We substitute the expression of $y_i$ in the Lagrangian (\ref{59}) to get
\begin{equation}
L=-\frac{\omega_+}{2}z_i^2-\frac{\omega_-}{2\chi ^2}(x_i-\chi z_i)^2+\frac{1}{2\chi ^2}\epsilon_{ij}(x_i\dot x_j-2\chi z_i\dot x_j).
\label{6226}
\end{equation}
Since $z_i$ is an auxiliary variable it can be eliminated from the above Lagrangian using its equation of motion to yield,
\begin{eqnarray}
L=\frac{1}{2}m\dot x_i^2+\frac{1}{2}m(\omega_+-\omega_-)\epsilon_{ij}x_i\dot x_j-\frac{1}{2}m\omega_+\omega_-x_i^2.
\label{lag2}
\end{eqnarray}
We observe that the Lagrangians \,(\ref{lag2}) and\,(\ref{lag1}) are same under the previous identification \,(\ref{kB}).
\subsection{Case 2}
When momenta are commuting variables, using (\ref{49}) we have from (\ref{430})
\begin{eqnarray}
L=-x_i\dot q_i+\frac{B}{2mk}\epsilon_{ij}q_i\dot q_j-\frac{1}{2}\left(\frac{q_i^2}{m}+kx_i^2\right).
\label{l.}
\end{eqnarray}
Now we can eliminate $x_i$ from (\ref{l.}) using its equation of motion to get
\begin{eqnarray}
L=\frac{1}{2k}\dot q_i^2+\frac{B}{2km}\epsilon_{ij}q_i\dot q_j-\frac{1}{2m}q_i^2.
\label{la1}
\end{eqnarray}
Following the method of previous subsection we can calculate the composite Lagrangian from (\ref{am}) and (\ref{59})

\begin{eqnarray}
L=\frac{1}{2m\omega_+\omega_-}\dot q_i^2+\frac{m(\omega_+-\omega_-)}{2m(m\omega_+\omega_-)}\epsilon_{ij}q_i\dot q_j-\frac{1}{2m}q_i^2.
\label{la2}
\end{eqnarray}
We note that Lagrangians \,(\ref{la2}) and \,(\ref{la1}) are again identical with the mapping (\ref{kB}). 
\section{Discussion}
Batalin--Tyutin extended space framework of generalized Landau problem clearly shows the dual nature of different types of noncommutativity. The usual (commutative) and noncommutative Landau models are obtained as different gauge fixed versions of a parent gauge invariant model.

 Moreover chiral oscillators and generalized Landau problem are related very closely. Here we have established a mapping between them. It is interesting that in the generalized Landau problem noncommutativity may appear in position variables or in momenta or in both. By suitably defining variables from chiral coordinates each type of noncommutativity can be reproduced. The important point is that this mapping exists not only at the algebraic level but also at the Lagrangian level. In the later case, we have shown the important role played by soldering method\cite{c,wot,ra} to understand the mapping properly.

\chapter{\label{chap:well}Noncommutative gravitational quantum well}

In the previous chapter we have discussed the generalized Landau problem where the shifting of noncommutativity was shown from the coordinates to the momenta and vice-versa. Also, the implications of noncommutativity in both phase space and configuration space variables were discussed. In this chapter we study the phenomenology of a quantum mechanical model with constant noncommutativity in both coordinates and momenta.

 The energy eigenstates of a particle confined in a potential well have been calculated for different force fields. Various experiments have been done for the atomic and nuclei models to observe the effects of electromagnetic and strong forces on elementary particle but performing a quantum mechanical experiment in the gravitational field is extremely difficult due to its weak nature compared to other force fields. Few years back, Nesvizhevsky {\it et. al.}\cite{nes,vvnes,nes1} completed an experiment at Grenoble and observed the first few energy states for the gravitational well. In order to build a potential well gravitational field alone is not sufficient since it forces a particle to fall along field lines. They put a reflecting plane beneath to confine the particle in a finite region of space. The neutron was used as the quantum particle since it is charge less and hence indifferent to the electromagnetic noise. The experiment was found to be in excellent agreement with the theoretical computations.

  Here we set the model in a planar noncommutative phase space background to determine the effects of noncommutativity on the energy spectrum. We use the WKB approximation to analyze the problem analytically. Finally the experimental findings of \cite{nes,vvnes,nes1} are used to put an upper bound on the noncommutative parameters. The numerical analysis of a similar model was given in \cite{ba}. In that paper the model was defined on the commutative space and exploiting an inverse phase space transformation it was expressed in terms of the noncommuting variables. A perturbative expansion of the Hamiltonian was then carried out to find the energy correction numerically. Here our results are obtained by using analytical computations and these agree with \cite{ba}. The study of this problem from other perspectives may be found in \cite{zhang,Ani}

 In section 3.1 we define the noncommutative space and give a general phase space transformation to connect the noncommutative space variables and the commutative counterparts. In the next section we discuss the position space and momentum space representation of the noncommutative algebra. In section 3.3 the quantum gravitational well is introduced. After summarizing the theoretical and experimental\cite{nes1} results of the energy spectrum in usual commutative space, we define the corresponding Hamiltonian in noncommutative space. The structure of this Hamiltonian is explicitly obtained in both noncommutative and commutative descriptions, leading to completely equivalent results. In section 3.4 the energy spectrum is computed by using the WKB approximation. An upper bound on the noncommutative parameter is derived by comparing with the recent experimental results given in \cite{nes1}. Finally some remarks are given in section 3.5.
\section{Noncommutative phase space}
We consider a planar phase space, where the standard Heisenberg algebra is obeyed
\begin{eqnarray}
\begin{array}{rcl}
&&[\hat x_i,\hat x_j]=0\\
&&[\hat p_i,\hat p_j]=0\\
&&[\hat x_i,\hat p_j]=i\hbar\delta_{ij}.
\end{array}
\label{1}
\end{eqnarray}
This algebra is invariant under the following symmetry transformation
\begin{eqnarray}
\begin{array}{rcl}
&&\hat x_i\rightarrow \hat p_i\\
&&\hat p_i\rightarrow \hat x_i\\
&&i\rightarrow -i.
\end{array}
\label{sym}
\end{eqnarray}

Now we define two operators $\hat y$ and $\hat q$ in the following way
\begin{eqnarray}
&&\hat y_i=\hat x_i+\alpha_1\epsilon_{ij}\hat p_j+\alpha_2\epsilon_{ij}\hat x_j
\label{100}\\
&&\hat q_i=\hat p_i+\beta_1\epsilon_{ij}\hat x_j+\beta_2\epsilon_{ij}\hat p_j
\label{200}
\end{eqnarray}
where $\alpha, \ \beta$ are arbitrary constants. Here we enforce a symmetry leading to $\hat y_i\rightarrow \hat q_i$ and $\hat q_i\rightarrow \hat y_i$ under the transformation (\ref{sym}). Clearly this is possible if we introduce the following transformation
\begin{eqnarray}
&&\alpha_i\rightarrow \beta_i\\
&&\beta_i\rightarrow \alpha_i.
\end{eqnarray}
Thus a symmetry transformation, analogous to (\ref{sym}), in the modified $y-q$ plane is given by
\begin{eqnarray}
\begin{array}{rcl}
&&\hat y_i\rightarrow \hat q_i\\
&&\hat q_i\rightarrow \hat y_i\\ 
&&\alpha_i\rightarrow \beta_i\\
&&\beta_i\rightarrow \alpha_i\\
&&i\rightarrow -i.
\end{array}
\label{sym1}
\end{eqnarray}
Using (\ref{1}) we can show that the new coordinates $\hat y$ and momenta $\hat q$ satisfy the algebra
\begin{eqnarray}
&&[\hat y_i,\hat y_j]=-2i\hbar\alpha_1\epsilon_{ij}\\
&&[\hat q_i,\hat q_j]=2i\hbar\beta_1\epsilon_{ij}\\
&&[\hat y_i,\hat q_j]=i\hbar(1+\alpha_2\beta_2-\alpha_1\beta_1)\delta_{ij}+i\hbar(\alpha_2-\beta_2)\epsilon_{ij}.
\end{eqnarray}
Under the symmetry transformation (\ref{sym1}) the above algebra is invariant. 

So far we did not associate any specific values to the coefficients $\alpha$ and $\beta$'s. Now if we set
\begin{eqnarray*}
&&\alpha_1=-\frac{\theta}{2\hbar}\\
&&\beta_1=\frac{\eta}{2\hbar}\\
&&\alpha_2=\beta_2=0
\end{eqnarray*}
we obtain
\begin{eqnarray}
\begin{array}{rcl}
&&[\hat y_i,\hat y_j]=i\theta\epsilon_{ij}\\
&&[\hat q_i,\hat q_j]=i\eta\epsilon_{ij}\\
&&[\hat y_i,\hat q_j]=i(1+\frac{\theta\eta}{4\hbar^2})\hbar\delta_{ij}=i\hbar_{eff}\delta_{ij}
\end{array}
\label{3}
\end{eqnarray}
which reproduces the noncommutative structure given in \cite{ba}. The term $\frac{\theta\eta}{4\hbar^2}$ is interpreted\cite{ba} as a correction to the Planck constant. However by taking the following values of $\alpha$ and $\beta$ one can keep the Planck constant unchanged:
\begin{eqnarray}
\begin{array}{rcl}
&&\alpha_1=-\frac{\theta}{2\hbar}\\
&&\beta_1=\frac{\eta}{2\hbar}\\
&&\alpha_2=\beta_2=\frac{1}{2\hbar}\sqrt{-\theta\eta}
\end{array}
\label{param}
\end{eqnarray}
which yields the noncommutative algebra
\begin{eqnarray}
\begin{array}{rcl}
&&[\hat y_i,\hat y_j]=i\theta\epsilon_{ij}\\
&&[\hat q_i,\hat q_j]=i\eta\epsilon_{ij}\\
&&[\hat y_i,\hat q_j]=i\hbar\delta_{ij}
\end{array}
\label{4}
\end{eqnarray}
so that the Planck constant is not modified. Physical applications of this type of noncommutative algebra may be found in \cite{c}.

 The inverse phase space transformation is given by
\begin{eqnarray}
\begin{array}{rcl}
&&\hat x_i=A\hat y_i+B\epsilon_{ij}\hat y_j+C\hat q_i+D\epsilon_{ij}\hat q_j\\
&&\hat p_i=E\hat y_i+F\epsilon_{ij}\hat y_j+A\hat q_i+B\epsilon_{ij}\hat q_j
\end{array}
\label{5}
\end{eqnarray}
where
\begin{eqnarray}
&&A=\frac{2\hbar^2-\theta\eta}{2(\hbar^2-\theta\eta)}, \  \  \  \  \ B=-\frac{\hbar\sqrt{-\theta\eta}}{2(\hbar^2-\theta\eta)}\nonumber\\
&&C=\frac{\theta\sqrt{-\theta\eta}}{2(\hbar^2-\theta\eta)}, \  \  \  \  \ D=\frac{\theta\hbar}{2(\hbar^2-\theta\eta)}
\label{A}\\
&&E=-\frac{\eta\sqrt{-\theta\eta}}{2(\hbar^2-\theta\eta)}, \  \  \  \  \ F=-\frac{\hbar\eta}{2(\hbar^2-\theta\eta)}.\nonumber
\end{eqnarray}
Observe that $\theta$ and $\eta$ must have different signs so that the various coefficients are real and well defined which guarantees the hermitian nature of physical operators $\hat x, \hat p$ and $\hat y,\hat q$.
\section{Representation of the algebra}
The differential representation of the Heisenberg algebra is easy to find. In a coordinate space representation, $\hat x_i$ are diagonal and $\hat p_i=-i\hbar\frac{\partial}{\partial x_i}$ and in a momentum space description, the momenta $\hat p_i$ are diagonal while $\hat x_i=i\hbar\frac{\partial}{\partial p_i}$. In order to find a differential representation of the noncommutative algebra (\ref{4}), we consider a general representation of the form
\begin{eqnarray}
\hat y_i\rightarrow \hat y_i, \  \  \ \hat q_i\rightarrow -ia\hbar\frac{\partial}{\partial \hat y_i}+b\hbar\epsilon_{ij}\frac{\partial}{\partial \hat y_j}+c\frac{\eta}{\hbar} \hat y_i+d\frac{\eta}{\hbar}\epsilon_{ij}\hat y_j
\end{eqnarray}
where $a, b, c$ and $d$ are dimensionless constants. Now consistency with algebra (\ref{4}) demands 
\begin{eqnarray}
a+d\frac{\eta\theta}{\hbar^2}=1, \  \  \ b+ic\frac{\eta\theta}{\hbar^2}=0, \  \  \ iad+2bc+i\frac{\theta\eta}{\hbar^2}(c^2+d^2)=i.
\end{eqnarray}
Since three equations are not sufficient to fix all the parameters, we find the solutions in terms of $a$ to obtain the following representation of the phase space variables
\begin{eqnarray}
\hat y_i&\rightarrow& \hat y_i\nonumber\\
\hat q_i&\rightarrow& -ia\hbar\frac{\partial}{\partial \hat y_i}\mp i\hbar\sqrt{1-a^2-\frac{\eta\theta}{\hbar^2}}\epsilon_{ij}\frac{\partial}{\partial \hat y_j}
\label{10}\\
&&\pm\frac{\sqrt{1-a^2-\frac{\eta\theta}{\hbar^2}}}{\theta}\hbar \hat y_i+\frac{1-a}{\theta}\epsilon_{ij}\hbar \hat y_j\nonumber
\end{eqnarray}
This representation should have a smooth commutative limit when $(\theta,\eta)\rightarrow 0$. The natural choice $a=1$ does not satisfy this condition. On the other hand if we take $a=\sqrt{1-\frac{\eta\theta}{\hbar^2}}$ then the representation,
\begin{eqnarray}
\hat q_i&\rightarrow& -i\hbar\sqrt{1-\frac{\eta\theta}{\hbar^2}}\frac{\partial}{\partial \hat y_i}+\frac{1-\sqrt{1-\frac{\eta\theta}{\hbar^2}}}{\theta}\hbar\epsilon_{ij}\hat y_j
\label{re}
\end{eqnarray}
has a smooth limit, which is
\begin{eqnarray}
\lim_{\eta\rightarrow 0}\lim_{\theta\rightarrow 0} \hat q_i=-i\hbar\frac{\partial}{\partial \hat y_i}=\lim_{\theta\rightarrow 0}\lim_{\eta\rightarrow 0} \hat q_i
\end{eqnarray}

Noting that the algebra (\ref{4}) is invariant under the transformation $(\hat y,\hat q,\theta,\eta)\rightarrow (\hat q,-\hat y,\eta,\theta)$, we can make this transformation in (\ref{re}) to get the momentum space representation.
\section{Gravitational well}
Before discussing the problem in the noncommutative space setting we first study it in usual commutative space. We consider a two dimensional plane where a particle of mass $m$ is subjected to the Earth's gravitational field in one direction; the vertical, taken to be described by the coordinate $x_1$. We assume that the gravitational acceleration $g$ is constant near the surface of the earth. The commutative Hamiltonian is given by
\begin{eqnarray}
\hat H=\frac{1}{2m}(\hat p_1^2+\hat p_2^2)+mg\hat x_1.
\label{ha}
\end{eqnarray}
Since the particle is free in the $x_2$ direction, its energy spectrum is continuous in that direction and the wave function can be written as
\begin{eqnarray}
\psi(x_2)=\int g(k)e^{ikx_2}dk.
\end{eqnarray}
In the other direction the wave function is the well known Airy function $\phi(\xi)$ with appropriate normalization\cite{lan},
\begin{eqnarray}
\psi_n(x_1)=A_n\phi(\xi)  \  ; \   \xi=\left(\frac{2m^2g}{\hbar^2}\right)^{\frac{1}{3}}(x_1-\frac{E_n}{mg}).
\label{psi1}
\end{eqnarray}
The zeroes of the Airy function, $\beta_n$ give the energy eigenvalues
\begin{eqnarray}
E_n=-\left(\frac{mg^2\hbar^2}{2}\right)^{\frac{1}{3}}\beta_n \ ; \ n=1, \ 2, \ 3...
\label{energy}
\end{eqnarray}
Below the classical turning point $x_n=\frac{E_n}{mg}$ the wave function oscillates and above $x_n$ it decays exponentially. This was observed experimentally by Nesvizhevsky {\it et al.}\cite{nes,vvnes}. They used neutron as the quantum particle because of its charge neutrality and long life time ($\tau\simeq 885.7$s)\cite{par}. They allowed the particles to fall towards a horizontal mirror which, combined with the gravitational field forms the potential well. By placing an absorber above the mirror they allow a cold neutron beam to flow with a horizontal velocity $v_2=6.5{\textrm{ms}}^{-1}$. Then they measure the number of transmitted neutrons as a function of absorber height: this was shown to be a step like function which establishes the quantum nature of the problem.

The energy levels are also written below simply by the WKB approximation where the error is $\sim \ 1\%$ compared to the results derived from (\ref{energy}). 
\begin{eqnarray}
E_n&=&\left(\frac{9m}{8}[\pi\hbar g(n-\frac{1}{4})]^2\right)^{\frac{1}{3}}
\label{ener}\\
&=&\alpha_{n}g^{\frac{2}{3}} \ ; \ n=1, \ 2, \ 3...
\label{wkb}
\end{eqnarray}
where
\begin{eqnarray}
\alpha_n&=&\left(\frac{9m}{8}[\pi\hbar (n-\frac{1}{4})]^2\right)^{\frac{1}{3}}.
\end{eqnarray}
A summary of both theoretical and experimental results is given. Taking the values of constants as
\begin{eqnarray}
&&\hbar=\frac{1}{2\pi}({\textrm {Planck constant}})=10.59\times 10^{-35} \ {\textrm {Js}}\\
&&g={\textrm {gravitational acceleration}}=9.81 \ {\textrm {ms}}^{-2}\\
&&m={\textrm {mass of neutron}}=167.32\times 10^{-29} \ {\textrm {Kg}}
\end{eqnarray}
the first two energy levels found from (\ref{ener}) are,
\begin{eqnarray}
&&E_1=1.392 \ {\textrm {peV}}=2.23\times 10^{-31}{\textrm{J}}
\label{ener1}\\
&&E_2=2.447 \ {\textrm {peV}}=3.92\times 10^{-31}{\textrm{J}}.
\label{ener2}
\end{eqnarray}
From $E_1$ and $E_2$ the classical turning points are calculated to be
\begin{eqnarray}
&&x_1=\frac{E_1}{mg}=13.59 \mu{\textrm m}\\
&&x_2=\frac{E_1}{mg}=23.88 \mu{\textrm m}.
\end{eqnarray}
These are in reasonable agreement with the experimental results\cite{nes1}
\begin{eqnarray}
&&x_1^{\textrm {exp}}=12.2\pm1.8({\textrm {syst}})\pm0.7({\textrm {stat}}) \ (\mu{\textrm {m}})\\
&&x_2^{\textrm {exp}}=21.6\pm2.2({\textrm {syst}})\pm0.7({\textrm {stat}}) \ (\mu{\textrm {m}}).
\end{eqnarray}
Error bars for the above mentioned energy levels are
\begin{eqnarray}
&&\Delta E_1^{\textrm {exp}}=6.55\times 10^{-32} \ {\textrm{J}}=0.41 \ {\textrm {peV}},
\label{exp1}\\
&&\Delta E_2^{\textrm {exp}}=8.68\times 10^{-32} \ {\textrm{J}}=0.54 \ {\textrm {peV}}.
\label{exp2}
\end{eqnarray}
\subsection{Noncommutative space description}
In the noncommutative space (\ref{4}) the analogue of the Hamiltonian (\ref{ha}) is defined as
\begin{eqnarray}
\hat H=\frac{1}{2m}(\hat q_1^2+\hat q_2^2)+mg\hat y_1
\label{label}
\end{eqnarray}
To find the spectrum, two approaches are possible. One can directly work in the noncommutative space variables or use the phase space transformations to reduce the problem on the usual commutative space. We first discuss the second approach. Using the maps (\ref{100},\ref{200}) together with the parametrization (\ref{param}), we find,
\begin{eqnarray}
\hat H&=&\frac{1}{2m}(\hat p_1^2+\hat p_2^2)+mg\hat x_1+\frac{\eta}{2m\hbar}\epsilon_{ij}\hat p_i\hat x_j+mg(-\frac{\theta}{2\hbar}\hat p_2+\frac{\sqrt{-\theta\eta}}{2\hbar}\hat x_2)\nonumber\\
&&+\frac{\eta^2}{8m\hbar^2}(\hat x_1^2+\hat x_2^2)+\frac{\eta\sqrt{-\theta\eta}}{8m\hbar^2}(\hat x_i\hat p_i+\hat p_i\hat x_i)-\frac{\theta\eta}{8m\hbar^2}(\hat p_1^2+\hat p_2^2).
\label{kunal}
\end{eqnarray}
Defining a new constant
\begin{eqnarray}
\gamma=\frac{2\hbar\theta}{4\hbar^2-\theta\eta}
\end{eqnarray}
and a new variable
\begin{eqnarray}
\bar{\hat p}_2=\hat p_2-m^2g\gamma
\end{eqnarray}
we can write the above Hamiltonian in the form
\begin{eqnarray}
\hat H&=&\frac{1}{2m}(1-\frac{\theta\eta}{4\hbar^2})(\hat p_1^2+\bar{\hat p}_2^2)+\frac{\eta^2}{8m\hbar^2}(\hat x_1^2+\hat x_2^2)+\frac{\eta}{2m\hbar}(\hat p_1\hat x_2-\bar{\hat p}_2\hat x_1)\nonumber\\
&&+\frac{\eta\sqrt{-\theta\eta}}{8m\hbar^2}(\hat x_1\hat p_1+\hat x_2\bar{\hat p}_2+\hat p_1\hat x_1+\bar{\hat p}_2\hat x_2)\nonumber\\
&&+mg\{(1-\frac{\eta\gamma}{2\hbar})\hat x_1+\frac{\sqrt{-\theta\eta}}{2\hbar}
(1+\frac{\gamma\eta}{2\hbar})\hat x_2\}-\frac{m^3g^2\theta^2}{2(4\hbar^2-\theta\eta)}.
\label{newh}
\end{eqnarray}
Since the difference between $\bar{\hat p}_2$ and $\hat p_2$ is just a constant, they satisfy the same commutation relations. The eigenvalues of $\bar{\hat p}_2$ are translated by an equal amount vis a vis those for $\hat p_2$ and hence these are not distinguished. Also neglecting the additive constant in the Hamiltonian (\ref{newh}) we get
\begin{eqnarray}
\hat H&=&\frac{1}{2m}(1-\frac{\theta\eta}{4\hbar^2})(\hat p_1^2+\hat p_2^2)+\frac{\eta\sqrt{-\theta\eta}}{8m\hbar^2}(\hat x_i\hat p_i+\hat p_i\hat x_i)+\frac{\eta}{2m\hbar}\epsilon_{ij}\hat p_i\hat x_j\nonumber\\
&&+\frac{\eta^2}{8m\hbar^2}(\hat x_1^2+\hat x_2^2)+mg\{(1-\frac{\eta\gamma}{2\hbar})\hat x_1+\frac{\sqrt{-\theta\eta}}{2\hbar}
(1+\frac{\gamma\eta}{2\hbar})\hat x_2\}.
\end{eqnarray}
Now we define a modified gravitational acceleration $\tilde{g}$ in the following way
\begin{eqnarray}
&&m\tilde{g} \ {\textrm {cos}}\epsilon=mg(1-\frac{\eta\gamma}{2\hbar})\\
&&m\tilde{g} \ {\textrm {sin}}\epsilon=mg\frac{\sqrt{-\theta\eta}}{2\hbar}
(1+\frac{\gamma\eta}{2\hbar}).
\end{eqnarray}
The tilting angle with the $x_1$ axis is given by
\begin{eqnarray}
\epsilon={\textrm{tan}}^{-1}\frac{\sqrt{-\theta\eta}}{2\hbar}\left(\frac{2\hbar+\eta\gamma}{2\hbar-\eta\gamma}\right)
\label{ang}
\end{eqnarray}
while, 
\begin{eqnarray}
\tilde{g}=g\{(1-\frac{\eta\gamma}{2\hbar})^2-\frac{\theta\eta}{4\hbar^2}(1+\frac{\gamma\eta}{2\hbar})^2\}^{\frac{1}{2}}
\label{X}
\end{eqnarray}
Since the product $\theta\eta$ is negative, $\tilde{g}$ is always positive definite. Now we rotate in the $x_1-x_2$ plane by an angle $\epsilon$, so that the coordinate of a point in the rotated frame is given by
\begin{eqnarray}
\begin{array}{rcl}
&&x_{1}'={\textrm {cos}}\epsilon \  x_{1}+{\textrm {sin}}\epsilon  \ x_{2}\\
&&x_{2}'={\textrm {cos}}\epsilon \  x_{2}-{\textrm {sin}}\epsilon  \  x_{1}.
\end{array}
\label{i}
\end{eqnarray}
Correspondingly, the momenta are transformed :
\begin{eqnarray}
\begin{array}{rcl}
&&p_{1}'={\textrm {cos}}\epsilon \  p_{1}+{\textrm {sin}}\epsilon  \ p_{2}\\
&&p_{2}'={\textrm {cos}}\epsilon \  p_{2}-{\textrm {sin}}\epsilon \  p_{1}.
\end{array}
\label{ii}
\end{eqnarray}
Using (\ref{i}) and (\ref{ii}) it is easy to show that
\begin{eqnarray}
&&p_{1}'^{2}+p_{2}'^{2}=p_{1}^{2}+p_{2}^{2}
\label{iii}\\
&&x_{1}'p_{2}'-x_{2}'p_{1}'=x_{1}p_{2}-x_{2}p_{1}
\label{iv}\\
&&x_{1}'^{2}+x_{2}'^{2}=x_{1}^{2}+x_{2}^{2}
\label{v}\\
&&x'_ip'_i+p'_ix'_i=x_ip_i+p_ix_i.
\label{vi}
\end{eqnarray}
We use the operator versions of (\ref{iii},\ref{iv},\ref{v},\ref{vi}) to write the noncommutative Hamiltonian in the rotated frame as,
\begin{eqnarray}
\hat H&=&\frac{1}{2m}(\hat p_1'^2+\hat p_2'^2)+m\tilde{g}x_1'+\frac{\eta}{2m\hbar}\epsilon_{ij}\hat p_i'\hat x_j'+\frac{\eta^2}{8m\hbar^2}(\hat x_1'^2+\hat x_2'^2)\nonumber\\
&&+\frac{\eta\sqrt{-\theta\eta}}{8m\hbar^2}(\hat x_i'\hat p_i'+\hat p_i'\hat x_i')
-\frac{\theta\eta}{8m\hbar^2}(\hat p_1'^2+\hat p_2'^2).
\label{223}
\end{eqnarray}
The primed and unprimed variables satisfy the same algebra; henceforth the primes are all dropped. Then we can identify the first three terms of the Hamiltonian (\ref{223}) exactly as the commutative Hamiltonian given in (\ref{ha}). This should be considered as the unperturbed Hamiltonian. The term $\frac{\eta}{2m\hbar}\epsilon_{ij}\hat p_i\hat x_j$ is effectively a Landau problem like term, where a magnetic field is present perpendicular to the $x_1-x_2$ plane. The term $\frac{\eta^2}{8m\hbar^2}(\hat x_1^2+\hat x_2^2)$ is practically an oscillating potential.

Since the noncommutative effects are rather small we first confine to the leading order approximation in $\theta$ and $\eta$. Moreover (\ref{X}) shows that in the leading order,
\begin{eqnarray}
\tilde{g}=g[1+O(\theta\eta)].
\end{eqnarray}
Hence the Hamiltonian (\ref{223}) in the first order approximation is given by, 
\begin{eqnarray}
\hat H&=&\frac{1}{2m}(\hat p_1^2+\hat p_2^2)+mg\hat x_1-\frac{\eta}{2m\hbar}(\hat x_1\hat p_2-\hat x_2\hat p_1)\\
&=&\hat H_0-\frac{\eta}{2m\hbar}(\hat x_1\hat p_2-\hat x_2\hat p_1)
\label{hamil}
\end{eqnarray}
where $\hat H_0$ is nothing but the commutative Hamiltonian already given in (\ref{ha}). The energy spectrum pertaining to this Hamiltonian will be computed in section 3.4.
\subsection{Alternative formulation}
Here we analyze the structure of the Hamiltonian directly in terms of the noncommuting space variables. 

Using the representation (\ref{re}), this Hamiltonian can be written in the form
\begin{eqnarray}
\hat H&=&\frac{\hbar^2}{2m}\left[-(1-\theta\eta)\left(\frac{\partial^2}{\partial \hat y_1^2}+\frac{\partial^2}{\partial \hat y_2^2}\right)+\left(\frac{1-\sqrt{1-\frac{\theta\eta}{\hbar^2}}}{\theta}\right)^2(\hat y_1^2+\hat y_2^2)\right]\nonumber\\
&&+\frac{\hbar^2}{2m}\left[2i(\sqrt{1-\frac{\theta\eta}{\hbar^2}})\frac{1-\sqrt{1-\frac{\theta\eta}{\hbar^2}}}{\theta}\left(\hat y_1\frac{\partial}{\partial \hat y_2}-\hat y_2\frac{\partial}{\partial \hat y_1}\right)\right]+mg\hat y_1\nonumber
\end{eqnarray}
Making use of the formula $\left(1-a\right)^{1/2}=1-\frac{1}{2}a$ for small (compared to unity) $a$, we simplify the Hamiltonian to get
\begin{eqnarray}
\hat H&=&\frac{1}{2m}\left[-(1-\theta\eta)\hbar^2\left(\frac{\partial^2}{\partial \hat y_1^2}+\frac{\partial^2}{\partial \hat y_2^2}\right)+\left(\frac{\eta}{2\hbar}\right)^2(\hat y_1^2+\hat y_2^2)\right]\nonumber\\
&&+\frac{1}{2m}\left[2i(1-\frac{\theta\eta}{2\hbar^2})\frac{\eta}{2}\left(\hat y_1\frac{\partial}{\partial \hat y_2}-\hat y_2\frac{\partial}{\partial \hat y_1}\right)\right]+mg\hat y_1
\end{eqnarray}
Keeping terms only upto first order in the noncommutative parameters, this further reduces to
\begin{eqnarray}
\hat H&=&\frac{1}{2m}\left[-\hbar^2\left(\frac{\partial^2}{\partial \hat y_1^2}+\frac{\partial^2}{\partial \hat y_2^2}\right)+i\eta\left(\hat y_1\frac{\partial}{\partial \hat y_2}-\hat y_2\frac{\partial}{\partial \hat y_1}\right)\right]+mg\hat y_1.
\label{thes}
\end{eqnarray}
Since $\theta$ does not appear in the leading order expression of the Hamiltonian, we drop it from the algebra (\ref{4}). In that case $\hat y$ and $-i\hbar\frac{\partial}{\partial \hat y}$ are nothing but the canonical pairs of ordinary quantum mechanics and (\ref{thes}) is identified with (\ref{hamil}).
\section{Bounds on noncommutative parameters}
Here the energy spectrum is computed and therefrom bounds on the noncommutative parameters are determined. Consider the Hamiltonian (\ref{hamil}) in the first order approximation. Now the term proportional to $\eta$ in the above Hamiltonian can be treated perturbatively. The unperturbed part $\hat H_0$ is known to be exactly solvable in terms of Airy functions\cite{lan}. Furthermore, using the property that Airy function (or any bound state wave function vis a vis motion in the direction $x_1$) is real, it is easily seen that
\begin{eqnarray}
<\hat p_1>_n=\int_0^{+\infty}dx_1\psi_n^*(-i\hbar\frac{\partial}{\partial x_1}\psi_n)=0.
\end{eqnarray}
This can also be understood physically from the fact that, for a bound state system, the average current flow in a particular direction is zero. So effectively the Hamiltonian turns out to be
\begin{eqnarray}
\hat H=\hat H_0-\frac{\eta}{2m\hbar}\hat x_1\hat p_2.
\label{54}
\end{eqnarray}
In this way we see that, in the leading order, the noncommutative corrections are entirely encoded in the term
\begin{eqnarray}
\hat H_I=-\frac{\eta}{2m\hbar}\hat x_1\hat p_2.
\end{eqnarray}

 We write the complete Hamiltonian (\ref{54}) in the form
\begin{eqnarray}
\hat H&=&\frac{1}{2m}(\hat p_1^2+\hat p_2^2)+m(g-\frac{\eta}{2m^2\hbar}\hat p_2)\hat x_1
\label{Ha}\\
&=&\frac{1}{2m}(\hat p_1^2+\hat p_2^2)+mg'\hat x_1
\end{eqnarray}
where $g'=g-\frac{\eta}{2m^2\hbar}\hat p_2$. Since in the $x_2$ direction the particle is free, $p_2$ is a constant of motion. In the experiment painstakingly performed by Nesvizhevsky {\it et al.}\cite{nes,vvnes} the expectation value of $p_2$ was
\begin{eqnarray}
<\hat p_2>=10.91\times 10^{-27} \ {\textrm {Kg m s}}^{-1}.
\label{p2}
\end{eqnarray}
Now we can use (\ref{wkb}) to write the corrected energy values of the Hamiltonian (\ref{Ha}) as,
\begin{eqnarray}
E_n+\Delta E_n&=&\alpha_n(g')^{\frac{2}{3}}\nonumber\\
&=&\alpha_n(g-\frac{\eta}{2m^2\hbar}<\hat p_2>)^{\frac{2}{3}}
\label{delt}
\end{eqnarray}
where $E_n$ corresponds to the unperturbed energy and $\Delta E_n$ is the correction. It is possible to find an analytic expression for $\Delta E_n$ from (\ref{delt}) by an expansion,
\begin{eqnarray}
E_n+\Delta E_n=\alpha_ng^{\frac{2}{3}}(1-\frac{\eta}{2gm^2\hbar}<\hat p_2>)^{\frac{2}{3}}.
\end{eqnarray}
Retaining the leading $\eta$-order term we find,
\begin{eqnarray}
\Delta E_n=-\frac{\eta}{3gm^2\hbar}<\hat p_2>E_n.
\label{bid}
\end{eqnarray}
The same functional form can be obtained from the virial theorem \cite{brau}.
 
Taking the values of $E_1$ and $E_2$ from (\ref{ener1}, \ref{ener2}) and $<p_2>$ from (\ref{p2}) we get on using (\ref{bid}),
\begin{eqnarray}
&&|\Delta E_1|=2.79\times 10^{29}\eta \ ({\textrm {J}})
\label{ar1}\\
&&|\Delta E_2|=4.90\times 10^{29}\eta \ ({\textrm {J}}).
\label{ar2}
\end{eqnarray}
Finally, using the experimental input from (\ref{exp1}, \ref{exp2}) leads to the following upper bounds on $\eta$;
\begin{eqnarray}
&&|\eta|\lesssim 2.35\times 10^{-61} \ {\textrm {kg}}^2{\textrm {m}}^2{\textrm {s}}^{-2} \  \ (n=1)
\label{up1}\\
&&|\eta|\lesssim 1.77\times 10^{-61} \ {\textrm {kg}}^2{\textrm {m}}^2{\textrm {s}}^{-2} \  \ (n=2)
\label{up2}
\end{eqnarray}
The upper bound on $\eta$ (\ref{up1}, \ref{up2}) are in excellent agreement with the numerical results obtained in \cite{ba} by perturbing about the exact Airy function solutions. 
\section{Discussion}
We have studied a model of a particle in the quantum well of the Earth's gravitational field and a perfectly reflecting horizontal plane beneath, defined in a space with noncommuting coordinates and momenta. By considering a phase space transformation we reduced the problem on the commutative space. We have shown that in this process it is not necessary to modify the Planck constant as is generally believed\cite{ba,bert}. The commutative space Hamiltonian was found to be indifferent to the $\theta$ (i. e. coordinate) noncommutativity in the leading order. The energy spectrum in this model was computed analytically by exploiting the WKB approximation. Comparison with the experimental findings of \cite{nes,vvnes,nes1} placed an upper bound on the $\eta$- noncommutativity parameter appearing in the algebra of momenta.


\chapter{\label{chap:deformsym}Deformed symmetry in noncommutative spaces}

So far we were discussing two dimensional noncommutativity which is known to satisfy some special properties. In fact in the previous chapter rotational symmetry was used extensively to simplify the Hamiltonian, but that symmetry does not hold for an arbitrary dimension. To see this we take the following general structure for arbitrary n-dimensions,
\begin{eqnarray}
[\hat{y}_i,\hat{y}_j]=i\theta_{ij}, \  \ i,j=1,2,...n
\label{bha}
\end{eqnarray} 
Under coordinate rotations $\delta \hat y_i=\omega_{ij}\hat y_j$ with $\omega_{ij}=-\omega_{ji}$, infinitesimal version of (\ref{bha}) gives
\begin{eqnarray}
\omega_{ik}\theta_{kj}+\omega_{jk}\theta_{ik}=0.
\end{eqnarray}
This is not true in general. But for $d=2$, $\omega_{ij}=\omega\epsilon_{ij}$ and $\theta_{ij}=\theta\epsilon_{ij}$ under which the above condition holds. Thus two dimensional noncommutativity is restricted in some sense. So to study deformed symmetries we take a general noncommutative relation 
\begin{equation}
[\hat y_{\mu},\hat y_{\nu}] = i\theta_{\mu\nu}(\hat y,\hat q), \  \ \mu,\nu=0,1,2,3\label{1.1} 
\end{equation}
where the phase space is spanned by the variables ($\hat{y},\hat{q}$).

There are some important issues related with the application of (\ref{1.1}). In relativistic theory, even a constant $\theta_{\mu\nu}$ breaks Poincar$\acute{\textrm e}$ symmetries\cite{car,ban}. Likewise for massless models, conformal symmetries are affected. However it might be possible to introduce quantum deformations of the generators such that the particular form of the commutator algebra remains covariant. This has been discussed in great detail, for a constant $\theta_{\mu\nu}$, using either higher order differential operators\cite{wessju,Aschierie,koch,rabin,lee} or twist functions following from quantum group arguments 
\cite{chainew,cha}. In this chapter we study deformed symmetries of two important noncommutative spaces. First one is the nonrelativistic constant $\theta$ noncommutativity
\begin{eqnarray}
&&\left[\hat{y}^i, \hat{y}^j\right] =i \theta^{ij},\nonumber\\
&&\left[\hat{q}^i, \hat{q}^j\right] = 0, \  \  \  \  \ i,j=1,2,...,n\nonumber\\
&&\left[\hat{y}^i, \hat{q}^j\right] = i \delta^{ij},\label{106}
\end{eqnarray}
and the other one is the Snyder algebra
\begin{eqnarray}
\begin{array}{rcl}
\left[\hat y^{\mu},\hat y^{\nu}\right] &=& i\theta(\hat y^{\mu}\hat q^{\nu} - \hat y^{\nu}\hat q^{\mu})\\
\left[\hat y^{\mu},\hat q_{\nu}\right] &=& i(\delta^{\mu}_{\nu} +\theta \hat q^{\mu}\hat q_{\nu}) \  \  \  \  \  \  \  \ \mu,\nu=0,1,2,3\\
\left[\hat q_{\mu},\hat q_{\nu}\right] &=& 0
\end{array}
\label{2.1}   
\end{eqnarray}
where $\theta$ is a measure of the noncommutativity. This is an example of relativistic nonconstant noncommutativity. Other cases like nonrelativistic nonconstant noncommutativity (an example of which is Lie algebraic noncomutativity which has been analyzed in \cite{luk}) are not discussed here
   
In the first part of this chapter we exploit the results of \cite{rabin} to write the deformed Schr\"odinger generators which satisfy the standard algebra. These are used to get the deformed transformations. We also construct  a non relativistic model which generates the algebra (\ref{106}) and is invariant under the deformed transformations. The N\"other theorem is then used to get back the deformed generators form this model. This shows the self consistency.

 In the second part, a deformed translation is defined under which Snyder algebra remains covariant. Next, a dynamical model invariant under such deformation is formulated. Using Dirac's\cite{n} constraint analysis the deformed brackets are computed which yield Snyder brackets. We therefore provide a dynamical realization of the algebra (\ref{2.1}). N\"other's theorem is then used to find the deformed generators from the dynamical model. The algebraic map between the variables of Snyder algebra and the standard commutative algebra is obtained. Finally, the analysis is extended to the conformal sector.

\section{Deformed Schr\"odinger symmetry and transformations}
The generators of the Schr\"odinger group consists of rotations ($\hat{\mathcal{J}}^{ij}$), translations ($\hat{\mathcal{P}}^i$), boosts ($\hat{\mathcal{G}}^i$), dilatations ($\hat{\mathcal{D}}$) and special conformal transformations ($\hat{\mathcal{K}}$). The standard representation of these generators are given by,
\begin{align}
\nonumber &\hat{\mathcal{J}}^{ij} = \hat{x}^{i}\hat{p}^{j} - \hat{x}^{j}\hat{p}^{i} \\
\nonumber &\hat{\mathcal{D}} = \frac{1}{2}\left(\hat{p}^{i}\hat{x}^i + \hat{x}^{i}\hat{p}^i\right) - \frac{\vec{\hat{p}}^2}{m}t \\
 &\hat{\mathcal{K}} = \frac{1}{2}m\left(\hat{x}^i-\frac{\hat{p}^i}{m}t\right)^2 \label{104}\\
\nonumber &\hat{\mathcal{G}}^i = m\hat{x}^i-\hat{p}^i t \\
 &\hat{\mathcal{P}}^i = \hat{p}^i \nonumber
\end{align}
where $m$ is the mass of the particle. The complete set of algebra, satisfied by these generators in a commutative space
\begin{eqnarray}
\left[\hat{x}^i, \hat{x}^j\right] = \left[\hat{p}^i, \hat{p}^j\right] = 0; \  \ 
 \left[\hat{x}^i, \hat{p}^j\right] = i \delta^{ij} \  \  \ ({\textrm {in the unit of $\hbar$}}),
\label{105}
\end{eqnarray}
is found to be,
\begin{align}%
&\nonumber \left[\hat{\mathcal{J}}^{ij}, \hat{\mathcal{J}}^{k\ell}\right] = - i\left(\delta^{kj}\hat{\mathcal{J}}^{i\ell} - \delta^{ki}\hat{\mathcal{J}}^{j\ell} + \delta^{\ell j}\hat{\mathcal{J}}^{ki} - \delta^{\ell i}\hat{\mathcal{J}}^{kj}\right)\\
\nonumber &\left[\hat{\mathcal{P}}^i, \hat{\mathcal{P}}^j\right] = 0 & &\left[\hat{\mathcal{G}}^i, \hat{\mathcal{G}}^j\right] = 0\\
\nonumber &\left[\hat{\mathcal{P}}^i, \hat{\mathcal{J}}^{jk}\right] = - i\left(\delta^{ij}\hat{\mathcal{P}}^{k}-\delta^{ik}\hat{\mathcal{P}}^{j}\right) & &\left[\hat{\mathcal{G}}^i, \hat{\mathcal{J}}^{jk}\right] = - i\left(\delta^{ij}\hat{\mathcal{G}}^{k}-\delta^{ik}\hat{\mathcal{G}}^{j}\right)\\
\nonumber &\left[\hat{\mathcal{P}}^i, \hat{\mathcal{G}}^j\right] = - i\delta^{ij}m & &\\
\nonumber &\left[\hat{\mathcal{J}}^{ij}, \hat{\mathcal{D}}\right] = 0 & &\left[\hat{\mathcal{J}}^{ij}, \hat{\mathcal{K}}\right] = 0\\
\nonumber &\left[\hat{\mathcal{D}}, \hat{\mathcal{G}}^i\right] = - i\hat{\mathcal{G}}^i & &\left[\hat{\mathcal{K}}, \hat{\mathcal{G}}^i\right] = 0\\
\nonumber &\left[\hat{\mathcal{D}}, \hat{\mathcal{P}}^i\right] = i\hat{\mathcal{P}}^i & &\left[\hat{\mathcal{K}}, \hat{\mathcal{P}}^i\right] = i\hat{\mathcal{G}}^i\\
\label{103} &\left[\hat{\mathcal{D}}, \hat{\mathcal{K}}\right] = - 2i\hat{\mathcal{K}} & &
\end{align}

Now in the noncommutative space (\ref{106}), one has to appropriately modify the expressions of generators so that they satisfy the standard algebra (\ref{103}). Taking general forms of these generators from dimensional arguments, it has been shown in \cite{rabin} that consistency with (\ref{103}) and various Jacobi identities fixes the deformed representation of the generators. These are given by,
\begin{eqnarray}
&&\nonumber {\hat{\cal J}}^{ij} = \hat{y}^i\hat{q}^j - \hat{y}^j\hat{q}^i + \frac{1}{2}\theta^{im}\hat{q}^m\hat{q}^j - \frac{1}{2}\theta^{jm}\hat{q}^m\hat{q}^i\\
&&\nonumber {\hat{\cal P}}^i = \hat{q}^i,\\
&&{\hat{\cal G}}^i = m\hat{y}^i - t\hat{q}^i + \frac{m}{2}\theta^{ij}\hat{q}^j,\label{112}\\
&&\nonumber {\hat{\cal D}} = \frac{1}{2}\left(\hat{q}^i\hat{y}^i + \hat{y}^i\hat{q}^i\right) - \frac{1}{m}\vec{\hat{q}}^2 t,\\
 &&{\hat{\cal K}} = \frac{m}{2}\left(\hat{y}^i - \frac{\hat{q}^i}{m}t\right)^2 + \frac{m}{2}\theta^{ij}\hat{y}^i\hat{q}^j - \frac{m}{8}\theta^{i\ell}\theta^{\ell m}\hat{q}^i\hat{q}^m.\nonumber
\end{eqnarray}
It can be easily verified that in the noncommutative space (\ref{106}) the deformed generators (\ref{112}) satisfy the undeformed algebra (\ref{103}). Expectedly, in the limit $\theta\rightarrow 0$ (\ref{112}) reduces to the undeformed generators (\ref{104}) under the identification $(\hat y, \hat q)\rightarrow (\hat x, \hat p)$.

 It is possible to give a map which connect the deformed generators (\ref{112}) with the undeformed generators (\ref{104}). The maps between the noncommutative variables ($\hat{y}_i,\hat{q}_i$) and the commutative variable ($\hat{x}_i,\hat{p}_i$) are given by,
\begin{eqnarray}
&&\hat{x}_i=\hat{y}_i+\frac{1}{2}\theta_{ij}\hat{q}_j.\nonumber\\
&&\hat{p}_i=\hat{q}_i
\label{dabu}
\end{eqnarray}
The noncommutative phase space algebra (\ref{106}) and the Heisenberg algebra (\ref{105}) are consistent with the above map.

Knowing the expressions of deformed generators (\ref{112}) it is straightforward to obtain the deformed transformation rules. These are obtained from the general 
relation 
\begin{eqnarray}
\delta {\hat r}_i=-i[\hat r_i,\hat G];\hat r_i=\hat y_i,\hat q_i
\end{eqnarray}
where $\hat G$ is the generator. Thus for translation generator $\hat P=a_i\hat q_i$ 
\begin{eqnarray}
\delta \hat y_i=-i[\hat y_i,\hat P]=a_i\\
\delta \hat q_i=-i[\hat q_i,\hat P]=0.
\end{eqnarray}
For rotation generator $\hat{J}=\omega_{ij}{\hat{\cal J}}_{ij}$
\begin{eqnarray}
\delta \hat y_i
&=&\omega_{ab}\theta_{ia}\hat q_b+2\omega_{ai}(\hat y_a+\frac{1}{2}\theta_{am}\hat q_m)\\
\delta \hat q_i&=&2\omega_{ai}\hat q_a.
\end{eqnarray}
For the Galilean boost $\hat{G}=\epsilon_i{\hat{\cal G}}_i$
\begin{eqnarray}
\delta \hat y_i&=&m\epsilon_j\theta_{ij}-t\epsilon_i-\frac{m}{2}\epsilon_j\theta_{ij}\label{def1a}\\
\delta \hat q_i&=& -m\epsilon_i.
\label{def2b}
\end{eqnarray}
For the dilatation generator ${\hat{\cal D}}$ 
\begin{eqnarray}
\delta \hat y_i&=&\theta_{ij}\hat q_j+\hat y_i-\frac{2}{m}\hat q_it\\
\delta \hat q_i&=&-\hat q_i.
\end{eqnarray}
For the expansion ${\hat{\cal K}}$  
\begin{eqnarray}
\delta \hat y_i&=&\frac{m}{2}\hat y_j\theta_{ij}+\frac{m}{4}\theta_{il}\theta_{lm}\hat q_m-t\theta_{ij}\hat q_j-t(\hat y_i-\frac{\hat q_i}{m}t)\\
\delta \hat q_i&=&-\frac{m}{2}\theta_{ij}\hat q_j-m(\hat y_i-\frac{\hat q_i}{m}t)
\end{eqnarray}
These are the complete expressions of the deformed transformation rules which keep the noncommutative algebra (\ref{106}) invariant.
\section{Construction of a dynamical model}
In order to construct a model which will generate the noncommutative algebra (\ref{106}) in a natural manner we take the following non relativistic action for a free particle of mass m
\begin{equation}
S_0=\int dt \ (p_i\dot x_i-\frac{p^2}{2m})
\label{mod1q}
\end{equation}
and use the classical version of the transformation (\ref{dabu}) to get the following form of the action,
\begin{equation}
S=\int dt L=\int dt \ (q_i\dot y_i+\frac{1}{2}\theta_{ij}q_i\dot q_j-\frac{q^2}{2m})
\label{q_i}
\end{equation}
This model in $2+1$ dimension has been studied in \cite{duvalhorvathy}. Discussions on the relativistic generalization of (\ref{q_i}) are given in \cite{deri,wor}. The action (\ref{q_i}) shows some interesting properties which we study now. Interpreting $y$ and $q$ of the above first order action as the configuration space variables, the canonical momenta conjugate to $y$ and $q$ are written as,
\begin{equation}
\pi_i^y=\frac{\partial L}{\partial \dot y_i}=q_i\\
\pi_i^q=\frac{\partial L}{\partial \dot q_i}=\frac{1}{2}\theta_{ji}q_j.
\label{pypq}
\end{equation}
Thus we get the primary constraints
\begin{equation}
\Omega^1_i=\pi_i^y-q_i\approx 0\\
\Omega^2_i=\pi_i^q-\frac{1}{2}\theta_{ji}q_j
\end{equation}
which satisfy the following Poisson algebra
\begin{equation}
\{\Omega^1_i,\Omega^1_j\}=0\\
\{\Omega^1_i,\Omega^2_j\}=-\delta_{ij}\\
\{\Omega^2_i,\Omega^2_j\}=\theta_{ij}.
\end{equation}
These are second class constraints which can be eliminated by the use of Dirac brackets. The first step is to construct the constraint matrix 
\begin{eqnarray}
\Lambda^{ab}=\left(\begin{matrix}\{\Omega^1_i,\Omega^1_j\}&\{\Omega^1_i,\Omega^2_j\}\cr\{\Omega^2_i,\Omega^1_j\}&\{\Omega^2_i,\Omega^2_j\}
\end{matrix}\right)=\left(\begin{matrix}0&-\delta_{ij}\cr\delta_{ij}&\theta_{ij}\end{matrix}\right).
\end{eqnarray}
We write the inverse of $\Lambda^{ab}$ as $\Lambda^{(-1)ab}$ such that $\Lambda^{ab}\Lambda^{(-1)bc}=\delta^{ac}$. It is given by,
\begin{equation}
\Lambda^{(-1)ab}=\left(\begin{matrix}\theta_{ij}&\delta_{ij}\cr-\delta_{ij}&0\end{matrix}\right).
\end{equation}
Dirac brackets can now be calculated from the definition (\ref{diracbracket}). For the action (\ref{mod1q}), the configuration space variables satisfy the following algebra
\begin{eqnarray}
&&\{y_i,y_j\}_{DB}=\theta_{ij}\nonumber\\
&&\{y_i,q_j\}_{DB}=\delta_{ij}\nonumber\\
&&\{q_i,q_j\}_{DB}=0
\label{p0}
\end{eqnarray}
This algebra is the classical version of the noncommutative commutator algebra (\ref{106}). Though there is no noncommutativity in the momentum ($q$) sector it is possible to construct an action from (\ref{q_i}) which will generate the momentum noncommutativity. We show it in the following way.

The equations of motion for the action (\ref{q_i}) are given by,
\begin{eqnarray}
&&\dot q_i=0\\
&&\dot y_i+\theta_{ij}\dot q_j-\frac{q_i}{m}=0
\end{eqnarray}
Combining above two equations one obtains $q_i=m\dot y_i$ which is used in (\ref{q_i}) to write the action in terms of $y$ alone
\begin{equation}
S=\int dt \frac{m}{2}\dot y^2+\frac{m^2}{2}\theta_{ij}\dot y_i\ddot y_j
\label{act}
\end{equation}
The two dimensional version of the above action has been studied thoroughly in \cite{LUKE}. As the Lagrangian contains second order derivative, we introduce two sets of momenta\cite{LUKE}
\begin{eqnarray}
&&q_i=\frac{\partial L}{\partial \dot y_i}-\frac{d}{dt}\frac{\partial L}{\partial \ddot y_i}=m \dot y_i+m^2\theta_{ij}\ddot y_j\\
&&\tilde{q_i}=\frac{\partial L}{\partial \ddot y_i}=-\frac{m^2}{2}\theta_{ij}\dot y_j.
\label{tilq}
\end{eqnarray}
The non zero Poisson bracket exists only for canonical pairs
\begin{eqnarray}
\{y_i,q_j\}=\{\dot y_i,\tilde q_j\}=\delta_{ij}.
\end{eqnarray}
All other brackets are zero. The second set of momenta (\ref{tilq}) forms the primary constraints for the model (\ref{act})
\begin{equation}
\Phi_i=\tilde{q_i}+\frac{m^2}{2}\theta_{ij}\dot y_j\approx 0
\end{equation}
which satisfy the following constraint algebra
\begin{equation}
\{\Phi_i,\Phi_j\}=m^2\theta_{ij}.
\end{equation}
Having obtained the constraint matrix we can calculate various Dirac brackets from the formula
\begin{equation}
\{A,B\}_{DB}=\{A,B\}-\{A,\Phi_i\}\frac{1}{m^2}\theta_{ij}^{-1}\{\Phi_j,B\}
\label{ref}
\end{equation}
Interestingly the second set of momenta are now noncommutative. The algebra we obtain is
\begin{eqnarray}
\{{\tilde q}_i,{\tilde q}_j\}_{DB}=-\frac{m^2}{4}\theta_{ij}.
\end{eqnarray}
Other Dirac brackets are obtained easily from (\ref{ref})
\begin{eqnarray}
&&\{y_i,y_j\}_{DB}=0= \ \{q_i,q_j\}_{DB}\\
&&\{y_i,q_j\}_{DB}=\delta_{ij}\\
&&\{\dot y_i,\dot y_j\}_{DB}=\frac{1}{m^2}\theta^{-1}_{ij}\\
&&\{\dot y_i,{\tilde q}_j\}_{DB}=\frac{1}{2}\delta_{ij}
\end{eqnarray}
This shows the dual nature of position and momenta noncommutativity.
\section{N\"other's theorem and generators}
It is possible to reproduce the deformed Schr\"odinger generators from a N\"other analysis of (\ref{q_i}). This shows the consistency between the dynamical approach of section 4.2 and the algebraic approach of section 4.1.

  In general, the invariance of an action $S$ under an infinitesimal symmetry transformation,
\begin{eqnarray}
\delta Q_i=\{Q_i,G\}
\label{5.1}
\end{eqnarray}
is given by
\begin{eqnarray}
\delta S=\int d \tau \frac{d}{d\tau}(\delta Q^iP_i-G)
\label{5.2}
\end{eqnarray}
where $G$ is the generator of the transformation and $P_{i}$ is the canonical momenta conjugate to $Q^{i}$. If the quantity inside the parentheses is denoted by $B(Q,P)$, then the generator is defined as,
\begin{eqnarray}
G=\delta Q^{i}P_{i}-B.
\label{5.3}
\end{eqnarray}
For the model (\ref{q_i}) both $y,q$ are interpreted as configuration space variables so that,
\begin{eqnarray}
G=\delta q_i\pi_i^q+\delta y_i\pi_i^y-B.
\end{eqnarray}
This is simplified using the expressions (\ref{pypq}) to get,
\begin{eqnarray}
G=\delta q_i(-\frac{1}{2}\theta_{ij}q_j)+\delta y_iq_i-B.
\label{generatorqw}
\end{eqnarray}
This relation is now used to find the deformed generators knowing the deformed transformation rules. For example we consider the Galilean boost. Using (\ref{generatorqw}) and the deformed transformations (\ref{def1a},\ref{def2b}) we write
\begin{eqnarray}
G=-m\epsilon_i(-\frac{1}{2}\theta_{ij}q_j)+(\frac{m}{2}\epsilon_{j}\theta_{ij}-t\epsilon_i)q_i-B
\label{qwe}
\end{eqnarray}
Now in order to calculate the variation of the Lagrangian(\ref{q_i}) we use (\ref{def1a},\ref{def2b}) to obtain the following results
\begin{eqnarray}
&&\delta(q_i\dot y_i)=-m\epsilon_i\dot y_i-q_i\epsilon_i\\
&&\delta(q_i\dot q_j)=-m\epsilon_i\dot q_j\\
&&\delta (q^2)=-2m q_i\epsilon_i
\end{eqnarray}
Using these expressions we find the variation of the Lagrangian
\begin{eqnarray}
\delta L=\frac{dB}{dt}=-m\epsilon_i\frac{d}{dt}(y_i+\frac{1}{2}\theta_{ij}q_j)
\end{eqnarray}
Extracting $B$ from the above equation we put it in (\ref{qwe}) to get the deformed generator
\begin{eqnarray}
G=\epsilon_i(my_i-tq_i+\frac{m}{2}\theta_{ij}q_j)
\end{eqnarray}
which is same as the boost generator given in (\ref{112}) without the parameter $\epsilon$. The other deformed generators obtained in this manner are identical with (\ref{112}).
             
\section{The Snyder space and its symmetries}
The study of deformed symmetries, presented in the previous sections can be done for other cases. A simple generalization is to take the relativistic version of (\ref{106}) and study deformed Poincare-conformal symmetry. This has been analyzed extensively in \cite{kk}. Instead we study the Snyder\cite{sny} algebra(\ref{2.1}). It was originally obtained by a dimensional descent from five dimensions and involves the angular momentum in the algebra of the non commuting coordinates. Taking the momentum operators commuting, as in the usual space, naturally leads to a deformed algebra among $\hat y-\hat q$, therefore ensuring the validity of the various Jacobi identities. It leads to a discrete space time compatible with Lorentz symmetry. Apart from its intrinsic interest this algebra has relevance in various contexts. For instance, a similar algebra is also obtained from quantum gravity in $2+1$ dimensions\cite{tooft}. There also exists a mapping between the Snyder space and $\kappa$-Minkowski space-time\cite{glik} which is frequently used in analyzing doubly special relativity. 

We now study the different symmetries associated with the Snyder Algebra(\ref{2.1}).
 
\subsection{Lorentz symmetry}
By its very construction the algebra (\ref{2.1}) is compatible with standard Lorentz transformations,
\begin{eqnarray}
\delta \hat y_{\mu}&=& \omega_{\mu\alpha}\hat y^{\alpha}\label{2.2}\\
\delta \hat q_{\mu}&=& \omega_{\mu\alpha}\hat q^{\alpha}. \label{2.3}
\end{eqnarray} 
with $\omega_{\mu\alpha} =-\omega_{\alpha\mu}$. This is checked in the following way. Consider the variation in the first relation,
\begin{eqnarray}
\delta[\hat y^{\mu},\hat y^{\nu}]&=&[\delta \hat y^{\mu},\hat y^{\nu}] + [\hat y^{\mu},\delta \hat y^{\nu}]\nonumber\\
&=& i\theta \omega^{\mu\alpha}(\hat y_{\alpha}\hat q^{\nu} -\hat y^{\nu}\hat q_{\alpha})  
 - i\theta \omega^{\nu\alpha}(\hat y_{\alpha}\hat q^{\mu} -\hat y^{\mu}\hat q_{\alpha}).\label{2.4}  
\end{eqnarray}
 The same expression is obtained by considering the variation on 
 the r.h.s of that relation,
\begin{eqnarray}
i\theta \delta(\hat y^{\mu}\hat q^{\nu}-\hat y^{\nu}\hat q^{\mu}) &=& i\theta \omega^{\mu\alpha}(\hat y_{\alpha}\hat q^{\nu} -\hat y^{\nu}\hat q_{\alpha})  
 - i\theta \omega^{\nu\alpha}(\hat y_{\alpha}\hat q^{\mu} -\hat y^{\mu}\hat q_{\alpha}). \label{2.5} 
\end{eqnarray}

An identical treatment follows for the other two relations. This 
is sufficient to ensure consistency of the Lorentz transformations.
 Expectedly, the generator retains its primitive (undeformed) structure,
\begin{equation}
\hat J_{\mu\nu} = \hat y_{\mu}\hat q_{\nu} - \hat y_{\nu}\hat q_{\mu}\label{2.6}\end{equation} 
so that,
\begin{eqnarray}
\delta \hat y_{\mu} = \frac{i}{2}\omega^{\alpha\beta}\left[\hat J_{\alpha\beta},\hat y_{\mu}\right] = \omega_{\mu\alpha}\hat y^{\alpha}\label{2.7}
\end{eqnarray} 
 and similarly for $\hat q_{\mu}$.
\subsection{Translation symmetry}
The commutative space transformation rules for translation $\delta \hat y_{\mu}=a_{\mu},\delta \hat q_{\mu} = 0$ are not compatible with the first relation in the Snyder algebra. So we take the following general expressions of $\delta \hat y_{\mu}$ and $\delta \hat q_{\mu}$ which are dimensionally consistent     
\begin{eqnarray}
&&\delta \hat y_{\mu}=a_{\mu} + \alpha \theta a_{\mu}\hat q^2 + \beta \theta a_{\rho}\hat q^{\rho}\hat q_{\mu} \label{2.11} \\
&&\delta \hat q_{\mu} = 0.\label{2.12} 
\end{eqnarray}

 Compatibility with the Snyder algebra fixes $\alpha=0$ and $\beta=1$. So the deformed transformation rule for the translation operator in Snyder space is given by,
\begin{eqnarray}
&&\delta \hat y_{\mu} = a_{\mu} + \theta a_{\rho}\hat q^{\rho}\hat q_{\mu}.\label{2.13}\\
&&\delta \hat q_{\mu} = 0.\label{2.14}
\end{eqnarray}

      Although we have a deformed transformation rule for translation, the generator remains the same as in the commutative space. To emphasize this point we note that  
\begin{eqnarray}
\delta \hat y^{\mu} &=& i\left[\hat G,\hat y^{\mu}\right] = ia^{\rho}[\hat q_{\rho},\hat y^{\mu}]\nonumber\\
&=& a^{\mu} + \theta a^{\rho}\hat q_{\rho}\hat q^{\mu}.\label{2.15}
\end{eqnarray}
and likewise for $\hat q_{\mu}$.
   
      Thus the Poincare generators in Snyder space and usual commutative space are form invariant. However, whereas Lorentz transformation remains undeformed,
translation get deformed. 

     Finally, in spite of the involved algebra (\ref{2.1}) these generators satisfy the usual Poincare algebra,
\begin{eqnarray}
\begin{array}{rcl}
&&[\hat q_{\mu},\hat q_{\nu}]=0\\
&&\left[\hat J_{\mu\nu},\hat q_{\lambda}\right]=i(\delta_{\mu\lambda}\hat q_{\nu}-\delta_{\nu\lambda}\hat q_{\mu})\\
&&\left[\hat J_{\mu\nu},\hat J_{\rho\sigma}\right]=-i(\delta_{\nu\rho}\hat J_{\mu\sigma}+\delta_{\mu\sigma}\hat J_{\nu\rho}-\delta_{\mu\rho}\hat J_{\nu\sigma}-\delta_{\nu\sigma}\hat J_{\mu\rho}).
\end{array}
\label{2.16}
\end{eqnarray}
\section{Dynamical model invariant under deformation and the Snyder algebra} 
Here we discuss a method by which a dynamical model is constructed to reproduce the Snyder algebra (\ref{2.1}) from the constraint analysis of the model. Several authors\cite{rom,gir,ghosh,rome} have suggested various models leading to this algebra but those results were obtained in a specific gauge. In our analysis we do not fix any gauge. Instead, we calculate the Dirac brackets of the configuration space variables of a dynamical model which is invariant under the deformed (translation) symmetry. The results lead to the Snyder algebra.

 Consider the following first order form of the action for a relativistic free particle of mass $m$, 
\begin{eqnarray}
S=\int d\tau[-\dot q^{\mu}y_{\mu}-e(q^2-m^2)]
\label{3.1}
\end{eqnarray}
where $e$ is a Lagrange multiplier enforcing the Einstein condition $q^2-m^2=0$.

  Since the Lorentz transformation is undeformed, obviously (\ref{3.1}) remains invariant. Under translation however,
\begin{eqnarray}
\delta S&=&\int d\tau[-\dot q^{\mu}(a_{\mu}+\theta a_{\rho}q^{\rho}q_{\mu})]\\
 &=&\int d\tau[-\frac{d}{d\tau}(q^{\mu}a_{\mu})- \theta a_{\rho}q^{\rho}q_{\mu}\dot q^{\mu}]\label{3.3}     
\end{eqnarray} 
obtained on exploiting (\ref{2.13}), (\ref{2.14}). The additional symmetry breaking term can be written as, 
\begin{eqnarray}
\theta a_{\rho}q^{\rho}q_{\mu}\dot q^{\mu}&=&\theta(\delta y_{\rho}-\theta a_{\sigma}q^{\sigma}q_{\rho})q^{\rho}q_{\mu}\dot q^{\mu}\label{3.4}\\
&=&\theta\delta[ y_{\rho}q^{\rho}q_{\mu}\dot q^{\mu}]-\theta^2\delta[y_{\sigma}q^{\sigma}q^2\dot q^{\mu}q_{\mu}]+\theta^3\delta[y_{\sigma}q^{\sigma}(q^2)^2\dot q^{\mu}q_{\mu}]+\cdot\cdot\cdot\label{3.5}\\
&=&\theta\delta[\frac{1}{1+\theta q^2}(y\cdot q)\dot q^{\mu}q_{\mu}]\label{3.6}
\end{eqnarray}
where recursive use of (\ref{2.13}) and (\ref{2.14}) has been done. Thus inclusion of the term $\frac{\theta}{1+\theta q^2}(y\cdot q)\dot q^{\mu}q_{\mu}$ in the action (\ref{3.1})

\begin{eqnarray}
S=\int d\tau[-\dot q^{\mu}y_{\mu}+\frac{\theta}{1+\theta q^2}(y.q)\dot q^{\mu}q_{\mu}-e(q^2-m^2)].\label{3.7}
\end{eqnarray}
makes it quasi-invariant under deformed translations,
\begin{eqnarray}
\delta S=\int d\tau[-\frac{d}{d\tau}(a^{\mu}q_{\mu})].\label{w}
\end{eqnarray}
and invariant under Lorentz transformation.

We interpret $y$ and $q$ of the first order action (\ref{3.7}) as the configuration variables in an extended space. The canonical momentum conjugate to $e$, $y$ and $q$ are denoted by $\pi_{e},\pi_{\mu}^{y},\pi_{\mu}^{q}$. They do not contain the velocity term and hence are interpreted as primary constraints. These are given by,
\begin{eqnarray}
&&\Phi=\pi_{e}\approx0\label{3.8}\\
&&\Phi_{1,\mu}=\pi_{\mu}^{y}\approx0\label{3.9}\\
&&\Phi_{2,\mu}=\pi_{\mu}^{q}+y_{\mu}-\frac{\theta}{1+\theta q^2}(y.q)q_{\mu}\approx0.\label{3.10}
\end{eqnarray}
The non vanishing Poisson brackets of the constraints are given by
\begin{eqnarray}
&&\{\Phi_{1,\mu},\Phi_{2,\nu}\}=-\eta_{\mu\nu}+\frac{\theta}{1+\theta q^2}q_{\mu}q_{\nu}\label{3.13} \\
&&\{\Phi_{2,\mu},\Phi_{2,\nu}\}=\frac{\theta}{1+\theta q^2}(q_{\nu}y_{\mu}-q_{\mu}y_{\nu})\label{3.14}. 
\end{eqnarray}
All other brackets are zero. The canonical Hamiltonian of the system is read off from the first order action (\ref{3.7})
\begin{eqnarray}
H_C=e(q^2-m^2).\label{3.15}
\end{eqnarray}
The total Hamiltonian is given by the sum of canonical Hamiltonian and the primary constraints with Lagrange multipliers
\begin{eqnarray}
H_T=e(q^2-m^2)+\lambda \Phi+\lambda_{1,\mu}\Phi_{1,\mu}+\lambda_{2,\mu}\Phi_{2,\mu}.\label{3.17}
\end{eqnarray}
Time consistency of the constraint (\ref{3.8}) leads to the following secondary constraint
\begin{eqnarray}
\Psi=\{H_T,\pi_e\}=q^2-m^2\approx0.\label{3.18}
\end{eqnarray}

 In order to eliminate the second class constraint sector $\Phi_1,\Phi_2$ by the use of Dirac brackets we follow the same method discussed in section 2.2. The inverse constraint matrix,
\begin{eqnarray}
\left(\begin{matrix}\theta(y^{\mu}q^{\nu}-y^{\nu}q^{\mu})& \eta^{\mu\nu}+\theta q^{\mu} q^{\nu}\cr -\eta^{\mu\nu}-\theta q^{\mu} q^{\nu}&0  \end{matrix}\right)
\end{eqnarray}
leads to the following Dirac brackets among the configuration space variables
\begin{eqnarray}
\begin{array}{rcl}
&&\{y^{\mu},y^{\nu}\}_{DB}=\theta\left(y^{\mu} q^{\nu}- y^{\nu} q^{\mu}\right )\\
&&\{q_{\mu},q_{\nu}\}_{DB}=0\\
&&\{y^{\mu},q_{\nu}\}_{DB}=\delta^{\mu}_{ \nu}+\theta q^{\mu}q_{\nu}.
\end{array}
\label{3.23}
\end{eqnarray}

It is straight forward to  elevate this algebra at the operator level. Since $q$ s commute among themselves there is no problem in the algebra between $y^{\mu}$ and $q_{\nu}$. Furthermore from the last algebra it is clear that $y^{\mu} q^{\nu}-q^{\nu}y^{\mu}$ is symmetrical in $\mu,\nu$. Since the bracket between $y^{\mu}$ and $y^{\nu}$ should be antisymmetric in $\mu,\nu$ no ordering problem appears. Thus without any ambiguity the Dirac brackets (\ref{3.23}) get lifted to the commutators (\ref{2.1}).  

   Since the secondary constraint $\Psi$ has vanishing Dirac brackets with all constraints,
\begin{eqnarray}
&&\{\Psi,\Psi\}_{DB}=\{\Psi,\Phi\}_{DB}=0\label{3.24}\\
&&\{\Psi,\Phi_{1,\mu}\}_{DB}=\{\Psi,\Phi_{2,\mu}\}_{DB}=0.\label{3.25}
\end{eqnarray}
it is first class in nature and generates the gauge (reparametrisation) transformations. The momenta $\Phi$ canonically conjugate to the Lagrange multiplier $e$ is not physically important.

\section{Translation and rotation generators from N\"other's theorem}
In this section we reproduce the deformed translation and rotation generator from a N\"other analysis. For the model (\ref{3.7}) the generator is written from (\ref{5.3}) as,
\begin{eqnarray}
G=\delta q^{\mu}\pi_{\mu}^q+\delta y^{\mu}\pi_{\mu}^y-B.\label{5.4}
\end{eqnarray}
where $q$ and $y$ are both interpreted as configuration space variables. Using the constraints (\ref{3.9}), (\ref{3.10}), the expression (\ref{5.4}) is further simplified to yield,
\begin{eqnarray}
G=\delta q^{\mu}(\frac{\theta}{1+\theta q^2}(y\cdot q)q_{\mu}-y_{\mu})-B.\label{5.5}
\end{eqnarray}
\vspace{0.40cm}\\
{\bf Translations } \ : \vspace{0.40cm}\\
   For translations we get from (\ref{2.14}) and (\ref{w}) 
\begin{eqnarray}
\delta q^{\mu}=0, \ B=-a^{\sigma}q_{\sigma}\label{5.6}
\end{eqnarray}
which when substituted in (\ref{5.5}) gives the translation generator
\begin{eqnarray}
G=a^{\sigma}q_{\sigma}\label{5.7}
\end{eqnarray}
\vspace{0.40cm}\\
 {\bf Rotations} \ : \vspace{0.40cm}\\ 
  Under rotation the Lagrangian itself is manifestly invariant ($\delta L=0$). Hence using (\ref{2.3}) we get,
\begin{eqnarray}
G=\omega^{\mu\alpha}q_{\alpha}(\frac{\theta}{1+\theta q^2}(y\cdot q)q_{\mu}-y_{\mu})=\frac{\omega^{\mu\alpha}}{2}J_{\alpha\mu}\label{5.8}
\end{eqnarray}
which is the cherished expression.
\section{Mapping between deformed and usual symmetries}
We give an algebraic map between the commutative and noncommutative variables by comparing the actions (\ref{3.1}) and (\ref{3.7}). At first (\ref{3.1}) is rewritten in terms of the commutative space variables $(x,p)$ (this is just a change in nomenclature)
\begin{eqnarray}
S=\int d\tau[-\dot p_{\mu}x^{\mu}-e(p^2-m^2)]\label{6.1}
\end{eqnarray}
with,
\begin{eqnarray}
\{x^{\mu},p_{\nu}\}=\delta_{\mu\nu}, \ \{x^{\mu},x^{\nu}\}=\{p_{\mu},p_{\nu}\}=0\label{6.2}
\end{eqnarray}
Then the actions (\ref{6.1}) and (\ref{3.7}) are mapped by the transformations,
\begin{eqnarray}
&& x_{\mu}=y_{\mu}-\frac{\theta}{1+\theta q^2}(y\cdot q)q_{\mu}\label{6.5}\\
&& p_{\mu}=q_{\mu}.\label{6.6}
\end{eqnarray}
The inverse map is given by,
\begin{eqnarray}
&&y_{\mu}=x_{\mu}+\theta(x\cdot p)p_{\mu}\label{6.8}\\
&&q_{\mu}=p_{\mu}.\label{6.9}
\end{eqnarray}
The classical Snyder algebra follows from the above relations by using the canonical algebra (\ref{6.2}),
\begin{eqnarray}
\{y^{\mu},q_{\nu}\}&=&\{x^{\mu}+\theta(x\cdot p)p^{\mu},p_{\nu}\}\label{6.10}\\
&=&\delta^{\mu}_{\nu}+\theta q^{\mu}q_{\nu}\label{6.11}
\end{eqnarray}
and likewise for the other brackets.

  It is feasible to construct an operator analogue of the maps (\ref{6.5}), (\ref{6.6}) by giving an ordering prescription. Using the Weyl (symmetric) ordering, we get,
\begin{eqnarray}
\hat x^{\mu}=\hat y^{\mu}&-&\frac{\theta}{8}[\frac{ \hat q^{\mu} \hat  q_{\rho}}{1+\theta \hat q^2}\hat y^{\rho} +  \hat q^{\mu} \hat q_{\rho} \hat  y^{\rho}\frac{1}{1+\theta  \hat q^2}\nonumber\\
&& + \frac{ \hat q^{\mu}}{1+\theta \hat q^2}\hat y^{\rho} \hat q_{\rho}
+ \frac{\hat q_{\rho}}{1+\theta  \hat q^2}\hat y^{\rho} \hat q^{\mu}\nonumber\\ 
&& +  \hat q^{\mu} \hat y^{\rho}\frac{ \hat q_{\rho}}{1+\theta  \hat q^2}
 + \hat q_{\rho}\hat y^{\rho} \frac{ \hat q^{\mu}}{1 +\theta  \hat q^2}\label{6.12}\\
&& + \frac{1}{1+\theta  \hat q^2}\hat y^{\rho}  \hat q^{\mu} \hat q_{\rho} 
+ \hat y^{\rho} \frac{ \hat q^{\mu}  \hat q_{\rho}}{1+\theta  \hat q^2}].\nonumber\\
\hat p_{\mu}=\hat q_{\mu}.\nonumber
\end{eqnarray}

The inverse transformation is found to be,
\begin{eqnarray}
\begin{array}{rcl}
&&\hat y^{\mu}=\hat x^{\mu}+\frac{\theta}{4}[\hat x^{\rho}\hat p_{\rho}\hat p^{\mu}+\hat p^{\mu}\hat p_{\rho}\hat x^{\rho}+ \hat p^{\mu}\hat x^{\rho}\hat p_{\rho}+\hat p_{\rho}\hat x^{\rho}\hat p^{\mu}]\\
&&\hat q_{\mu}=\hat p_{\mu}
\end{array}
\label{6.13}                              
\end{eqnarray}
which is just the Weyl ordered form of (\ref{6.8}), (\ref{6.9}). 

       A slightly lengthy computation reveals that the quantum Snyder algebra (\ref{2.1}) as a commutator algebra. This follows from (\ref{6.13}) by using the standard canonical commutators involving $x$ and $p$. 

\section{Deformed conformal symmetry}
In order to study the deformed symmetry associated with the dilatation and the special conformal transformation we first give an algebraic analysis followed by a dynamical treatment related to the action (\ref{3.7}).
\subsection{Dilatation symmetry}
Under the usual transformations for dilatation $\delta \hat y_{\mu} = \epsilon \hat y_{\mu},\delta \hat q_{\mu} = - \epsilon \hat q_{\mu}$, covariance of only the last relation in (\ref{2.1}) is preserved. Thus as an ansatz, we take
\begin{eqnarray}
&&\delta \hat q_{\mu} =-\epsilon \hat q_{\mu}
\label{7.2}\\
&&\delta \hat y_{\mu} = \epsilon \hat y_{\mu} + \epsilon \hat Q_{\mu}(\theta).\label{7.3}
\end{eqnarray} 
The covariance of the second relation in (\ref{2.1}) under (\ref{7.3}) yields,
\begin{equation}
[\hat Q_{\mu},\hat q_{\nu}] = -2i\theta \hat q_{\mu}\hat q_{\nu}.\label{7.6}
\end{equation}
Up to an ordering ambiguity a solution for $\hat Q_{\mu}$ is given by
\begin{equation}
\hat Q_{\mu} = -\frac{2\theta(\hat y\cdot \hat q)\hat q_{\mu}}{1+\theta \hat q^2}.\label{7.7}
\end{equation}
Requiring covariance of the $\hat y_{\mu}-\hat y_{\nu}$ bracket
in (\ref{2.1}), this ambiguity is fixed. It leads to the transformation law,
\begin{eqnarray}
\delta \hat y_{\mu} &=& \epsilon[\hat y_{\mu}-\hat y_{\rho}\frac{\theta \hat q^{\rho}\hat q_{\mu}}{1+\theta \hat q^2} - \frac{\theta \hat q^{\rho}\hat q_{\mu}}{1+\theta \hat q^2}\hat y_{\rho}].\label{7.8} 
\end{eqnarray}
The dilatation generator yielding the deformed transformations is given by,
\begin{equation}
\hat D = \frac{1}{2}[\hat y_{\rho}\frac{\hat q^{\rho}}{1+\theta \hat q^2} + \frac{\hat q^{\rho}}{1+\theta \hat q^2}\hat y_{\rho}]\label{7.9}
\end{equation}
In the limit $\theta\rightarrow {0}$, it reduces to the standard expression.
   
It is straightforward to check that although $D$ in (\ref{7.9}) is deformed,
the corresponding algebra of generators remains the same
\begin{eqnarray}
&&\left[\hat D,\hat D\right]=0\label{7.14}\\
&&\left[\hat D,\hat q_{\mu}\right]=i \hat q_{\mu}\label{7.15}\\
&&\left[\hat D,\hat J_{\mu\nu}\right]=0.\label{7.16}
\end{eqnarray}
\vspace{0.40cm}\\
{\bf Generator from N\"other's theorem} \ :\vspace{0.10cm}\\

 We take the massless form of (\ref{3.7}),
\begin{eqnarray}
S=\int d\tau[-\dot q^{\mu}y_{\mu}+\frac{\theta}{1+\theta q^2}(y.q)\dot q^{\mu}q_{\mu}-e q^2].
\label{7.2.1}
\end{eqnarray}
to study its invariance under the classical version of the deformed dilatation transformations (\ref{7.2},\ref{7.8}).
 
The total variation of the Lagrangian 
\begin{eqnarray}
\delta L &=&2\epsilon e q^2. \label{7.2.2} 
\end{eqnarray}
cannot be expressed as a total time derivative. However on the constraint shell ($q^2 = 0$), invariance is achieved, $\delta L = 0$.

Using (\ref{5.5}), the variation (\ref{7.2}) and $B=0$ (since $\delta L=0$), we obtain,
\begin{eqnarray}
G &=& -\epsilon q^{\mu}(\frac{\theta (y\cdot q)q_{\mu}}{1+\theta q^2} - y_{\mu})
\nonumber\\
&=& \frac{\epsilon (y\cdot q)}{1+\theta q^2}.\label{7.2.3}
\end{eqnarray}

     It is possible to construct the operator analogue of the above generator by following the Weyl ordered prescription,
\begin{equation}
\hat D = \frac{\epsilon}{4}[\hat y_{\rho}\frac{\hat q^{\rho}}{1+\theta \hat q^2} + \frac{\hat q^{\rho}}{1+\theta \hat q^2}\hat y_{\rho} + \hat q_{\rho}\hat y^{\rho}\frac{1}{1+\theta \hat q^2} +\frac{1}{1+\theta \hat q^2} \hat y^{\rho}\hat q_{\rho}].\label{7.2.4}
\end{equation}
The last two terms combine to give the first two terms so that the final 
expression exactly agrees with (\ref{7.9}). 
 
\subsection{Special conformal symmetry}
In order to discuss the deformed special conformal transformations we adopt a classical treatment. Making use of the transformations (\ref{6.5}), (\ref{6.6}) we construct the deformed generator from the usual expression.
After getting the deformed generator, covariance of the Snyder algebra is
shown.  

      In the ordinary commutative space the generator for the special conformal transformation is given by,
\begin{equation}
K_{\mu} = 2x_{\mu}(x^{\rho}p_{\rho}) -x^2 p_{\mu}.\label{7.3.1}
\end{equation}
Using (\ref{6.5}), (\ref{6.6}) the deformed generator is obtained
\begin{eqnarray}
K_{\mu}&=& -y^2q_{\mu} + 2y_{\mu}(y\cdot q) \frac{1}{1+\theta q^2}.\nonumber\\
&& + \theta ^2 (y\cdot q)^2 \frac{1}{(1+\theta q^2)^2}q^2 q_{\mu}.\label{7.3.2} 
\end{eqnarray}
The transformation rules for the deformed conformal transformation are given by 
\begin{eqnarray}
\delta y_{\mu} &=& \epsilon^{\nu}\{y_{\mu},K_{\nu}\}\\
&=& \{\theta y^2 q_{\mu}q_{\nu} - y^2 \delta_{\mu\nu} -6\theta
(y_{\nu}q_{\mu})(y\cdot q) \frac{1}{1+\theta q^2}\nonumber\\
&& + 2 y_{\mu}y_{\nu} + \theta ^2 (y\cdot q)^2 \frac{1}{(1+\theta q^2)^2}(q^2\delta_{\mu\nu}-\theta q_{\mu}q_{\nu}q^2 + 2q_{\mu}q_{\nu})\}\epsilon^{\nu}\label{7.3.4}\\
\delta q_{\mu} &=& \epsilon^{\nu}\{q_{\mu},K_{\nu}\}\\
&=& \{2y_{\mu}q_{\nu} -2y_{\nu}q_{\mu} -2(y\cdot q) \frac{1}{1+\theta q^2}\delta_{\mu\nu}\}\epsilon^{\nu}.\label{7.3.5}
\end{eqnarray}
These deformed conformal transformations have a smooth limit $\theta \rightarrow 0$ in which they reduce to the familiar structures in commutative
space. From these transformations it is found that 
\begin{eqnarray}
\begin{array}{rcl}
\delta\{y^{\mu},y^{\nu}\}_{DB} &=& \{\delta y^{\mu},y^{\nu}\} + \{y^{\mu},\delta y^{\mu}\}\\
&=&\theta\delta\left(y^{\mu} q^{\nu}- y^{\nu} q^{\mu}\right)\\
\delta\{q_{\mu},q_{\nu}\}_{DB}&=&0\\
\delta\{y^{\mu},q_{\nu}\}_{DB}&=&\delta\left(\delta^{\mu}_{ \nu}+\theta q^{\mu}q_{\nu}\right).
\end{array}
\label{7.3.6}
\end{eqnarray}
This is sufficient to prove the compatibility of the deformed transformation laws with the classical Snyder algebra, manifested in the form of Dirac brackets (\ref{3.23})
\vspace{0.40cm}\\
{\bf Generator from N\"other's theorem} \ :\vspace{0.40cm}\\
 We first calculate the variation of the individual pieces in the action (\ref{7.2.1})
\begin{eqnarray}
-\delta({\dot q_{\mu}y^{\mu}})&=& \{\frac{d}{d\tau}(y^2 q_{\mu} - 2D y_{\mu}) + 
 2 \dot y_{\mu}\frac{y\cdot q}{1+\theta q^2} - 2\dot y_{\mu}\nonumber\\   
&& -\theta(\dot q\cdot q)K_{\mu} + \frac{d}{d\tau}(-2\theta y_{\mu}q^2 D + 
\theta^2 D^2 q^2 q_{\mu})\nonumber\\
&& + [\theta^2 D^2 \dot q_{\mu} - \frac{d}{d\tau}(-2\theta y_{\mu}D + \theta^2
D^2 q_{\mu})]q^2\}\epsilon^{\mu},\nonumber\\
\theta \delta (\frac{y\cdot q}{1+\theta q^2})(\dot q\cdot q) &=& -\{\theta K_{\mu} (\dot q\cdot q)\}\epsilon^{\mu},\nonumber\\
\theta (\frac{y\cdot q}{1+\theta q^2}) \delta (\dot q\cdot q) &=& \{\frac{d}{d\tau}(2D^2 \theta^2 q^2 q_{\mu}- 2Dy_{\mu}\theta q^2)- 2q^2(\theta^2q_{\mu} - 
 \theta  y_{\mu})\frac{dD}{d\tau}\}\epsilon^{\mu},\nonumber\\
-e\delta (q^2)&=& \{4 e q^2 (\theta D q_{\mu} - y_{\mu})\}\epsilon^{\mu}.\label{7.3.7}       
\end{eqnarray}
Some terms are expressible as a total time derivative which are retained since these will be useful in obtaining the generator. Terms not expressible in this way drop out due to the mass shell constraint $q^2 = 0$. Combining all terms, we obtain,
\begin{eqnarray}
\delta L &=& \frac{d}{d\tau}[K_{\mu}\epsilon^{\mu}]\label{7.3.8}
\end{eqnarray} 
where $K_{\mu}$ is given in (\ref{7.3.2}). 


The variation $\delta q^{\mu}$ is obtained from (\ref{7.3.5}) while B is abstracted from (\ref{7.3.8}). From the definition (\ref{5.5}) we find,
\begin{eqnarray}
G&=&2\epsilon^{\nu}(y_{\mu}q_{\nu} - y_{\nu}q_{\mu} -\frac{y\cdot q}{1+\theta q^2}
\delta_{\mu\nu})(\theta \frac{y\cdot q}{1+\theta q^2}q_{\mu} - y_{\mu})-\epsilon^{\nu}K_{\nu}\nonumber\\
&=&2\epsilon^{\nu}K_{\nu} -\epsilon^{\nu}K_{\nu}\nonumber\\
&=& \epsilon^{\nu}K_{\nu}\label{7.3.9}   
\end{eqnarray}     
 thereby reproducing the desired definition of the deformed generator given in (\ref{7.3.2}).
\section{Discussion}
Deformed Schr\"odinger symmetry for nonrelativistic constant noncommutativity and deformed conformal-Poincar$\acute{\textrm e}$ symmetries for relativistic Snyder type noncommutativity have been studied in this chapter. The method, we followed here is quite general and can be applied in other noncommutative spaces.  We also constructed dynamical models invariant under the deformed symmetries. Using Dirac's constraint analysis, we obtain the noncommutative algebras (both constant and Snyder type) from the respective dynamical models. It is noteworthy that we did not fix any gauge to get the algebra. In this sense our method is different from some other approaches where gauge fixing was mandatory. In addition, for constant noncommutativity we have analyzed a Lagrangian involving second order time derivative. This reproduced momentum noncommutativity. Maps between the variables of commutative and noncommutative algebra are also given. Finally, we use the N\"other's theorem to derive the deformed generators from the dynamical models. These are shown to be identical with the deformed generators obtained from the algebraic approach.

As a future prospect we could develop a differential calculus involving higher order derivatives to find the differential representations of the deformed generators. The modified coproduct rule and the associated Hopf algebra can then be obtained. However the algebra of these deformed generators should remain same as the usual undeformed one.


\chapter{\label{chap:lagrangian}Noncommutative gauge theory: Lagrangian analysis}

In order to construct field theory on a noncommutative space there are two approaches. In one approach fields are treated as operators in some Hilbert space. In the other method fields are taken as some functions of commutative space variables and noncommutativity among these variables is introduced by an appropriate star ($*$) product\cite{zachos}. In this chapter we follow the second approach where a noncommutative Lagrangian and its equations of motion consist of ordinary fields and their derivatives with the replacement of ordinary product by the $*$ product. Introduction of gauge symmetry in noncommutative field theory is also possible. Recent analysis\cite{vas,wess,chai,gaume,jwess,wes} reveals that it can be done in two different ways. In one approach star deformed gauge transformations are taken, keeping the comultiplication (Leibniz) rule unchanged and in the other approach gauge transformations are taken as in the commutative case at the expense of a modified Leibniz rule.

While both approaches preserve gauge invariance of the action, an important distinction between these two approaches has been mentioned in \cite{chai,zet,gaume}. In the case of star gauge transformations, gauge symmetries act only on the fields in a similar way as in theories on commutative space time. Star gauge symmetry can thus be interpreted as a physical symmetry in the usual sense. On the other hand if ordinary gauge transformations with a twisted Leibniz rule is taken, then the transformations do not act only on the fields. Consequently it is not a physical symmetry in the conventional sense and it is obscure whether N\"other charges and Ward identities can be derived.

In this chapter both these approaches are studied within a common Lagrangian framework. To do this we recall that Lagrangian analysis of gauge symmetries for commutative space theories has been discussed using certain gauge identities\cite{gitman}. These gauge identities involve the Euler derivatives and the generators of gauge transformations. This type of analysis has been applied in various contexts \cite{shirzad,rothe,rothe1}.

In this chapter we first briefly discuss the general formulation with the Einstein-Hilbert action as an example. Then the analysis is used for the noncommutative non-Abelian gauge theory. Both the approaches are considered. In the first case, gauge generators, obtained from the gauge identity are found to be star deformation of the commutative space relations. In the other approach generators of the undeformed gauge transformations are shown to be similar with the commutative space relations. Furthermore, we find that the relation connecting the gauge  generator with the gauge identity (which is form invariant under the star and twisted gauge transformations) is neither the undeformed result nor its star deformation, as obtained in the previous treatment. Rather, it is a twisted form of the conventional (undeformed) result.

\section{General formulation}
To study the dynamics of a field from an action principle we consider a general Lagrangian involving only upto first order derivatives of the field of the form\footnote{We use the notation $x$ for the four vector $x^{\mu}=({\bf{x}},t)$.},
\begin{eqnarray}
S=\int\textrm{d}t \ L=\int \textrm{d}^4x \  \mathcal{L}
\left(q_{\alpha}({\bf{x}},t), \ \partial_iq_{\alpha}({\bf{x}},t), \ \partial_tq_{\alpha}({\bf{x}},t)\right)
\label{LLL}
\end{eqnarray}
where $\alpha$ denotes the number of fields. Also, it contains all other (e. g. symmetry) indices relevant for the problem. An arbitrary variation of this action is written as
\begin{eqnarray}
\delta S=-\int \textrm{d}^4x \ \delta q^{\alpha}({\bf{x}},t)L_{\alpha}({\bf{x}},t).
\label{lll}
\end{eqnarray}
The equations of motion are obtained by setting the Euler derivative $L$ to be zero,
\begin{eqnarray}
L_{\alpha}=0.
\end{eqnarray}
Now we vary the field $q^{\alpha}$ in the following way
\begin{eqnarray}
\delta q^{\alpha}({\bf{x}},t)=\sum_{s=0}^n(-1)^s\int\textrm{d}^3{\bf{z}} \ \frac{\partial^s\eta^b({\bf{z}},t)}{\partial t^s}\rho^{\alpha b}_{(s)}(x,z)
\label{aaa}
\end{eqnarray}
with $\eta$ and $\rho$ being the parameter and generator respectively, of the transformation. Under this variation of the field, the variation of the action is written from (\ref{lll}) as
\begin{eqnarray}
\delta S&=&-\int\textrm{d}^4x \ \int\textrm{d}^3{\bf{z}} \ \eta^b({\bf{z}},t)\rho^{\alpha b}_{(0)}(x,z)L_{\alpha}({\bf{x}},t)-\nonumber\\
&&\int \textrm{d}^4x \  \sum_{s=1}^n(-1)^s\int \textrm{d}^3{\bf{z}} \ \frac{\partial}{\partial t}\left(\frac{\partial^{s-1}\eta^b({\bf{z}},t)}{\partial t^{s-1}}\right)\rho^{\alpha b}_{(s)}(x,z)L_{\alpha}({\bf{x}},t)\nonumber\\
&=&-\int\textrm{d}^4x \ \int\textrm{d}^3{\bf{z}} \ \eta^b({\bf{z}},t)\rho^{\alpha b}_{(0)}(x,z)L_{\alpha}({\bf{x}},t)-\nonumber\\
&&\int \textrm{d}^4x \sum_{s=1}^n(-1)^{s-1}\int \textrm{d}^3{\bf{z}} \ \frac{\partial^{s-1}\eta^b({\bf{z}},t)}{\partial t^{s-1}}\frac{\partial}{\partial t}\left(\rho^{\alpha b}_{(s)}(x,z)L_{\alpha}({\bf{x}},t)\right)\nonumber\\
&=&-\int \textrm{d}^4z \ \eta^b({\bf{z}},t)\left(\int \textrm{d}^3{\bf{x}} \ \rho^{\alpha b}_{(0)}(x,z)L_{\alpha}({\bf{x}},t)\right)-\nonumber\\
&&\int \textrm{d}^4z \ \eta^b({\bf{z}},t)\left(\int \textrm{d}^3{\bf{x}} \ \frac{\partial}{\partial t}(\rho^{\alpha b}_{(1)}(x,z)L_{\alpha}({\bf{x}},t))\right)-\cdot\cdot\cdot
\label{mannmann}
\end{eqnarray}
We define a quantity\cite{gitman,shirzad}
\begin{eqnarray}
\Lambda^a({\bf{z}},t)=\left[\sum_{s=0}^n\int \textrm{d}^3{\bf{x}} \ \frac{\partial^s}{\partial t^s}\left(\rho^{\alpha a}_{(s)}(x,z)L_{\alpha}({\bf{x}},t)\right)\right].
\label{lamlam}
\end{eqnarray}
to write (\ref{mannmann}) in a compact form
\begin{eqnarray}
\delta S=-\int\textrm{d}^4z \ \eta^a({\bf{z}},t)\Lambda^a({\bf{z}},t).
\label{bakbak}
\end{eqnarray} 
If the action remains unchanged ($\delta S=0$) under the field transformation (\ref{aaa}) then it implies,
\begin{eqnarray}
\Lambda^a({\bf{z}},t)=0.
\end{eqnarray}

The last equality, which is called the gauge identity, must be true without use of any equation of motion. The expression (\ref{aaa}) defines the gauge transformations of the fields with $\rho$ being the generator of gauge transformation. 
\vspace{0.40cm}\\
{\bf Diffeomorphism symmetry as an example} \ :\vspace{0.40cm}\\
Now we use the general analysis to study the diffeomorphism symmetry of the general theory of relativity. We start from the gauge identity and compute the generators $\rho$ from (\ref{lamlam}). Then we obtain the explicit structure of the diffeomorphism transformation of the metric from (\ref{aaa}).

 The Einstein-Hilbert action is given by, 
\begin{eqnarray}
S&=&\int d^4x\mathcal{L}(g)\nonumber \\
&=&\int d^4x\sqrt{-g}R=\int d^4x\sqrt{-g}g^{\mu\nu}R_{\mu\nu}(g)
\label{cs}
\end{eqnarray}
where $R_{\mu\nu}$ is the Ricci tensor
\begin{eqnarray}
R_{\mu\nu}=\Gamma^{\lambda}_{\nu\mu,\lambda}-\Gamma^{\lambda}_{\lambda\mu,\nu}
+\Gamma^{\lambda}_{\nu\mu}\Gamma^{\sigma}_{\sigma\lambda}-\Gamma^{\sigma}_{\lambda\mu}\Gamma^{\lambda}_{\nu\sigma}.
\label{ricci}
\end{eqnarray}
The definition of the Christoffel connection in terms of the metric components,
\begin{eqnarray}
\Gamma^{\rho}_{\mu\nu}=\frac{1}{2}g^{\rho\sigma}(g_{\nu\sigma,\mu}+g_{\mu\sigma,\nu}-g_{\mu\nu,\sigma})
\label{Gamma}
\end{eqnarray}
comes from the metric compatibility condition
\begin{eqnarray}
\nabla_{\rho}g_{\mu\nu}\equiv \partial_{\rho}g_{\mu\nu}-\Gamma^{\alpha}_{\rho\mu}g_{\alpha\nu}-\Gamma^{\alpha}_{\rho\nu}g_{\mu\alpha}=0.
\label{metric}
\end{eqnarray}
Varying the action (\ref{cs}) with respect to the metric $g_{\mu\nu}$ we get the Euler derivative $L_{\mu\nu}$ i. e.
\begin{eqnarray}
\delta S=\int L^{\mu\nu}\delta g_{\mu\nu}
\end{eqnarray}
where the explicit form of  $L^{\mu\nu}$ is written as,
\begin{eqnarray}
L^{\mu\nu}=\sqrt{-g}G^{\mu\nu}=\sqrt{-g}(R^{\mu\nu}-\frac{1}{2}g^{\mu\nu}R)
\label{euler}
\end{eqnarray}
leading to the usual Einstein's equation, $L^{\mu\nu}=0$. Now to find the gauge identity we recall the Bianchi identity\cite{wineberg}
\begin{eqnarray}
\nabla_{\eta}R_{\lambda\mu\nu\kappa}+\nabla_{\nu}R_{\lambda\mu\kappa\eta}+\nabla_{\kappa}R_{\lambda\mu\eta\nu}=0
\label{bianchi}
\end{eqnarray}
where the Riemann tensor is defined as,
\begin{eqnarray}
R_{\lambda\mu\nu\kappa}=g_{\lambda\sigma}R^{\sigma}_{ \ \mu\nu\kappa}=g_{\lambda\sigma}(\Gamma^{\sigma}_{\mu\kappa,\nu}
-\Gamma^{\sigma}_{\mu\nu,\kappa}+\Gamma^{\eta}_{\mu\kappa}\Gamma^{\sigma}_{\nu\eta}-\Gamma^{\eta}_{\mu\nu}\Gamma^{\sigma}_{\kappa\eta}).
\end{eqnarray}
Contracting $\lambda$ with $\nu$ and $\mu$ with $\kappa$, in the above identity, using (\ref{metric}) we get
\begin{eqnarray}
\nabla_{\mu}G^{\mu\nu}=0.
\end{eqnarray}
This is referred as the gauge identity \cite{ortin}. Since the Euler derivative we defined in (\ref{euler}) is not $G^{\mu\nu}$ but $\sqrt{-g}G^{\mu\nu}$ we take our gauge identity as,
\begin{eqnarray}
\Lambda_{\alpha}\equiv2\nabla_{\beta}L^{\beta}_{\alpha}=0.
\label{iden}
\end{eqnarray}
The extra factor 2 is introduced for later convenience. In order to write the above equation (\ref{iden}) in a more convenient way we note that the definition of $\Gamma$ (\ref{Gamma}) can be used to write the divergence of Einstein tensor 
\begin{eqnarray}
\nabla_{\mu}G^{\mu}_{\nu}=\partial_{\mu}G^{\mu}_{\nu}+\Gamma^{\mu}_{\mu\alpha}G^{\alpha}_{\nu}-\Gamma^{\alpha}_{\mu\nu}G^{\mu}_{\alpha}
\end{eqnarray}
in the following form
\begin{eqnarray}
\nabla_{\mu}G^{\mu}_{\nu}=(\partial_{\mu}G^{\mu}_{\nu}+\frac{1}{2}g^{\mu\beta}\partial_{\alpha}g_{\mu\beta}G^{\alpha}_{\nu}-\frac{1}{2}G^{\mu\beta}\partial_{\nu}g_{\beta\mu}).
\label{nabG}
\end{eqnarray}
Now using (\ref{metric}) and (\ref{nabG}) we write the gauge identity (\ref{iden}) as,
\begin{eqnarray}
\Lambda_{\nu}=2\nabla_{\mu}L^{\mu}_{\nu}=
2\nabla_{\mu}\sqrt{-g}G^{\mu}_{\nu}&=&2\sqrt{-g}(\partial_{\mu}G^{\mu}_{\nu}+\frac{1}{2}g^{\mu\beta}\partial_{\alpha}g_{\mu\beta}G^{\alpha}_{\nu}-\frac{1}{2}G^{\mu\beta}\partial_{\nu}g_{\beta\mu})\nonumber\\
&=&2\partial_{\mu}\sqrt{-g}G^{\mu}_{\nu}-\partial_{\nu}g_{\alpha\beta}\sqrt{-g}G^{\alpha\beta}\nonumber\\
&=&2\partial_{\mu}L^{\mu}_{\nu}-\partial_{\nu}g_{\alpha\beta}L^{\alpha\beta}
\label{ident}
\end{eqnarray}
where we have used the important relation
\begin{eqnarray}
\partial_{\mu}g=gg^{\alpha\beta}\partial_{\mu}g_{\alpha\beta}.
\end{eqnarray}
In the metric formulation of gravity, (\ref{lamlam}) is rewritten as,
\begin{eqnarray}
\Lambda_{\alpha}(z)=\sum_{s=0}^{n}\int \textrm{d}^3 {\bf x}\frac{\partial^s}{\partial t^s}\left(\rho_{\mu\nu\alpha(s)}(x,z)L^{\mu\nu}(x)\right).
\label{47}
\end{eqnarray}
Comparing this equation with the identity (\ref{ident}), the generators are obtained. The expression for the nonvanishing generators are given below
\begin{eqnarray}
&&\rho_{000(0)}=-\partial_0g_{00}\delta(x-z)
\label{00}\\
&&\rho_{000(1)}=2g_{00}\delta(x-z)
\label{01}\\
&&\rho_{00k(0)}=-\partial_kg_{00}\delta(x-z)
\label{k0}\\
&&\rho_{00k(1)}=2g_{0k}\delta(x-z)
\label{k1}\\
&&\rho_{0i0(0)}=-\partial_0g_{0i}\delta(x-z)+\partial_i^z\left(g_{00}\delta(x-z)\right)\\
&&\rho_{0i0(1)}=g_{0i}\delta(x-z)\\
&&\rho_{0ik(0)}=-\partial_kg_{0i}\delta(x-z)+\partial_i^z\left(g_{0k}\delta(x-z)\right)\\
&&\rho_{0ik(1)}=g_{ki}\delta(x-z)\\
&&\rho_{ij0(0)}=-\partial_0g_{ij}\delta(x-z)+\partial_j^z\left(g_{i0}\delta(x-z)\right)+\partial_i^z\left(g_{j0}\delta(x-z)\right)\\
&&\rho_{ijk(0)}=-\partial_kg_{ij}\delta(x-z)+\partial_j^z\left(g_{ik}\delta(x-z)\right)+\partial_i^z\left(g_{jk}\delta(x-z)\right).
\label{0000}
\end{eqnarray}
We now calculate the diffeomorphism transformation of the metric $g_{\mu\nu}$ from (\ref{aaa}). It is first rewritten as,
\begin{eqnarray}
\delta g_{\mu\nu}(x)=\sum_{s=0}^{n}(-1)^s\int {\textrm d}^3 {\bf z} \ \frac{\partial^s\eta^{\alpha}(z)}{\partial t^s}\rho_{\mu\nu\alpha (s)}(x,z).
\label{58}
\end{eqnarray}
Using the generators (\ref{00}--\ref{0000}) in the above expression, the diffeomorphism law is obtained
\begin{eqnarray}
\delta g_{\mu\nu}=-\partial_{\alpha}g_{\mu\nu}\eta^{\alpha}-g_{\mu\alpha}\partial_{\nu}\eta^{\alpha}-g_{\alpha\nu}\partial_{\mu}\eta^{\alpha}.
\label{metric9}
\end{eqnarray}
This can be written in the covariant notation
\begin{eqnarray}
\delta g_{\mu\nu}=-\nabla_{\mu}\eta_{\nu}-\nabla_{\nu}\eta_{\mu}.
\end{eqnarray}
The above result expresses the diffeomorphism transformation of the metric field $g_{\mu\nu}$. It can also be obtained from the first order formulation of general relativity (Palatini formulation) for which our general formulation is applicable.
\section{Noncommutative gauge theory}
We consider a noncommutative space, where the coordinates $\hat{x}^{\mu}$ satisfy the following canonical relation{\footnote{From now we shall denote the noncommutative coordinates by $\hat x.$}}
\begin{eqnarray}
[\hat x^{\mu},\hat x^{\nu}] = i \theta^{\mu\nu} 
\label{cano}
\end{eqnarray}
where $\theta^{\mu\nu}$ is a constant antisymmetric matrix. The noncommutative coordinates satisfying (\ref{cano}) are the generators of an associative algebra $\mathcal{A}_x$\cite{W1,W2,W3}. According to the Weyl correspondence, we can associate an element of ${\cal{A}}_x$ ($W$)with a function ($f$) of classical variables $x^{\mu}$\cite{Wigner} by the unique prescription
\begin{equation}
W(f) = \frac{1}{(2\pi)^2} \int d^4 k e^{ik_{\mu}{\hat {x}}^{\mu}}{\tilde{f}}(k)
\label{Weylll}
\end{equation}
where ${\tilde{f}}(k)$ is the Fourier transform of $f(x)$.
\begin{equation}
{\tilde{f}}(k)= \frac{1}{(2\pi)^2} \int d^4 x e^{-ik_{\mu}{{x}}^{\mu}}{{f}}(x)
\label{fourier}
\end{equation}
New operators can be obtained by multiplication of $W$'s defined in (\ref{Weylll}). The classical function corresponding to this new operator is denoted by $f*g$. So the requirement $W(f)W(g) = W(f*g)$ is written as
\begin{equation}
W(f)W(g) = W(f*g) = \frac{1}{(2\pi)^{4}} \int d^4 k d^4 pe^{ik_{\mu}{\hat {x}}^{\mu}}e^{ip_{\nu}{\hat {x}}^{\nu}}{\tilde{f}}(k){\tilde{g}}(p)
\label{WWeyl}
\end{equation}
The product of two exponentials in the integral is obtained by the Baker-Campbell-Hausdorff formula
\begin{equation}
e^Ae^B = e^{A+B+\frac{1}{2}[A,B]+\frac{1}{12}([A,[A,B]]-[B,[A,B]])-\frac{1}{48}([B,[A,[A,B]]]+[A,[B,[A,B]]])...}
\label{BCH}
\end{equation}
to write (\ref{WWeyl}) as
\begin{equation}
W(f*g) = \frac{1}{(2\pi)^{4}} \int d^4 k d^4 pe^{i(k_{\mu}+p_{\mu}){\hat {x}}^{\mu}-\frac{i}{2}k_{\mu}p_{\nu}\theta^{\mu\nu}}
{\tilde{f}}(k){\tilde{g}}(p)
\end{equation}
Comparing the above equation with (\ref{Weylll}) we get the expression for $\widetilde{f*g}(k,p)$
\begin{equation}
\widetilde{f*g}(k,p)=e^{-\frac{i}{2}k_{\mu}p_{\nu}\theta^{\mu\nu}}
{\tilde{f}}(k){\tilde{g}}(p)
\end{equation}
$f*g$ can now be read off from (\ref{fourier})
\begin{equation}
f*g=\frac{1}{(2\pi)^{4}} \int d^4 k d^4 pe^{i(k_{\mu}+p_{\mu}){{x}}^{\mu}-\frac{i}{2}k_{\mu}p_{\nu}\theta^{\mu\nu}}
{\tilde{f}}(k){\tilde{g}}(p)=e^{\frac{i}{2}\frac{\partial}{\partial x^{\mu}}\theta^{\mu\nu}\frac{\partial}{\partial y^{\nu}}}f(x)g(y)|_{y\rightarrow x}
\label{str}
\end{equation}
This is the Moyal--Weyl $*$ product\cite{moyal}. Star product of functions within an integral satisfy the well known property 
\begin{eqnarray}
\int \textrm{d}^4 x \ A(x)*B(x)=\int \textrm{d}^4 x \ A(x)B(x)=\int \textrm{d}^4x \ B(x)*A(x)
\label{b1}
\end{eqnarray}
and the trace like property,
\begin{eqnarray}
\int \textrm{d}^4 x \ (A*B*C)=\int \textrm{d}^4 x \ (B*C*A)=\int\textrm{d}^4 x \ (C*A*B)
\label{b2}
\end{eqnarray}
\vspace{0.40cm}\\
{\bf Noncommutative action and the gauge transformations}
 \ :\vspace{0.40cm}\\
Inserting the $*$ product in place of the ordinary product we construct the noncommutative free Dirac action in four dimension
\begin{eqnarray}
S_F=\int \textrm{d}^4x \ [{\hat {\bar{\psi}}}(x)*(i\gamma^{\mu}\partial_{\mu}-m){\hat\psi}(x)].
\end{eqnarray}
To introduce the connection, the commutative covariant derivative 
\begin{eqnarray}
D_{\mu}=\partial_{\mu}+igA_{\mu}
\end{eqnarray}
is replaced by the noncommutative covariant derivative
\begin{eqnarray}
D_{\mu}*=\partial_{\mu}+ig{\hat A}_{\mu}*.
\end{eqnarray}
This leads to the field strength tensor
\begin{eqnarray}
[D_{\mu}*,D_{\nu}*]=ig{\hat F}_{\mu\nu}
\end{eqnarray}
with
\begin{eqnarray}
{\hat F}_{\mu\nu}(x)\equiv\partial_{\mu}{\hat A}_{\nu}(x)-\partial_{\nu}{\hat A}_{\mu}(x)+ig[{\hat A}_{\mu}(x),{\hat A}_{\nu}(x)]_*.
\label{f}
\end{eqnarray}
Here the star commutator is given by
\begin{eqnarray}
[{\hat A}_{\mu}(x),{\hat A}_{\nu}(x)]_*={\hat A}_{\mu}(x)*{\hat A}_{\nu}(x)-{\hat A}_{\nu}(x)*{\hat A}_{\mu}(x).
\end{eqnarray}
The gauge field defined in this way is coupled with the Dirac field to give the complete action for the noncommutative non-Abelian theory  
\begin{eqnarray}
S=\int \textrm{d}^4x \ [-\frac{1}{2}\textrm{Tr}({\hat F}_{\mu\nu}(x)*{\hat F}^{\mu\nu}(x))+{\hat {\bar{\psi}}}(x)*(i\gamma^{\mu}D_{\mu}*-m){\hat \psi}(x)].
\label{lag}
\end{eqnarray}

The action (\ref{lag}) is invariant under both deformed gauge transformations,
\begin{eqnarray}
&&\delta {\hat A}_{\mu}=\mathcal{D}_{\mu}*\hat\alpha=\partial_{\mu}\hat\alpha+ig({\hat A}_{\mu}*\hat\alpha-\hat\alpha*{\hat A}_{\mu}),\label{Amu}\\
&&\delta {\hat F}_{\mu\nu}=ig[{\hat F}_{\mu\nu},\hat\alpha]_*=ig({\hat F}_{\mu\nu}*\hat\alpha-\hat\alpha*{\hat F}_{\mu\nu})\label{Fmunu}\\
&&\delta {\hat \psi}=-ig\hat\alpha*{\hat \psi}\label{si}\\
&&\delta {\hat{\bar{\psi}}}=ig{\hat{\bar{\psi}}}*\hat\alpha\label{sibar}
\end{eqnarray}
with the usual Leibniz rule,
\begin{eqnarray}
\delta (f*g)=(\delta f)*g+f*(\delta g)
\label{tX}
\end{eqnarray}
as well as the undeformed gauge transformations
\begin{eqnarray}
&&\delta_{\hat\alpha} {\hat A}_{\mu}=\mathcal{D}_{\mu}\hat\alpha=\partial_{\mu}\hat\alpha+ig({\hat A}_{\mu}\hat\alpha-\hat\alpha {\hat A}_{\mu}),\label{YY1}\\
&&\delta_{\hat\alpha} {\hat F}_{\mu\nu}=ig[{\hat F}_{\mu\nu},\hat\alpha]=ig({\hat F}_{\mu\nu}\hat\alpha-\hat\alpha {\hat F}_{\mu\nu})\\
&&\delta_{\hat\alpha} \hat \psi=-ig\hat\alpha\hat \psi\\
&&\delta_{\hat\alpha} {\hat{\bar{\psi}}}=ig{\hat{\bar{\psi}}}\hat\alpha
\label{YY}
\end{eqnarray}
with the twisted Leibniz rule\cite{vas,wess,jwess},
\begin{eqnarray}
\delta_{\hat\alpha}(f*g)&=&\sum_n(\frac{-i}{2})^n\frac{\theta^{\mu_1\nu_1}\cdot \cdot \cdot\theta^{\mu_n\nu_n}}{n!}\nonumber\\
&&(\delta_{\partial_{\mu_1}\cdot \cdot \cdot\partial_{\mu_n}\hat\alpha}f*\partial_{\nu_1}\cdot \cdot \cdot\partial_{\nu_n}g+\partial_{\mu_1}\cdot \cdot \cdot\partial_{\mu_n}f*\delta_{\partial_{\nu_1}\cdot \cdot \cdot\partial_{\nu_n}\hat\alpha}g)
\label{co}
\end{eqnarray}

For deformed gauge symmetry (\ref{Amu}--\ref{tX}), it is obvious from the definition of the field strength tensor (\ref{f}) and the gauge transformations (\ref{Amu}) that, in general, both ${\hat A}_{\mu}$ as well as ${\hat F}_{\mu\nu}$ are enveloping algebra valued for deformed gauge symmetry. For the case of twisted gauge symmetry (\ref{YY1}--\ref{YY},\ref{co}), however, one has to consider the equation of motion derived later (see (\ref{eqn})), interpreted as equations for the gauge field ${\hat A}_{\mu}$, to conclude that here also ${\hat A}_{\mu}$ is enveloping algebra valued. The field tensor ${\hat F}_{\mu\nu}$, by its very definition (\ref{f}), is of course enveloping algebra valued. Thus, in both treatments of gauge symmetry, ${\hat A}_{\mu}$ and ${\hat F}_{\mu\nu}$ are enveloping algebra valued\cite{jwess}. This implies that the gauge potential ${\hat A}_{\mu}$ has to be expanded over a basis of the vector space spanned by the homogeneous polynomials in the generators of the Lie algebra,
\begin{eqnarray}
{\hat A}^{\mu}(x)&=&{\hat A}^{\mu}_a(x):T^a:+{\hat A}^{\mu}_{a_1a_2}(x):T^{a_1}T^{a_2}:\nonumber\\&&+...{\hat A}^{\mu}_{a_1a_2...a_n}(x):T^{a_1}T^{a_2}...T^{a_n}:+...
\label{amu}
\end{eqnarray}
where the double dots indicate totally symmetrised products,
\begin{eqnarray}
:T^a:&=&T^a\nonumber\\
:T^{a_1}T^{a_2}:&=&\frac{1}{2}\{T^{a_1},T^{a_2}\}=\frac{1}{2}(T^{a_1}T^{a_2}+T^{a_2}T^{a_1})
\label{ta}\\
:T^{a_1}...T^{a_n}:&=&\frac{1}{n!}\sum_{\pi\in S_n}T^{a_{\pi(1)}}...T^{a_{\pi(n)}}\nonumber
\end{eqnarray}
These symmetrised products may be simplified by using the basic Lie algebraic relation,
\begin{eqnarray}
&&[T^a,T^b]=if^{abc}T^c
\label{fabc}
\end{eqnarray}
where $f^{abc}$ are the structure constants.

Apart from forming a Lie algebra (\ref{fabc}) the generators (\ref{ta}) also close under anti commutation\cite{amorim,rabinb},
\begin{eqnarray}
&&\{T^a,T^b\}=d^{abc}T^c.
\label{dabc}
\end{eqnarray}
The simpler nontrivial algebra that matches these conditions is $U(N)$ in the representation given by $N\times N$ hermitian matrices.

Following \cite{banora,armoni} it is feasible to choose $T^1=\frac{1}{\sqrt{2N}}\mathbb{I}_{(N\times N)}$ and the remaining $N^2-1$ of the $T$'s as in $SU(N)$. Then the trace condition also follows as,
\begin{eqnarray}
{\textrm{Tr}}(T^aT^b)=\frac{1}{2}\delta^{ab}
\label{trace}
\end{eqnarray}
and $f^{abc}$, $d^{abc}$ are completely antisymmetric and completely symmetric respectively.

From now we will work with these simplifications. The gauge potential and the field strength will be explicitly written as,
\begin{eqnarray}
{\hat A}_{\mu}&=&{\hat A}_{\mu}^aT^a\\
{\hat F}_{\mu\nu}&=&{\hat F}_{\mu\nu}^aT^a
\end{eqnarray}
where the $T^a$'s are the $N^2$ hermitian generators of $U(N)$ that satisfy the conditions (\ref{fabc}), (\ref{dabc}) and (\ref{trace}).

In order to derive the field equations we first vary the gauge field ${\hat A}$ to get the equation of motion for the gauge field,
\begin{eqnarray}
\partial_{\mu}{\hat F}^{\mu\nu}+ig[{\hat A}_{\mu},{\hat F}^{\mu\nu}]_*+\hat{j}^{\nu}=0
\label{eqn}
\end{eqnarray}
where $\hat{j}^{\nu}$ is the fermionic current
\begin{eqnarray}
\hat{j}^{\nu}=g\hat\psi_j(\gamma^{\nu})_{ij}*{\hat{\bar{\psi}}}_i.
\label{14a1}
\end{eqnarray}
The variation of the matter field ${\hat{\bar{\psi}}}$ and $\hat\psi$ in the action (\ref{lag}) 
gives the equation of motion
\begin{eqnarray}
i\gamma^{\mu}\partial_{\mu}\hat\psi-g\gamma^{\mu}{\hat A}_{\mu}*\hat\psi-m\hat\psi=0.
\label{eqpsi}
\end{eqnarray}
\begin{eqnarray}
i\partial_{\mu}{\hat{\bar{\psi}}}\gamma^{\mu}+g{\hat{\bar{\psi}}}*\gamma^{\mu}{\hat A}_{\mu}+m{\hat{\bar{\psi}}}=0.
\label{eqpsibar}
\end{eqnarray}
Operating $\partial_{\nu}$ on (\ref{eqn}) we get a current conservation law\cite{wes}
\begin{eqnarray}
\partial_{\nu}\hat{J}^{\nu}=0; \ \hat{J}^{\nu}\equiv ig[{\hat A}_{\mu},{\hat F}^{\mu\nu}]_*+\hat{j}^{\nu}
\label{j}
\end{eqnarray}
It is also possible to obtain the current defined in (\ref{j}) from (\ref{lag}) by using a N\"other-like procedure\cite{gaume}. Making the following ``global" transformations, 
\begin{eqnarray} 
&&\delta {\hat A}_{\mu}(x)=ig[\omega(x),{\hat A}_{\mu}(x)]_*\\
&&\delta\hat\psi(x)=-ig\omega(x)*\hat\psi(x)\\
&&\delta{\hat{\bar{\psi}}}(x)=ig{\hat{\bar{\psi}}}(x)*\omega(x)
\end{eqnarray}
if we set $\omega(x)$ to a constant at the end of the calculation, the conserved current (\ref{j}) follows from (\ref{lag}).

\section{Analysis for star deformed gauge symmetry}
We discussed in section 5.1 that, the presence of gauge symmetry is characterized by an identity called the gauge identity. The gauge generators are related to this identity. As an application, the gauge identity is first abstracted by simple inspection, after which the generator is read off. The gauge transformations of the fields can then be computed from this generator.

 In order to apply that analysis in the noncommutative space we write the noncommutative version of the gauge identity
\begin{eqnarray}
\Lambda^a({\bf{z}},t)=\left[\sum_{s=0}^n\int \textrm{d}^3{\bf{x}} \ \frac{\partial^s}{\partial t^s}\left(\rho^{\alpha a}_{(s)}(x,z)*L_{\alpha}({\bf{x}},t)\right)\right].
\label{lam}
\end{eqnarray}
which is the analogue of (\ref{lamlam}) and the noncommutative gauge transformations of the fields
\begin{eqnarray}
\delta q^{\alpha}({\bf{x}},t)=\sum_{s=0}^n(-1)^s\int\textrm{d}^3{\bf{z}} \ \frac{\partial^s\eta^b({\bf{z}},t)}{\partial t^s}*\rho^{\alpha b}_{(s)}(x,z)
\label{a}
\end{eqnarray}
which is the analogue of (\ref{aaa})

In order to find the gauge identity we have to first derive the Euler derivatives. This is simply done by considering an arbitrary variation of the action (\ref{lag}), expressed in terms of the variations of the basic fields,
\begin{eqnarray}
\delta S=-\int \textrm{d}^4x \ \delta {\hat A}_{\mu}^{a}*L^{\mu a}+\delta\hat\psi_i*L_i+\delta{\hat{\bar{\psi}}}_i*L'_i
\label{tm}
\end{eqnarray}
where $L_{\mu}^a, \ L_i$ and $L'_i$ are the Euler derivatives 
\begin{eqnarray}
&&L^{\mu a}=-\left(\mathcal{D}_{\sigma}*{\hat F}^{\sigma\mu}\right)^a-g\hat\psi_j(\gamma^{\mu}T^a)_{ij}*{\hat{\bar{\psi}}}_i
\label{eu1}\\
&&L_i=-i\partial_{\mu}{\hat{\bar{\psi}}}_j(\gamma^{\mu})_{ji}-g{\hat{\bar{\psi}}}_j*(\gamma^{\mu}{\hat A}_{\mu}^aT^a)_{ji}-m{\hat{\bar{\psi}}}_i
\label{eu2}\\
&&L_i'=-i(\gamma^{\mu})_{ij}\partial_{\mu}{\hat\psi}_j+g(\gamma^{\mu}{\hat A}_{\mu}^aT^a)_{ij}*{\hat\psi}_j+m\hat\psi_i.
\label{eu3}
\end{eqnarray}
Here the noncommutative covariant derivative $\mathcal{D}*$ is defined in the adjoint representation (\ref{Amu}) i. e. 
\begin{eqnarray}
&&\mathcal{D}_{\mu}*\xi=\partial_{\mu}\xi+ig[{\hat A}_{\mu},\xi]_*;\\
&&(\mathcal{D}_{\mu}*\xi)^a=\partial_{\mu}\xi^{a}-\frac{g}{2}f^{abc}\{{\hat A}^b_{\mu},\xi^{c}\}_*+i\frac{g}{2}d^{abc}[{\hat A}^b_{\mu},\xi^{c}]_*
\label{D}
\end{eqnarray}
where we have used (\ref{fabc}) and (\ref{dabc}). We now define a quantity $\Lambda$, involving the various Euler derivatives of the system as,
\begin{eqnarray}
\Lambda^a\equiv-\left(\mathcal{D}^{\mu}*L_{\mu}\right)^a-igT^a_{ij}{\hat\psi}_j*L_i-igT^a_{ji}L_i'*{\hat{\bar{\psi}}}_j.
\label{lambdaa}
\end{eqnarray}
Exploiting (\ref{D}) and (\ref{eu1},\ref{eu2},\ref{eu3}) the above expression, by an explicit calculation, is found out to be zero, i. e. it vanishes identically without using any equations of motion,
\begin{eqnarray}
\Lambda^a\equiv-\left(\mathcal{D}^{\mu}*L_{\mu}\right)^a-igT^a_{ij}{\hat\psi}_j*L_i-igT^a_{ji}L_i'*{\hat{\bar{\psi}}}_j=0.
\label{lambda}
\end{eqnarray}

The above relation is the cherished gauge identity for the model (\ref{lag}). The structure of $\Lambda^a$ in (\ref{lambdaa}) is similar to the general form (\ref{lam}) in the sense that it involves the appropriate Euler derivatives. To find the generator $\rho$ let us write (\ref{lam}) in a convenient way which is more suitable for our particular model,
\begin{eqnarray}
\Lambda^a({\bf{z}},t)&=&\sum_s\int \textrm{d}^3{\bf{x}} \ \frac{\partial^s}{\partial t^s}\left(\rho^{b\mu a}_{(s)}(x,z)*L^b_{\mu}({\bf{x}},t)\right)+\nonumber\\
&&\sum_s\int \textrm{d}^3{\bf{x}} \ \frac{\partial^s}{\partial t^s}\left(\phi^a_i(x,z)*L_i({\bf{x}},t)+\phi'^a_i(x,z)*L'_i({\bf{x}},t)\right).
\label{eq}
\end{eqnarray}
The values of the generators $\rho$, $\phi$ and $\phi'$ are obtained by comparing (\ref{lambdaa}) and (\ref{eq}). Since the calculations involve some subtlety due to the noncommutative nature of the coordinates, few intermediate steps are presented here. The contribution coming from the space component of the gauge field Euler derivative $L_{\mu}$ is written from (\ref{lambda}) as
\begin{eqnarray}
\Lambda^a|_{L_i}&=&-\left(\mathcal{D}^{i}*L_{i}\right)^a\nonumber\\
&=&\frac{g}{2}f^{abc}\{{\hat A}^{ib},L_i^c\}_*-i\frac{g}{2}d^{abc}[{\hat A}^{ib},L_i^c]_*-\partial^{iz}L_{i}^a.
\label{ku}
\end{eqnarray}

Using the properties (\ref{b1},\ref{b2}), (\ref{ku}) is written in the following way
\begin{eqnarray}
&&\Lambda^a|_{L_i}({\bf{z}},t)\nonumber\\
&=&-\int \textrm{d}^3{\bf{x}} \ \frac{g}{2}\left(f^{abc}\{\delta^3({\bf{x}}-{\bf{z}}),{\hat A}^{ic}(x)\}_*+id^{abc}[\delta^3({\bf{x}}-{\bf{z}}),{\hat A}^{ic}(x)]_*\right)*L_i^b(x)\nonumber\\
&&-\int \textrm{d}^3{\bf{x}} \ \delta^{ab}\partial^{i{\bf{z}}}\delta^3({\bf{x}}-{\bf{z}})L_{i}^b(x).
\label{bdr}
\end{eqnarray}
This has to be identified with the $s=0$ contribution coming from (\ref{eq}) which is given by,
\begin{eqnarray}
\Lambda^a|_{L_i}({\bf{z}},t)=\int \textrm{d}^3{\bf{x}} \ \left(\rho^{bi a}_{(0)}(x,z)*L^b_{i}({\bf{x}},t)\right).
\label{mo}
\end{eqnarray}
Comparing (\ref{bdr}) and (\ref{mo}) we obtain,
\begin{eqnarray}
\rho^{bia}_{(0)}(x,z)&=&-\delta^{ab}\partial^{i{\bf{z}}}\delta^3({\bf{x}}-{\bf{z}})-\nonumber\\
 &&\frac{g}{2}f^{abc}\{\delta^3({\bf{x}}-{\bf{z}}),{\hat A}^{ic}(x)\}_*-i\frac{g}{2}d^{abc}[\delta^3({\bf{x}}-{\bf{z}}),{\hat A}^{ic}(x)]_*.
\label{37}
\end{eqnarray}
Other components of the gauge generator can be obtained in a similar way. Here we give the full expressions of these components which will be useful in finding the gauge transformations of the different fields.
\begin{eqnarray}
\rho^{b0a}_{(0)}(x,z)&=&-\frac{g}{2}f^{abc}\{\delta^3({\bf{x}}-{\bf{z}}),{\hat A}_0^c(x)\}_*-i\frac{g}{2}d^{abc}[\delta^3({\bf{x}}-{\bf{z}}),{\hat A}_0^c(x)]_*
\label{a1}\\
\rho^{b0a}_{(1)}(x,z)&=&-\delta^{ab}\delta^3({\bf{x}}-{\bf{z}})
\label{a2}\\
\phi^a_{i(0)}(x,z)&=&-igT^a_{ij}\delta^3({\bf{x}}-{\bf{z}})*\hat\psi_j(x)\\
\phi'^a_{i(0)}(x,z)&=&-igT^a_{ji}{\hat{\bar{\psi}}}_j(x)*\delta^3({\bf{x}}-{\bf{z}})
\end{eqnarray}

Let us next consider the gauge transformations. From (\ref{a}) we write the gauge transformation equation for the space component of the gauge field
\begin{eqnarray}
\delta {\hat A}^{i a}({\bf{x}},t)&=&\sum_s(-1)^s\int\textrm{d}^3{\bf{z}} \ \frac{\partial^s\hat\alpha^b({\bf{z}},t)}{\partial t^s}*\rho^{aib}_{(s)}(x,z)\nonumber\\
&=&\int\textrm{d}^3{\bf{z}} \ \left(\hat\alpha^b({\bf{z}},t)*\rho^{aib}_{(0)}(x,z)\right)
\label{mono}
\end{eqnarray}
where we have changed the notation $\eta$ by $\hat\alpha$. Exploiting the identity\cite{amorim,rabinb}
\begin{eqnarray}
A(x)*\delta(x-z)=\delta(x-z)*A(z)
\label{delta*}
\end{eqnarray}
 and interchanging $a, \ b$, the generator (\ref{37}) is recast as,
\begin{eqnarray}
\rho^{aib}_{(0)}(x,z)&=&-\delta^{ab}\partial^{i{\bf{z}}}\delta^3({\bf{x}}-{\bf{z}})+\nonumber\\
 &&\frac{g}{2}f^{abc}\{\delta^3({\bf{x}}-{\bf{z}}),{\hat A}^{ic}(z)\}_*+i\frac{g}{2}d^{abc}[\delta^3({\bf{x}}-{\bf{z}}),{\hat A}^{ic}(z)]_*
\label{a3}
\end{eqnarray}

Use of (\ref{a3}) along with the identities (\ref{b1}) and (\ref{b2}) in (\ref{mono}) implies that
\begin{eqnarray}
\delta {\hat A}^{ia}=\partial^{i}\hat\alpha^{a}-\frac{g}{2}f^{abc}\{{\hat A}^{ib},\hat\alpha^{c}\}_*+i\frac{g}{2}d^{abc}[{\hat A}^{ib},\hat\alpha^{c}]_*=(\mathcal{D}^{i}*\hat\alpha)^a
\label{r1}
\end{eqnarray}
where the operator $\mathcal{D}$ had already been defined in (\ref{D}). Similarly the generators (\ref{a1},\ref{a2}) lead to the zeroth component 
\begin{eqnarray}
\delta {\hat A}^{0a}=\partial^{0}\hat\alpha^{a}-\frac{g}{2}f^{abc}\{{\hat A}^{0b},\hat\alpha^{c}\}_*+i\frac{g}{2}d^{abc}[{\hat A}^{0b},\hat\alpha^{c}]_*=(\mathcal{D}^{0}*\hat\alpha)^a.
\label{r2}
\end{eqnarray}
Combining the two results (\ref{r1}) and (\ref{r2}) we get the following star covariant gauge transformation rule for the gauge field
\begin{eqnarray}
\delta {\hat A}^{\mu a}=(\mathcal{D}^{\mu}*\hat\alpha)^a
\label{chd}
\end{eqnarray}
Using the above equation the gauge variation of the field strength tensor is obtained from its definition (\ref{f})
\begin{eqnarray}
\delta {\hat F}_{\mu\nu}=ig[{\hat F}_{\mu\nu},\hat\alpha]_*
\end{eqnarray}
The star gauge transformation of the matter fields are obtained in a similar manner
\begin{eqnarray}
&&\delta\hat\psi_i(x)=-ig\hat\alpha^a(x)*T^a_{ij}\hat\psi_j(x)
\label{cd}\\
&&\delta{\hat{\bar{\psi}}}_i(x)=igT^a_{ji}{\hat{\bar{\psi}}}_j(x)*\hat\alpha^a(x)
\label{cdh}
\end{eqnarray}

Thus the star gauge transformations of all the fields have been systematically obtained. They reproduced the results (\ref{chd},\ref{cd},\ref{cdh}) previously stated in Section 5.2 (\ref{Amu}--\ref{sibar}) under which the action (\ref{lag}) is invariant.

 The generators $\rho$ are mapped with the gauge identity $\Lambda^a$ (\ref{lambda}) by the relation (\ref{lam}). If we set $\theta=0$, then these just correspond to the usual commutative space results for Yang--Mills theory in the presence of matter\cite{gitman}. This implies that, as it occurs in the gauge transformations, the mapping (\ref{lam}) is also a star deformation of the usual undeformed (commutative space) map.

The analysis presented above is very general and can be done for the first order action for noncommutative non-Abelian theory. Let us briefly summarize the results. The first order action for the pure gauge field is given by,
\begin{eqnarray}
S=&\int& \textrm{d}^4x \ [\frac{1}{2}\textrm{Tr}({\hat F}_{\mu\nu}(x)*{\hat F}^{\mu\nu}(x))-\nonumber\\
&&\textrm{Tr}{\hat F}_{\mu\nu}(x)*(\partial^{\mu}{\hat A}^{\nu}(x)-\partial^{\nu}{\hat A}^{\mu}(x)+ig[{\hat A}^{\mu}(x),{\hat A}^{\nu}(x)]_*
)]
\end{eqnarray}
Here ${\hat A}_{\mu}$ and ${\hat F}_{\mu\nu}$ are treated as independent fields. Variation of these two fields lead to the following equations of motion,
\begin{eqnarray}
&&L^{\mu}=-D_{\nu}*{\hat F}^{\nu\mu}=0\\
&&L^{\mu\nu}=-\frac{1}{2}[{\hat F}^{\mu\nu}-(\partial^{\mu}{\hat A}^{\nu}-\partial^{\nu}{\hat A}^{\mu}+ig[{\hat A}^{\mu},{\hat A}^{\nu}]_*
)]=0
\end{eqnarray}
The gauge identity containing the Euler derivatives $L_{\mu}$ and $L_{\mu\nu}$ is given by,
\begin{eqnarray}
-(D^{\mu}*L_{\mu})-ig[{\hat F}^{\mu\nu},L_{\mu\nu}]_*=0
\label{4.102}
\end{eqnarray}
Now (\ref{lam}) is written in the following way, 
\begin{eqnarray}
\Lambda^a({\bf{z}},t)=\left[\sum_{s=0}^n\int \textrm{d}^3{\bf{x}} \ \frac{\partial^s}{\partial t^s}\left(\rho^{b\mu a}_{(s)}(x,z)*L_{\mu}^b({\bf{x}},t)+\rho^{b\mu\nu a}_{(s)}(x,z)*L_{\mu\nu}^b({\bf{x}},t)\right)\right].
\label{4.105}
\end{eqnarray}
Comparison of (\ref{4.102}) and (\ref{4.105}) gives the generators (\ref{37}--\ref{a2}) and the new generator
\begin{eqnarray}
\rho^{b\mu\nu a}_{(0)}(x,z)=-\frac{g}{2}f^{abc}\{{\hat F}^{\mu\nu c}(x),\delta^3({\bf{x}}-{\bf{z}})\}_*+i\frac{g}{2}d^{abc}[{\hat F}^{\mu\nu c}(x),\delta^3({\bf{x}}-{\bf{z}})]_*
\label{robmu}
\end{eqnarray}
Using (\ref{robmu}) in (\ref{a}), gauge variation of the ${\hat F}^{\mu\nu}$ is obtained independently
\begin{eqnarray}
\delta {\hat F}^{\mu\nu a}(x)&=&\int \textrm{d}^3{\bf{z}}\hat\alpha^b(z)*\rho^{a\mu\nu b}_{(0)}(x,z)\\&=&ig[{\hat F}^{\mu\nu},\hat\alpha]^a.
\end{eqnarray}
The gauge variation of the $\hat A$ field can be obtained similarly.

Let us now mention a technical point. In obtaining the gauge transformations -- say (\ref{r1}) from (\ref{mono}), use is made of identities like (\ref{b1}), (\ref{b2}) which are strictly valid over the whole four dimensional space time. Since (\ref{mono}) involves only the space integral, manipulations based on these identities imply only space-space noncommutativity. This is quite reminiscent of the Hamiltonian formulation of gauge symmetries\cite{rabinb} where $\theta^{0i}=0$ from the beginning. 

\section{Analysis for twisted gauge symmetry}
For simplicity we take the pure gauge theory
\begin{eqnarray}
S=-\frac{1}{2}\int \textrm{d}^4x \ {\textrm {Tr}}({\hat F}_{\mu\nu}(x)*{\hat F}^{\mu\nu}(x))
\label{s}
\end{eqnarray}
where the field strength tensor was defined in (\ref{f}). 

Using the undeformed gauge transformation (\ref{YY1}) and the deformed coproduct rule (\ref{co}), the variation of the (star) product of gauge fields is also seen to be undeformed,
\begin{eqnarray}
\delta_{\hat\alpha}({\hat A}_{\mu}*{\hat A}_{\nu})=\partial_{\mu}\hat\alpha {\hat A}_{\nu}+{\hat A}_{\mu}\partial_{\nu}\hat\alpha-ig[\hat\alpha,({\hat A}_{\mu}*{\hat A}_{\nu})]
\label{gag}
\end{eqnarray}
which is the exact analogue of,
\begin{eqnarray}
\delta(A_{\mu}A_{\nu})&=&\left(\partial_{\mu}\alpha+ig[A_{\mu},\alpha]\right)A_{\nu}+A_{\mu}\left(\partial_{\nu}\alpha+ig[A_{\nu},\alpha]\right)\nonumber\\
&=&\partial_{\mu}\alpha {A}_{\nu}+{A}_{\mu}\partial_{\nu}\alpha-ig[\alpha,({A}_{\mu}{A}_{\nu})].
\end{eqnarray}
Here $A_{\mu}$ is the commutative space gauge field with normal gauge transformation. The above result is used to find the gauge transformation of the field strength tensor
\begin{eqnarray}
\delta_{\hat\alpha}{\hat F}_{\mu\nu}&=&\partial_{\mu}\delta_{\hat\alpha}{\hat A}_{\nu}-\partial_{\nu}\delta_{\hat\alpha}{\hat A}_{\nu}+ig\delta_{\hat\alpha}[{\hat A}_{\mu},{\hat A}_{\nu}]_*\\
&=&\partial_{\mu}(\partial_{\nu}\hat\alpha+ig[{\hat A}_{\nu},\hat\alpha])-\partial_{\nu}(\partial_{\mu}\hat\alpha+ig[{\hat A}_{\mu},\hat\alpha])\nonumber\\&&+ig\left([\partial_{\mu}\hat\alpha,{\hat A}_{\nu}]+[{\hat A}_{\mu},\partial_{\nu}\hat\alpha]-ig[\hat\alpha,[{\hat A}_{\mu},{\hat A}_{\nu}]_*]\right)
\label{GF}\\
&=&-ig[\hat\alpha,{\hat F}_{\mu\nu}].
\end{eqnarray}
Likewise one finds,
\begin{eqnarray}
\delta_{\hat\alpha}({\hat F}^{\mu\nu}*{\hat F}_{\mu\nu})=-ig[\hat\alpha,{\hat F}^{\mu\nu}*{\hat F}_{\mu\nu}]
\end{eqnarray}
Both ${\hat F}_{\mu\nu}$ and ${\hat F}_{\mu\nu}*{\hat F}^{\mu\nu}$ have the usual (undeformed) transformation properties. Thus the action (\ref{s}) is invariant under the gauge transformation (\ref{gag}) and the deformed coproduct rule (\ref{co}).

There is another way of interpreting the gauge invariance which makes contact with the gauge identity. Making a gauge variation of the action (\ref{s}) and taking into account the twisted coproduct rule (\ref{co}), we get
\begin{eqnarray}
\delta_{\hat\alpha}S&=&-\frac{1}{2}\int \textrm{d}^4x \ {\textrm {Tr}}\delta_{\hat\alpha}({\hat F}_{\mu\nu}*{\hat F}^{\mu\nu})\\
&=&-\frac{1}{2}\int \textrm{d}^4x \ [{\textrm {Tr}}(\delta_{\hat\alpha}{\hat F}_{\mu\nu}*{\hat F}^{\mu\nu}+{\hat F}_{\mu\nu}*\delta_{\hat\alpha}{\hat F}^{\mu\nu}\nonumber\\&&-\frac{i}{2}\theta^{\mu_1\nu_1}(\delta_{\partial_{\mu_1}\hat\alpha}{\hat F}_{\mu\nu}*\partial_{\nu_1}{\hat F}^{\mu\nu}+\partial_{\mu_1}{\hat F}_{\mu\nu}*\delta_{\partial_{\nu_1}\hat\alpha}{\hat F}^{\mu\nu})\nonumber\\&&-\frac{1}{8}\theta^{\mu_1\nu_1}\theta^{\mu_2\nu_2}(\delta_{\partial_{\mu_1}\partial_{\mu_2}\hat\alpha}{\hat F}_{\mu\nu}*\partial_{\nu_1}\partial_{\nu_2}{\hat F}^{\mu\nu}+\partial_{\mu_1}\partial_{\mu_2}{\hat F}_{\mu\nu}*\delta_{\partial_{\nu_1}\partial_{\nu_2}\hat\alpha}{\hat F}^{\mu\nu})\nonumber\\&&+\cdot\cdot\cdot)].
\label{delS}
\end{eqnarray}
Now using the result (\ref{GF}) each term of (\ref{delS}) can be computed separately. For example we concentrate on the first term. Using the identity (\ref{b1}) and the trace condition (\ref{trace}) we write the first term as 
\begin{eqnarray}
\delta_{\hat\alpha}S|_{{\textrm{1st term}}}&=&-\frac{1}{4}\int \textrm{d}^4x \ (\delta_{\hat\alpha}{\hat F}^{\mu\nu a}*{\hat F}_{\mu\nu}^a+{\hat F}^{\mu\nu a}*\delta_{\hat\alpha}{\hat F}_{\mu\nu}^a)\\
&=&-\frac{1}{2}\int \textrm{d}^4x \ \delta_{\hat\alpha}{\hat F}^{\mu\nu a}{\hat F}_{\mu\nu}^a.
\end{eqnarray}
Making use of (\ref{GF}) and dropping the surface terms the above expression is found out to be,
\begin{eqnarray}
\delta_{\hat\alpha}S|_{{\textrm{1st term}}}=-\int \textrm{d}^4x&\hat\alpha^a&(-\partial^{\mu}\partial^{\nu}{\hat F}_{\mu\nu}-ig\partial^{\mu}[{\hat A}^{\nu},{\hat F}_{\mu\nu}]-ig[{\hat A}^{\mu},\partial^{\nu}{\hat F}_{\mu\nu}]\nonumber\\&&+g^2[{\hat A}^{\mu}*{\hat A}^{\nu},{\hat F}_{\mu\nu}])^a.
\end{eqnarray}
The second term of (\ref{delS}) is identically zero due to the antisymmetric nature of $\theta^{\mu\nu}$. We write that as,
\begin{eqnarray}
\delta_{\hat\alpha}S|_{{\textrm{2nd term}}}&=&-\frac{1}{2}\int \textrm{d}^4x \ \hat\alpha^a\frac{i}{2}\theta^{\mu_1\nu_1}(-ig\{\partial_{\mu_1}{\hat F}^{\mu\nu},\partial_{\nu_1}{\hat F}_{\mu\nu}\})^a\\
&=&-\int \textrm{d}^4x \ \hat\alpha^a\frac{i}{2}\theta^{\mu_1\nu_1}(-ig\{\partial_{\mu_1}\partial^{\mu}{\hat A}^{\nu},\partial_{\nu_1}{\hat F}_{\mu\nu}\}\nonumber\\&&+g^2\{\partial_{\mu_1}({\hat A}^{\mu}*{\hat A}^{\nu}),\partial_{\nu_1}{\hat F}_{\mu\nu}\})^a\\
&=&-\int \textrm{d}^4x \ \hat\alpha^a\frac{i}{2}\theta^{\mu_1\nu_1}(-ig\partial^{\mu}\{\partial_{\mu_1}{\hat A}^{\nu},\partial_{\nu_1}{\hat F}_{\mu\nu}\}-\nonumber\\&&ig\{\partial_{\mu_1}{\hat A}^{\mu},\partial_{\nu_1}\partial^{\nu}{\hat F}_{\mu\nu}\}+g^2\{\partial_{\mu_1}({\hat A}^{\mu}*{\hat A}^{\nu}),\partial_{\nu_1}{\hat F}_{\mu\nu}\})^a.
\end{eqnarray}
The third term is written as
\begin{eqnarray}
\delta_{\hat\alpha}S|_{{\textrm{3rd term}}}=-\int \textrm{d}^4x&(\partial^{\mu_1}\partial^{\mu_2}\hat\alpha^a)&\frac{1}{2}(\frac{i}{2})^2\theta^{\mu_1\nu_1}\theta^{\mu_2\nu_2}(-\partial^{\mu}\partial^{\nu}\partial^{\nu_1}\partial^{\nu_2}{\hat F}_{\mu\nu}\nonumber\\&&-ig\partial^{\mu}[{\hat A}^{\nu},\partial^{\nu_1}\partial^{\nu_2}{\hat F}_{\mu\nu}]\nonumber\\&&-ig[{\hat A}^{\mu},\partial^{\nu}\partial^{\nu_1}\partial^{\nu_2}{\hat F}_{\mu\nu}]\nonumber\\&&+g^2[{\hat A}^{\mu}*{\hat A}^{\nu},\partial^{\nu_1}\partial^{\nu_2}{\hat F}_{\mu\nu}])^a.
\end{eqnarray}
Using the antisymmetry of $\theta^{\mu\nu}$ and dropping the various surface terms, we write the above expression as,
\begin{eqnarray}
\delta_{\hat\alpha}S|_{{\textrm{3rd term}}}=-\int \textrm{d}^4x&\hat\alpha^a&\frac{1}{2}(\frac{i}{2})^2\theta^{\mu_1\nu_1}\theta^{\mu_2\nu_2}(-ig\partial^{\mu}[\partial^{\mu_1}\partial^{\mu_2}{\hat A}^{\nu},\partial^{\nu_1}\partial^{\nu_2}{\hat F}_{\mu\nu}]\nonumber\\&&-ig[\partial^{\mu_1}\partial^{\mu_2}{\hat A}^{\mu},\partial^{\nu}\partial^{\nu_1}\partial^{\nu_2}{\hat F}_{\mu\nu}]\nonumber\\&&+g^2[\partial^{\mu_1}\partial^{\mu_2}({\hat A}^{\mu}*{\hat A}^{\nu}),\partial^{\nu_1}\partial^{\nu_2}{\hat F}_{\mu\nu}])^a.
\end{eqnarray}
Other terms can be obtained in a similar manner. Combining all these terms we finally get,
\begin{eqnarray}
\delta_{\hat\alpha}S&=&-\int \textrm{d}^4x \ \hat\alpha^a(-\partial^{\mu}\partial^{\nu}{\hat F}_{\mu\nu}-ig\partial^{\mu}[{\hat A}^{\nu},{\hat F}_{\mu\nu}]_*-ig[{\hat A}^{\mu},\partial^{\nu}{\hat F}_{\mu\nu}]_*\nonumber\\&&+g^2[{\hat A}^{\mu}*{\hat A}^{\nu},{\hat F}_{\mu\nu}]_*)^a\\
&=&-\int \textrm{d}^4x \ \hat\alpha^a\Lambda^a
\end{eqnarray}
where,
\begin{eqnarray}
\Lambda^a=-(\mathcal{D}^{\mu}*L_{\mu})^a=-(\mathcal{D}^{\mu}*\mathcal{D}^{\sigma}*{\hat F}_{\sigma \mu})^a
\label{70}
\end{eqnarray}
that vanishes identically. Note that this is exactly the same as the expression in the gauge identity (\ref{lambda}) without the fermionic fields. This proves the invariance of the action.

Let us now repeat the analysis of the previous section with appropriate modifications.
Since the gauge transformations are undeformed, the gauge generators are expected to have the same form
as in the commutative space. To see this note that the gauge variation of the zeroth component of the ${\hat A}_{\mu}$ field, following from (\ref{gag}), can be written as, 
\begin{eqnarray}
\delta_{\hat\alpha}{\hat A}_0^a(z)&=&\partial_0\hat\alpha^a(z)-gf^{abc}{\hat A}_0^b(z)\hat\alpha^c(z)\cr
&=&g\int \textrm{d}^3{\bf{z}} \ f^{abc}{\hat A}_0^c\hat\alpha^b\delta^3({\bf{x}}-{\bf{z}})+\int\textrm{d}^3{\bf{z}} \ \delta^{ab}\delta^3({\bf{x}}-{\bf{z}})\frac{\partial}{\partial t}\hat\alpha^b.
\end{eqnarray}
Clearly the above result can be expressed in our standard  form (\ref{a}),
\begin{eqnarray}
\delta_{\hat\alpha} {\hat A}_0^a(z)&=&\sum_s(-1)^s\int\textrm{d}^3{\bf{z}}\frac{\partial^s\hat\alpha^b({\bf{z}},t)}
{\partial t^s}\rho^{a0b}_{(s)}(x,z)\cr
&=&\int\textrm{d}^3{\bf{z}} \ \hat\alpha^b({\bf{z}},t)\rho^{a0b}_{(0)}(x,z)-\int\textrm{d}^3{\bf{z}} \ \frac{\partial \hat\alpha^b({\bf{z}},t)}{\partial t}\rho^{a0b}_{(1)}(x,z)
\end{eqnarray}
where
\begin{eqnarray}
&&\rho^{a0b}_{(0)}(x,z)=gf^{abc}{\hat A}_0^c\delta^3({\bf{x}}-{\bf{z}})
\label{r11}\\
&&\rho^{a0b}_{(1)}(x,z)=-\delta^{ab}\delta^3({\bf{x}}-{\bf{z}})
\label{r22}
\end{eqnarray}
is the gauge generator. Similarly
\begin{eqnarray}
\delta_{\hat\alpha} {\hat A}_i^a(z)&=&\partial_i\hat\alpha^a(z)-gf^{abc}{\hat A}_i^b(z)\hat\alpha^c(z)
\end{eqnarray}
is written in the form
\begin{eqnarray}
\delta_{\hat\alpha} {\hat A}_i^a(z)&=&\sum_s(-1)^s\int\textrm{d}^3{\bf{z}} \ \frac{\partial^s\hat\alpha^b({\bf{z}},t)}{\partial t^s}\rho^{aib}_{(s)}(x,z)
\end{eqnarray}
for the value
\begin{eqnarray}
\rho^{aib}_{(0)}(x,z)=-\delta^{ab}\partial^{i{\bf{z}}}\delta^3({\bf{x}}-{\bf{z}})+gf^{abc}{\hat A}_i^c\delta^3({\bf{x}}-{\bf{z}}).
\label{r33}
\end{eqnarray}
No star products appear in the gauge generators $\rho$ and their structure is similar to the undeformed commutative
space expressions. To identify the difference (both from the commutative space results and the star deformed
results) it is essential to look at the gauge identity and its connection with the corresponding gauge generator.

Now as already implied in (\ref{70}), we have a gauge identity for this system, exactly similar to the previous case,
\begin{eqnarray}
\Lambda^a=-\left(\mathcal{D}^{\mu}*L_{\mu}\right)^a=0
\label{su}
\end{eqnarray}
where $L_{\mu}$ is the Euler derivative defined in (\ref{70}). The Euler--Lagrange equation of motion is given by
\begin{eqnarray}
\mathcal{D}^{\sigma}*{\hat F}_{\sigma\mu}=0.
\end{eqnarray}

The gauge identity and the Euler derivatives are mapped by the relation,
\begin{eqnarray}
\Lambda^a({\bf{z}},t)=\sum_{s=0}^n\int \textrm{d}^3{\bf{x}} \ \frac{\partial^s}{\partial t^s}\left(\rho'^{b\mu a}_{(s)}(x,z)L^b_{\mu}({\bf{x}},t)\right)
\end{eqnarray}
where the values of $\rho'^{b\mu a}_{(0)}(x,z)$ and $\rho'^{b\mu a}_{(1)}(x,z)$ are equal to those of $\rho^{b\mu a}_{(0)}$ and $\rho^{b\mu a}_{(1)}$ of the previous example, given in (\ref{a1}), (\ref{a2}) and (\ref{37}). This happens since the Euler derivatives and the gauge identity are identical to those discussed in the previous section. Now we express $\rho'$ in terms of $\rho$. To do this, (\ref{a1}) is rewritten under the identification $\rho=\rho'$ as,
\begin{eqnarray}
\rho'^{b0a}_{(0)}(x,z)=-\frac{g}{2}f^{abc}\{\delta^3({\bf{x}}-{\bf{z}}),{\hat A}_0^c(x)\}_*-i\frac{g}{2}d^{abc}[\delta^3({\bf{x}}-{\bf{z}}),{\hat A}_0^c(x)]_*.
\end{eqnarray}
Now making use of the definition of star product (\ref{str}), the above expression is written in the following way
\begin{eqnarray}
\rho'^{b0a}_{(0)}(x,z)&=&-gf^{abc}{\hat A}^c_0\delta^3({\bf{x}}-{\bf{z}})-g\sum_{n=1}^{\infty}(\frac{i}{2})^n\frac{\theta^{\mu_1\nu_1}\cdot\cdot\cdot\theta^{\mu_n\nu_n}}{n!}\nonumber\\
&&[(\frac{f^{abc}}{2}+i\frac{d^{abc}}{2})\partial_{\mu_1}\cdot\cdot\cdot\partial_{\mu_n}\delta^3({\bf{x}}-{\bf{z}})\partial_{\nu_1}\cdot\cdot\cdot\partial_{\nu_n}{\hat A}^{0c}(x)\\
&&(+\frac{f^{abc}}{2}-i\frac{d^{abc}}{2})
\partial_{\mu_1}\cdot\cdot\cdot\partial_{\mu_n}{\hat A}^{0c}(x)\partial_{\nu_1}\cdot\cdot\cdot\partial_{\nu_n}\delta^3({\bf{x}}-{\bf{z}})].\nonumber
\end{eqnarray}
Note that the $\theta$ independent term is nothing but the gauge generator $\rho^{b0a}_{(0)}$ given in (\ref{r11}). Similarly calculating the other components $\rho'^{bia}_{(0)}$ and $\rho'^{b0a}_{(1)}$ from (\ref{a2}) and (\ref{37}) we obtain,
\begin{eqnarray}
\rho'^{b\mu a}_{(0)}(x,z)&=&\rho^{b\mu a}_{(0)}(x,z)-g\sum_{n=1}^{\infty}(\frac{i}{2})^n\frac{\theta^{\mu_1\nu_1}\cdot\cdot\cdot\theta^{\mu_n\nu_n}}{n!}\nonumber\\
&&[(\frac{f^{abc}}{2}+i\frac{d^{abc}}{2})\partial_{\mu_1}\cdot\cdot\cdot\partial_{\mu_n}\delta^3({\bf{x}}-{\bf{z}})\partial_{\nu_1}\cdot\cdot\cdot\partial_{\nu_n}{\hat A}^{\mu c}(x)
\label{tui}\\
&&(+\frac{f^{abc}}{2}-i\frac{d^{abc}}{2})
\partial_{\mu_1}\cdot\cdot\cdot\partial_{\mu_n}{\hat A}^{\mu c}(x)\partial_{\nu_1}\cdot\cdot\cdot\partial_{\nu_n}\delta^3({\bf{x}}-{\bf{z}})]\nonumber\\
\rho'^{b0a}_{(1)}(x,z)&=&\rho^{b0a}_{(1)}(x,z).
\end{eqnarray}

Here $\rho'$ is not the generator, rather it is $\rho$ (\ref{r11},\ref{r22},\ref{r33}). Although the generator remains undeformed, the relation mapping the gauge identity with the generator is neither the commutative space result nor its star deformation as found in the other approach. Rather, it is twisted from the undeformed result. The additional twisted terms are explicitly given in (\ref{tui}). Expectedly, in the limit $\theta\rightarrow 0$ the twisted terms vanish.
\section{Discussion}
Gauge symmetries on canonically deformed coordinate spaces were considered. Within a common Lagrangian framework both types of gauge symmetries in these noncommutative spaces were discussed. Explicit structures of the gauge generators were obtained in either case. The connection of these  generators with the gauge identity, which must exist whenever there is a gauge symmetry, was also established. In the former case, this connection was a star deformation of the commutative space result. In the latter case, on the other hand, the commutative space result was appropriately twisted.  A first order formulation of the deformed gauge transformation was also given where the gauge variation of the field strength tensor and the gauge field were found independently. The results thus obtained were shown to be consistent with the second order formulation. All results obtained here reduce to the usual commutative space expressions in the limit of vanishing $\theta$.

It is quite remarkable that these fundamental properties of gauge symmetries (i. e. occurrence of gauge identity and its connection with the corresponding generator through the Euler derivatives) were found in the noncommutative theory, adopting either of the two interpretations. This strongly suggests a meaningful interpretation for gauge transformation on noncommutative spaces.


\chapter{\label{chap:hamiltonian}Noncommutative gauge theory: Hamiltonian analysis}

Apart from the Lagrangian formulation there is also a Hamiltonian formulation of describing the gauge symmetries of a commutative space action\cite{rothe,rothe1,henn}. In this approach, Dirac's\cite{n} conjecture is followed to obtain the gauge generators from a linear combination of the first class constraints. The gauge variation of the fields are then found by Poisson bracketing the generator with the respective fields.

Here we provide a systematic Hamiltonian analysis of gauge theory on a canonical noncommutative space. The analysis is applied for both cases -- star deformed gauge transformation with usual coproduct rule and undeformed gauge transformation with twisted coproduct rule. In this sense this chapter is complementary to the previous chapter where Lagrangian analysis was performed for both the star deformed and the twisted gauge symmetry. As a specific model, the same noncommutative Yang--Mills action coupled to fermionic matter has been taken. The first class constraints (both primary and secondary) of the theory are identified. The gauge generator is formed by taking a linear combination of these first class constraints. The independent gauge parameters are identified to write the generator in an appropriate way. The Poisson brackets between this generator and the field variables give the star deformed gauge transformations. Subsequently by providing a ``twist" to the Poisson brackets, the twisted gauge transformations are obtained. This twist is dictated by a novel interpretation of the twisted coproduct of gauge transformations. We find that the twisted coproduct is the normal coproduct with the stipulation that the gauge parameter is pushed outside the star operation at the end of all computations. 

\section{General formulation}
We first give a general description of a field theoretical model defined on a canonical noncommutative space. The results are basically appropriate star deformations of the commutative space results. 

As a starting point we recall that, the Euler--Lagrange equation of motion is derived from the action principle by requiring the commutativity of an arbitrary $\delta$ variation with the time differentiation i. e.
\begin{eqnarray}
\delta \frac{d}{dt}q_i =  \frac{d}{dt}\delta  q_i.
\end{eqnarray}
In the following Hamiltonian analysis, based on \cite{rothe,rothe1,henn}, we impose this requirement to derive few important results.

 We consider a system with a canonical Hamiltonian $H_c$ and a set of first class constraints $\Phi_{a}\approx0$. In general $\Phi_{a}$ includes both the primary and secondary constraints and satisfy the following involutive Poisson algebra
\begin{eqnarray}
&&\{H_c,\Phi_a(x)\}=\int \textrm{d}y \ V^b_a(x,y)*\Phi_b(y),
\label{ve}\\
&&\{\Phi_a(x),\Phi_b(y)\}=\int \textrm{d}z \ C^c_{ab}(x,y,z)*\Phi_c(z)
\label{cc}
\end{eqnarray}
where $V$ and $C$ are structure functions. 

For such a system the total Hamiltonian is given by the sum of the canonical Hamiltonian and a linear combination of the primary first class constraints.
\begin{eqnarray}
H_T = H_c + \int \textrm{d}x \ v^{a_1}(x)*\Phi_{a_1}(x).
\label{H0T}
\end{eqnarray}
Here $v^{a_1}$ are Lagrange multipliers. The label $a_1(a_1\le a)$ denotes the primary first class constraints while $a_2$ is kept for the secondary sector. The Hamilton's equations are obtained by using (\ref{H0T})
\begin{eqnarray}
\dot q_i(x) = \{q_i(x),H_T\} = \{q_i(x),H_c\} + \int\textrm{d}y \ v^{a_1}(y)*\{q_i(x),\Phi_{a_1}(y)\}.
\label{eof}
\end{eqnarray}
The generator of the system, according to Dirac's algorithm is a linear combination of all the first class constraints,
\begin{eqnarray}
&&G=\int \textrm{d}x \ \epsilon^a(x)*\Phi_a(x).
\label{G6}
\end{eqnarray}
Though the number of gauge parameters appearing in the above equation is $a$, all of them are not independent. In fact the number of independent $\epsilon '$s is given by the number of independent primary first class constraints (labeled by `$a_1$'). To find the relations among these parameters, we review the method of \cite{rabinb} which is an adaptation of the commutative space approach discussed in\cite{rothe,rothe1,henn}.

The gauge transformation of a variable $F$ is obtained by Poisson bracketing it with the gauge generator $G$ defied in (\ref{G6})
\begin{eqnarray}
\delta F(x) =\int \textrm{d}y \  \epsilon_a(y)*\{F(x),\Phi^a(y)\}.
\label{gt}
\end{eqnarray}
This equation together with (\ref{eof}) yields,
\begin{eqnarray}
\delta \dot q_i(x) &=& \int \textrm{d}z \ \epsilon^a(z)*\{\{q_i(x),H_c\},\Phi_a(z)\} +\nonumber\\
&&\int \int\textrm{d}y \ \textrm{d}z \ \epsilon^b(z)*v^{a_1}(y)*\{\{q_i(x),\Phi_{a_1}(y)\},\Phi_b(z)\}+\nonumber\\
&&\int\textrm{d}y \ \delta v^{a_1}(y)*\{q_i(x),\Phi_{a_1}(y)\}
\label{deldot}
\end{eqnarray}
and
\begin{eqnarray}
\frac{d}{dt}\delta q_i(x)& =& \int\textrm{d}y \  \epsilon^a(y)*\{\{q_i(x),\Phi_a(y)\},H_c\}\nonumber\\&&
+ \int\textrm{d}y \  \textrm{d}z \ \epsilon^a(y)*v^{a_1}(z)*\{\{q_i(x),\Phi_a(y)\},\Phi_{a_1}(z)\}\nonumber\\&&
+\int\textrm{d}y \  \frac{d\epsilon^a}{dt}(y)*\{q_i(x),\Phi_a(y)\}.
\label{dotdel}
\end{eqnarray}
Equating (\ref{deldot}) with (\ref{dotdel}) and using the Jacobi identity we get
\begin{eqnarray}
&&\int\textrm{d}z \ \epsilon^a(z)*\{\{H_c,\Phi_a(z)\}, q_i\}\nonumber\\ 
&&+\int\textrm{d}y \ \textrm{d}z \  \epsilon^a(z)*
v^{a_1}(y)*\{\{\Phi_{a_1}(y),\Phi_{a}(z)\},q_i\}\nonumber\\&&
- \int\textrm{d}y \ \delta v^{a_1}(y)\{q_i,\Phi_{a_1}(y)\} + \int\textrm{d}y \ \frac{d\epsilon^a(y)}{dt}*\{q_i,\Phi_a(y)\}=0.
\end{eqnarray}
Making use of the algebra (\ref{ve}) and (\ref{cc}), we write the above equation as, 
\begin{eqnarray}
&&\int\textrm{d}z \ (\left[\frac{d\epsilon^b(z)}{dt}   - \int\textrm{d}y \ \epsilon^a(z)*[V_a^b(z,y)  + \int\textrm{d}u \ v^{a_1}(u)*C_{a_1a}^b(u,z,y)]\right]\nonumber\\&&*\frac{\partial\Phi_b(y)}{\partial p_i} - \delta v^{a_1}(z)*\frac{\partial\Phi_{a_1}(z)}{\partial p_i})
 = 0.\nonumber
\end{eqnarray}
Since the constraints are taken to be irreducible (i. e. independent) we get the following conditions, from the secondary and primary sectors, respectively,
\begin{eqnarray}
\frac{\textrm{d}\epsilon^{b_2}(x)}{\textrm{d}t}&=&\int \textrm{d}y \ \epsilon^a(y)*V^{b_2}_a(y,x)\nonumber\\
&&+\int \textrm{d}y \ \textrm{d}z \ \epsilon^a(y)*v^{a_1}(z)*C^{b_2}_{a_1a}(z,y,x)
\label{sl}
\end{eqnarray}
\begin{eqnarray}
\delta v^{b_1}(x)&=& \frac{d\epsilon^{b_1}(x)}{dt}   - \int\textrm{d}y \ \epsilon^a(y)*V_a^{b_1}(y,x) \nonumber\\
&&- \int\textrm{d}y \ \textrm{d}z \ \epsilon^a(y)*v^{a_1}(z)*C_{a_1a}^{b_1}(z,y,x).
\end{eqnarray}
The first relation expresses the fact that the gauge parameters $\epsilon^a$ are not all independent. In fact we find that, as stated earlier, the number of independent parameters of a gauge system is equal to the number of primary first class constraints. On the other hand, the second equation gives the variation of the Lagrange multipliers.

\section{Analysis for star deformed gauge symmetry}
The general analysis of a gauge theory on noncommutative space is now used here for the model (\ref{lag}) to study its Hamiltonian description. Throughout the chapter we assume $\theta^{0i}=0$ to avoid higher order time derivatives. Due to the presence of grassmanian variables in our model (\ref{lag}), the Poisson brackets in the previous section should be replaced by the graded brackets. The graded brackets between the fermionic variables are given by,
\begin{eqnarray}
\{\hat\psi_{\alpha}(x),{\hat\psi}_{\beta}^{\dagger}(y)\}=-i\delta_{\alpha\beta}\delta(x-y)
\label{psi}.
\end{eqnarray}
The canonical momenta of the Lagrangian (\ref{lag}),
\begin{eqnarray}
\hat{\pi}^a_{\sigma}=\frac{\partial\mathcal{L}}{\partial \dot{{\hat A}}^{\sigma a}}=\hat{F}^a_{\sigma 0}
\label{mmomenta}
\end{eqnarray}
satisfy the basic Poisson bracket relation
\begin{eqnarray}
\{{\hat A}^{\mu a}(x),\hat{\pi}^b_{\nu}(y)\}=\delta^{ab}\delta^{\mu}_{\nu}\delta(x-y)
\label{bbracket}
\end{eqnarray}
The zeroth component of the momenta (\ref{mmomenta}) leads to a primary constraint 
\begin{eqnarray}
\Phi_1^a=\hat{\pi}^a_0\approx0.
\label{con1001}
\end{eqnarray}
The canonical Hamiltonian of the system is given by, 
\begin{eqnarray}
H&=&\int \textrm{d}x \  [\frac{1}{2}\hat{\pi}^{ic}*\hat{\pi}^{ic}+\frac{1}{4}{\hat F}_{ij}^a*{\hat F}^{ija}-(\mathcal{D}_i*\hat{\pi}^i)^a*{\hat A}_0^a\nonumber\\
&&-i{\hat{\bar{\psi}}}*\gamma^i\partial_i\hat\psi+g{\hat{\bar{\psi}}}*\gamma^{\mu}{\hat A}_{\mu}*\hat\psi+m{\hat{\bar{\psi}}}*\hat\psi]
\label{ham001}
\end{eqnarray}
where the operator $\mathcal{D}*$ has already been defined in (\ref{D}). Now  using (\ref{bbracket}), the secondary constraints of the system are computed 
\begin{eqnarray}
&&\Phi_2^a=\{H,\Phi_1^a\}=\{H,\hat{\pi}^a_0\}=(\mathcal{D}_i*\hat{\pi}_i)^a-g\hat\psi_{\lambda}* (T^a)_{\sigma\lambda}(\hat\psi^{\dagger})_{\sigma}\approx 0.
\label{con2001}
\end{eqnarray}
Note that this constraint is the zeroth component of the equation of motion (\ref{eqn}) expressed in phase space variables. The algebra of the $\Phi_1$ constraints is trivial,
\begin{eqnarray}
&&\{\Phi_1^a(x),\Phi_1^b(y)\}=0
\label{f11}\\
&&\{\Phi_1^a(x),\Phi_2^b(y)\}=0.
\label{f12}
\end{eqnarray}
The algebra of the constraint $\Phi_2$ with itself is also found to close. Since this calculation involves some nontriviality, couple of intermediate steps are presented here. We write
\begin{eqnarray}
\Phi_2^a=\Xi^a+\chi^a
\end{eqnarray}
where
\begin{eqnarray}
\Xi^a&=&(\mathcal{D}_i*\hat{\pi}_i)^a\nonumber\\
&=&\partial_{i}\hat{\pi}^a_{i}-\frac{g}{2}f^{abc}\{{\hat A}_{i}^b,\hat{\pi}^{c}_{i}\}_*+i\frac{g}{2}d^{abc}[{\hat A}_{i}^b,\hat{\pi}^{c}_{i}]_*\nonumber\\
{\textrm{and}} \  \  \chi^a&=&-g\hat\psi_{\lambda}* (\Xi^a)_{\sigma\lambda}({\hat\psi}^{\dagger})_{\sigma}.
\end{eqnarray}
Here the star($*$) anticommutator is defined as,
\begin{eqnarray}
\{A,B\}_*=A*B+B*A
\end{eqnarray}
The graded brackets of the terms $\Xi^a$ and $\chi^a$ separately close among themselves. Let us show it first for $\Xi^a$\cite{amorim}. Using the identity (\ref{delta*}) we obtain
\begin{eqnarray}
\{\partial_i\hat{\pi}^{a}_i(x),-\frac{g}{2}f^{bcd}\{{\hat A}_j^c(y),\hat{\pi}^{d}_{j}(y)\}_*\}+\{-\frac{g}{2}f^{acd}\{{\hat A}_i^c(x),\hat{\pi}^{d}_{i}(x)\}_*,\partial_j\hat{\pi}^{b}_j(y)\}\nonumber\\=\frac{g}{2}f^{abc}\{\delta(x-y),\partial_i\hat{\pi}^{c}_{i}(x)\}_*
\label{1z}
\end{eqnarray}
and
\begin{eqnarray}
\{\partial_i\hat{\pi}^{a}_i(x),i\frac{g}{2}d^{bcd}[{\hat A}_j^c(y),\hat{\pi}^{d}_{j}(y)]_*\}+\{i\frac{g}{2}d^{acd}[{\hat A}_i^c(x),\hat{\pi}^{d}_{i}(x)]_*,\partial_j\hat{\pi}_j^{b}(y)\}\nonumber\\
=-i\frac{g}{2}d^{abc}[\delta(x-y),\partial_i\hat{\pi}^{c}_i(x)]_*
\label{2z}
\end{eqnarray}
Exploiting the Jacobi identity 
\begin{eqnarray}
[\hat{\pi}_i(x),[{\hat A}_i(x),\, T^b\delta(x-y)]_*]_*+[{\hat A}_i(x),[T^b\delta(x-y) ,\hat{\pi}_i(x)]_*]_*\nonumber\\
+[T^b\delta(x-y),[\hat{\pi}_i(x),{\hat A}_i(x)]_*]_*=0
\end{eqnarray}
the remaining terms of $\{\Xi^a(x),\Xi^b(y)\}$ are written as
\begin{equation}
\frac{i}{2}g^2f^{abc}\{\delta(x-y),[{\hat A}_i,\hat{\pi}_i]_*^c\}_*+\frac{1}{2}g^2d^{abc}[\delta(x-y),[{\hat A}_i,\hat{\pi}_i]_*^c]_*.
\label{3z}
\end{equation}

Combining the expressions (\ref{1z}), (\ref{2z}) and (\ref{3z}), we get the closed algebra
\begin{equation}
\{\Xi^a(x),\Xi^b(y)\}=\frac{g}{2}f^{abc}\{\delta(x-y),\Xi^c(x)\}_*-
i\frac{g}{2}d^{abc}[\delta(x-y),\Xi^c(x)]_*
\label{T}
\end{equation}
Now to show that the graded bracket $\{\chi^a(x),\chi^b(y)\}$ really closes we use the product rule
\begin{eqnarray}
\begin{array}{rcl}
\{A,BC\}=\{A,B\}C+(-1)^{\eta_A\eta_B}B\{A,C\}\\
\{AB,C\}=A\{B,C\}+(-1)^{\eta_B\eta_C}\{A,C\}B
\end{array}
\label{pro}
\end{eqnarray}
where
\begin{eqnarray}
&&\eta=0 \  \  \ {\textrm{for bosonic variable and}}\nonumber\\
&&\eta=1 \  \  \ {\textrm{for fermionic variable}}\nonumber
\end{eqnarray}
The graded bracket (\ref{psi}), together with the product rule (\ref{pro}) and the identity (\ref{delta*}) allow us to compute the following bracket
\begin{eqnarray}
\{\chi^a(x),\chi^b(y)\}=\frac{g}{2}f^{abc}\{\delta(x-y),\chi^c(x)\}_*-i\frac{g}{2}d^{abc}[\delta(x-y),\chi^c(x)]_*.
\label{ch}
\end{eqnarray}
Combination of (\ref{T}) and (\ref{ch}) gives the closure of $\Phi_2$,
\begin{eqnarray}
\{\Phi^a_2(x),\Phi^b_2(y)\}=\frac{g}{2}f^{abc}\{\delta(x-y),\Phi^c_2(x)\}_*-i\frac{g}{2}d^{abc}[\delta(x-y),\Phi^c_2(x)]_*.
\end{eqnarray}
In the limit $\theta\rightarrow 0$ the above expression reduces to the standard commutative space result $\{\Phi^a_2(x),\Phi^b_2(y)\}=gf^{abc}\delta(x-y)\Phi^c_2$. The involutive algebra of the canonical Hamiltonian with the constraints is found to be,
\begin{eqnarray}
&&\{H_c,\Phi_1^a\}=\Phi_2^a
\label{al}\\
&&\{H_c,\Phi_2^a\}=-\frac{g}{2}f^{abc}\{{\hat A}^{0b},\Phi_2^c\}_*+i\frac{g}{2}d^{abc}[{\hat A}^{0b},\Phi_2^c]_*.
\label{al1}
\end{eqnarray}
 The term $C^{b_2}_{a_1a}$ of (\ref{cc}) vanishes due to the algebra (\ref{f11}) and (\ref{f12}). So we simplify (\ref{sl}) as
\begin{eqnarray}
\frac{\textrm{d}\epsilon^{b_2}(x)}{\textrm{d}t}=\int \textrm{d}y \ \epsilon^a(y)*V^{b_2}_a(y,x).
\label{sl1}
\end{eqnarray}
The $V$ function defined in (\ref{ve}) is found from the algebra (\ref{al}) and (\ref{al1}) 
\begin{eqnarray}
(V^2_1)^{ab}(x,y)&=&\delta^{ab}\delta(x-y),
\label{50}\\
(V^2_2)^{ab}(x,y)&=&\frac{g}{2}f^{abc}\{\delta(x-y),{\hat A}^{0c}(y)\}_*\nonumber\\
&&+i\frac{g}{2}d^{abc}[\delta(x-y),{\hat A}^{0c}(y)]_*.
\label{51}
\end{eqnarray}
Now we write (\ref{sl1}) in its expanded form as,
\begin{eqnarray}
\frac{\textrm{d}\epsilon^{2a}(x)}{\textrm{d}t}=\int \textrm{d}y \ \epsilon^{1b}(y)*(V^2_1)^{ba}(y,x)+\int \textrm{d}y \ \epsilon^{2b}(y)*(V^2_2)^{ba}(y,x).
\end{eqnarray}
Using (\ref{50}) and (\ref{51}) in the above equation we get
\begin{eqnarray}
\dot \epsilon^{2a}=\epsilon^{1a}-\frac{g}{2}f^{abc}\{\epsilon^{2b}(x),{\hat A}^{0c}(x)\}_*+i\frac{g}{2}d^{abc}[\epsilon^{2b}(x),{\hat A}^{0c}(x)]_*
\end{eqnarray}
so that
\begin{eqnarray}
\epsilon^{1a}=(\mathcal{D}_0*\epsilon^2)^a.
\label{54001}
\end{eqnarray}
thereby eliminating $\epsilon^1$ in favour of $\epsilon^2$. This result is used in (\ref{G6}) to write the generator in terms of the single parameter ($\epsilon^2$) as,
\begin{eqnarray}
G=\int \textrm{d}x \ (\mathcal{D}_0*\epsilon^{2})^a*\Phi_1^a+\epsilon^{2a}*\Phi_2^a
\label{generator}
\end{eqnarray}
where the constraints $\Phi_1$ and $\Phi_2$ were defined in (\ref{con1001}) and (\ref{con2001}). After obtaining the complete form of the generator, we can now calculate the variation of the different fields from (\ref{gt}),
\begin{eqnarray}
\delta q_{\alpha}(x)&=&\int \textrm{d}y \ (\mathcal{D}_0*\epsilon^{2})^a(y)*\{q_{\alpha}(x), \Phi^a_1(y)\}\nonumber\\
&&+\int \textrm{d}y \ \epsilon^{2a}(y)*\{q_{\alpha}(x), \Phi^a_2(y)\}.
\label{eps2}
\end{eqnarray}
Let us first study the gauge transformation of the field ${\hat A}^{\mu}$. The variation of its time component is
\begin{eqnarray}
\delta {\hat A}^a_{0}(x)&=&\int \textrm{d}y \ (\mathcal{D}_0*\epsilon^{2})^b(y)*\{{\hat A}^a_{0}(x),\hat{\pi}^b_0(y)\}\nonumber\\
&=&\int \textrm{d}y \ (\mathcal{D}_0*\epsilon^{2})^b(y)\delta^{ab}*\delta(x-y)\nonumber\\
&=&\int \textrm{d}y \ (\mathcal{D}_0*\epsilon^{2})^a(y)\delta(x-y)\nonumber\\
&=&(\mathcal{D}_0*\epsilon^{2})^{a}
\label{A0}
\end{eqnarray}
where we have used the identity (\ref{b1}). The variation of the space component is likewise given by,
\begin{eqnarray}
\delta {\hat A}^a_{i}(x)&=&\int \textrm{d}y \ \epsilon^{2b}(y)*\{{\hat A}^a_{i}(x),\mathcal{D}_j*\hat{\pi}_j^b(y)\}\nonumber\\
&=&\int \textrm{d}y \ \epsilon^{2b}(y)*(-\partial_i^y\delta(x-y)\delta^{ab}+\frac{g}{2}f^{bca}\{{\hat A}_i^c(y),\delta(x-y)\}_*\nonumber\\
&&-i\frac{g}{2}d^{bca}[{\hat A}_i^c(y),\delta(x-y)]_*).\nonumber
\end{eqnarray}
Now dropping the boundary term and using the cyclicity property (\ref{b2}) we write the above expression as
\begin{eqnarray}
\delta {\hat A}^a_{i}(x)&=&\partial_i\epsilon^{2a}-\frac{g}{2}f^{abc}\{{\hat A}_i^b,\epsilon^{2c}\}_*+i\frac{g}{2}d^{abc}[{\hat A}_i^b,\epsilon^{2c}]_*\nonumber\\
&=&(\mathcal{D}_i*\epsilon^{2})^a(x).
\label{49001}
\end{eqnarray}
Combining (\ref{A0}) and (\ref{49001}) we obtain,
\begin{eqnarray}
\delta {\hat A}_{\mu}^a=(\mathcal{D}_{\mu}*\epsilon^2)^a
\end{eqnarray}
thereby reproducing (\ref{Amu}) with the identification $\epsilon^2\rightarrow\hat{\alpha}$. In a likewise manner the gauge transformation of the matter fields are also obtained,
\begin{eqnarray}
\delta{\hat\psi}_{\alpha}(x)&=&-ig\epsilon^{2a}(x)*\left(T^a\right)_{\alpha\beta}{\hat\psi}_{\beta}(x).
\label{deltsi}\\
\delta{\hat{\bar{\psi}}}_{\alpha}(x)&=&ig\left(T^a\right)_{\beta\alpha}{\hat{\bar{\psi}}}_{\beta}(x)*\epsilon^{2a}(x)
\label{deltsibar}
\end{eqnarray}
which reproduces (\ref{si}) and (\ref{sibar}).

It is also possible to compute the gauge variations of star composites in the same way. For example,
\begin{eqnarray}
\delta({\hat\psi}_{\alpha}(x)*{\hat\psi}_{\beta}(x))&=&\int\textrm{d}y \ \epsilon^{2a}(y)*\{{\hat\psi}_{\alpha}(x)*{\hat\psi}_{\beta}(x),\Phi^{2a}(y)\}\nonumber\\ 
&=&ig\int\textrm{d}y \ (T^a)_{\beta \lambda}\epsilon^{2a}(y)*{\hat\psi}_{\lambda}(y)*{\hat\psi}_{\alpha}(x)*\delta(x-y)\nonumber\\&&-ig\int\textrm{d}y \ (T^a)_{\alpha \lambda}\epsilon^{2a}(y)*{\hat\psi}_{\lambda}(y)*\delta(x-y)*{\hat\psi}_{\beta}(x).\nonumber
\end{eqnarray}
Using the identity (\ref{delta*}) the argument of ${\hat\psi}_{\alpha}$ in the first integral and that of ${\hat\psi}_{\beta}$ in the second integral is shifted from $x$ to $y$ so that star product is defined only at the same point ($y$). Finally, using (\ref{b1}) and (\ref{b2}), and keeping in mind the grassmanian nature of the fermionic field we get
\begin{eqnarray}
\delta({\hat\psi}_{\alpha}*{\hat\psi}_{\beta})=-ig\left((T^a)_{\beta \lambda}{\hat\psi}_{\alpha}*\epsilon^{2a}*{\hat\psi}_{\lambda}+(T^a)_{\alpha \lambda}\epsilon^{2a}*{\hat\psi}_{\lambda}*{\hat\psi}_{\beta}\right).
\end{eqnarray}
This is the result one also finds by using (\ref{deltsi}) and the standard coproduct rule,
\begin{eqnarray}
\delta({\hat\psi}_{\alpha}*{\hat\psi}_{\beta})&=&(\delta{\hat\psi}_{\alpha})*{\hat\psi}_{\beta}+{\hat\psi}_{\alpha}*(\delta{\hat\psi}_{\beta})\\
&=&-ig\left(\epsilon^{2a}(T^a)_{\alpha \lambda}*{\hat\psi}_{\lambda}*{\hat\psi}_{\beta}+{\hat\psi}_{\alpha}*\epsilon^{2a}(T^a)_{\beta \lambda}*{\hat\psi}_{\lambda}\right).
\end{eqnarray}
Likewise, other star composites can be treated identically. This completes the Hamiltonian analysis of star deformed gauge symmetry. Note that the standard coproduct rule (\ref{tX}) is necessary for the invariance of the action as well as the consistency of the analysis.
\section{Analysis for twisted gauge symmetry}
So far we were discussing about the star deformed gauge transformation from a general Hamiltonian formulation which obeys the normal coproduct rule (\ref{tX}). But as discussed in the previous chapter the action (\ref{lag}) is also invariant under the undeformed gauge transformations (\ref{YY1}--\ref{YY}) with the twisted coproduct rule (\ref{co}).

In fact from (\ref{co}), the twisted gauge variation of the composite fields (${\hat A}_{\mu}*{\hat A}_{\nu}$) was found in (\ref{gag}) to finally deduce $\delta_{\hat{\alpha}}{\hat F}_{\mu\nu}$.

 The gauge variation of the other star composites are similarly computed from (\ref{co}),
\begin{eqnarray}
&&\delta_{\hat{\alpha}}({\hat A}_{\mu}*{\hat\psi})=(\partial_{\mu}\hat{\alpha}){\hat\psi}-ig\hat{\alpha}({\hat A}_{\mu}*{\hat\psi})
\label{g.v.}\\
&&\delta_{\hat{\alpha}}(\hat\phi*{\hat\psi})=-ig\hat{\alpha}^a\left((T^a\hat\phi)*{\hat\psi}+\hat\phi*(T^a{\hat\psi})\right).
\label{73}
\end{eqnarray}

 It is noteworthy that not only the gauge transformations of the basic fields but also the transformation rules for the star products of variables are identical to the corresponding undeformed relations.

We now present an alternative interpretation of the twisted coproduct rule (\ref{co}). The results (\ref{gag}), (\ref{g.v.}), (\ref{73}) are seen to follow by using the standard coproduct rule (\ref{tX}) but pushing the gauge parameter $\hat{\alpha}$ outside the star operation at the end of the computations. Denoting this manipulation as,
\begin{eqnarray}
\delta_{\hat{\alpha}}(A*B)\sim(\delta_{\hat{\alpha}}A)*B+A*(\delta_{\hat{\alpha}}B)
\end{eqnarray}
we find
\begin{eqnarray}
\delta_{\hat{\alpha}}(\hat{\phi}*{\hat\psi})&\sim&(\delta_{\hat{\alpha}}\hat\phi)*{\hat\psi}+\hat\phi*(\delta_{\hat{\alpha}}{\hat\psi})\\
&\sim&-ig(\hat{\alpha}\hat\phi)*{\hat\psi}-ig\hat\phi*(\hat{\alpha}{\hat\psi})\\
&=&-ig\hat{\alpha}^a\{(T^a\hat \phi)*{\hat\psi}+\hat\phi*(T^a{\hat\psi})\}
\end{eqnarray}
which reproduces (\ref{73}). Likewise we see,
\begin{eqnarray}
\delta_{\hat{\alpha}}({\hat A}_{\mu}*{\hat\psi})&\sim&(\delta_{\hat{\alpha}}{\hat A}_{\mu})*{\hat\psi}+{\hat A}_{\mu}*(\delta_{\hat{\alpha}}{\hat\psi})\nonumber\\
&\sim&(\partial_{\mu}\hat{\alpha}-ig\hat{\alpha}^a[T^a,{\hat A}_{\mu}])*{\hat\psi}+{\hat A}_{\mu}*(-ig\hat{\alpha}^aT^a{\hat\psi})\nonumber\\
&=&\partial_{\mu}\hat{\alpha}{\hat\psi}-ig\hat{\alpha}^a([T^a,{\hat A}_{\mu}]*{\hat\psi})-ig\hat{\alpha}^a({\hat A}_{\mu}*T^a{\hat\psi})\nonumber\\
&=&\partial_{\mu}\hat{\alpha}{\hat\psi}-ig\hat{\alpha}({\hat A}_{\mu}*{\hat\psi})
\end{eqnarray}
which reproduces (\ref{g.v.}).  

We now suitably modify the Hamiltonian formulation of the previous section to systematically obtain the undeformed gauge transformations (\ref{YY1}--\ref{YY}) as well as the relations (\ref{gag}), (\ref{g.v.}), (\ref{73}) manifesting the twisted Leibniz rule. As far as the gauge generator is concerned the analysis is similar to the previous case and the same expression (\ref{generator}) is obtained. This is not unexpected since the Gauss constraint defining the generator is basically the time component of the field equations which are identical in both treatments. The difference can come only through the computation of the relevant Poisson brackets that lead to the gauge transformations. In our interpretation the twisted coproduct is just the standard coproduct with the proviso that the gauge parameter is pushed outside the star operation at the end of the computations. This motivates us to adopt a similar prescription for computing the modified Poisson brackets.

 The gauge variation of the time component of ${\hat A}^{\mu}$ field is found by suitably Poisson bracketing with (\ref{generator}) (renaming $\epsilon^{2}$ as $\hat{\alpha}$),
\begin{eqnarray}
\delta_{\hat{\alpha}} {\hat A}^a_{0}(x)&=&\int \textrm{d}y \ (\mathcal{D}_0*\hat{\alpha})^b(y)*\{{\hat A}^a_{0}(x),\hat{\pi}^b_0(y)\}\nonumber\\
&\sim&\int \textrm{d}y \ (\mathcal{D}_0*\hat{\alpha})^b(y)\delta^{ab}*\delta(x-y)\nonumber\\
&\sim&\int(\textrm{d}y \ \partial_0\hat{\alpha}^{a}-\frac{g}{2}f^{abc}\{{\hat A}_0^b,\hat{\alpha}^{c}\}_*+i\frac{g}{2}d^{abc}[{\hat A}_0^b,\hat{\alpha}^{c}]_*)(y)*\delta(x-y)\nonumber\\
&=&\partial_0\hat{\alpha}^{a}-gf^{abc}{\hat A}_0^b\hat{\alpha}^{c}\label{tw1}
\end{eqnarray}
where in the last step we put $\hat{\alpha}$ outside the star product following our prescription. 
The variation of the space component is also calculated in a similar way
\begin{eqnarray}
\delta_{\hat{\alpha}} {\hat A}^a_{i}(x)&=&\int \textrm{d}y \ \hat{\alpha}^{b}(y)*\{{\hat A}^a_{i}(x),\mathcal{D}_j*\hat{\pi}_j^b(y)\}\nonumber\\
&\sim&\int \textrm{d}y \ \hat{\alpha}^{b}(y)*(-\partial_i^y\delta(x-y)\delta^{ab}+\frac{g}{2}f^{bca}\{{\hat A}_i^c(y),\delta(x-y)\}_*\nonumber\\
&&-i\frac{g}{2}d^{bca}[{\hat A}_i^c(y),\delta(x-y)]_*)\nonumber\\
&\sim&\int \textrm{d}y \ \hat{\alpha}^{a}(y)*(-\partial_i^y\delta(x-y))+\nonumber\\&&\frac{g}{2}f^{bca}(\hat{\alpha}^{b}(y)*{\hat A}_i^c(y)*\delta(x-y)+\hat{\alpha}^{b}(y)*\delta(x-y)*{\hat A}_i^c(y))\nonumber\\
&&-i\frac{g}{2}d^{bca}(\hat{\alpha}^{b}(y)*{\hat A}_i^c(y)*\delta(x-y)-\hat{\alpha}^{b}(y)*\delta(x-y)*{\hat A}_i^c(y)).\nonumber
\end{eqnarray}
Now dropping the boundary term, using the relation (\ref{b1}) and the cyclicity property (\ref{b2}) we write the above expression as
\begin{eqnarray}
\delta_{\hat{\alpha}} {\hat A}^a_{i}(x)&\sim&\partial_i\hat{\alpha}^{a}(x)+\nonumber\\&&\frac{g}{2}f^{bca}(\hat{\alpha}^{b}(x)*{\hat A}_i^c(x)+{\hat A}_i^c(x)*\hat{\alpha}^{b}(x))\nonumber\\
&&-i\frac{g}{2}d^{bca}(\hat{\alpha}^{b}(x)*{\hat A}_i^c(x)-{\hat A}_i^c(x)*\hat{\alpha}^{b}(x)).\nonumber
\end{eqnarray}
Finally, keeping the gauge parameter $\hat{\alpha}$ outside the star product we obtain
\begin{eqnarray}
\delta_{\hat{\alpha}} {\hat A}^a_{i}(x)=\partial_i\hat{\alpha}^{a}-gf^{abc}{\hat A}_i^b\hat{\alpha}^{c}.
\label{tw2}
\end{eqnarray}
Combining (\ref{tw1}) and (\ref{tw2}) we write the gauge variation in a covariant notation
\begin{eqnarray}
\delta_{\hat{\alpha}} {\hat A}_{\mu}^a=(\mathcal{D}_{\mu}\hat{\alpha})^a.
\label{77}
\end{eqnarray}

The gauge variation of the fermionic field can be obtained in a similar way
\begin{eqnarray}
&&\delta_{\hat{\alpha}}{\hat\psi}_{\alpha}(x)=-ig\hat{\alpha}^{a}(x)\left(T^a\right)_{\alpha\beta}{\hat\psi}_{\beta}(x)
\label{78}\\
&&\delta_{\hat{\alpha}}{\hat{\bar{\psi}}}_{\alpha}(x)=ig\left(T^a\right)_{\beta\alpha}{\hat{\bar{\psi}}}_{\beta}(x)\hat{\alpha}^{a}(x).
\label{79}
\end{eqnarray}
The calculation of the gauge variation of composite fields needs some care. For example, consider the variation $\delta_{\hat{\alpha}}({\hat A}_{\mu}*{\hat\psi})$,
\begin{eqnarray}
\delta_{\hat{\alpha}}({\hat A}_{0}(x)*{\hat\psi}(x))&=&T^a\delta_{\hat{\alpha}}({\hat A}_{0}^a(x)*{\hat\psi}(x))\nonumber\\
&\sim&T^a\int \textrm{d}y \ (\mathcal{D}_0*\hat{\alpha}^{b})(y)*\{{\hat A}^a_{0}(x)*{\hat\psi}(x),\hat{\pi}^b_0(y)\}+\nonumber\\
&&T^b\int\textrm{d}y \ \hat{\alpha}^{c}(y)*\{{\hat A}^b_{0}(x)*{\hat\psi}(x),-g{\hat\psi}(y)*(T^c){\hat\psi}^{\dagger}(y)\}\nonumber\\
&\sim&T^a\int \textrm{d}y \ (\mathcal{D}_0*\hat{\alpha}^{a})(y)*\delta(x-y)*{\hat\psi}(x)\nonumber\\
&&-igT^b\int\textrm{d}y \ \hat{\alpha}^{c}(y)*T^c{\hat\psi}(y)*{\hat A}^b_{0}(x)*\delta(x-y).
\end{eqnarray}
As mentioned earlier, the star product for two functions is defined only at the same spacetime point. So using the identity (\ref{delta*}) we change the argument of ${\hat\psi}$ and ${\hat A}^b_{0}$ from $x$ to $y$ to obtain
\begin{eqnarray}
\delta_{\hat{\alpha}}({\hat A}_{0}(x)*{\hat\psi}(x))&\sim&T^a\int \textrm{d}y \ (\mathcal{D}_0*\hat{\alpha}^{a})(y)*{\hat\psi}(y)*\delta(x-y)\nonumber\\
&&-igT^b\int\textrm{d}y \ \hat{\alpha}^{c}(y)T^c*{\hat\psi}(y)*\delta(x-y)*{\hat A}^b_{0}(y).
\end{eqnarray}
Using the properties (\ref{b1}), (\ref{b2}) and finally removing the gauge parameter $\hat{\alpha}$ outside the star product we obtain
\begin{eqnarray}
\delta_{\hat{\alpha}}({\hat A}_{0}*{\hat\psi})&=&T^a(\partial_0\hat{\alpha}^{a}{\hat\psi}-gf^{abc}\hat{\alpha}^{c}({\hat A}_{0}^b*{\hat\psi}))-igT^bT^c\hat{\alpha}^{c}({\hat A}_{0}^b*{\hat\psi}).
\end{eqnarray}
Following the symmetry algebra (\ref{fabc},\ref{dabc}) we write the above result as
\begin{eqnarray}
\delta_{\hat{\alpha}}({\hat A}_{0}*{\hat\psi})&=&T^a(\partial_0\hat{\alpha}^{a}{\hat\psi}-gf^{abc}\hat{\alpha}^{c}({\hat A}_{0}^b*{\hat\psi}))\nonumber\\
&&-igT^a\hat{\alpha}^{c}({\hat A}_{0}^b*{\hat\psi})(\frac{1}{2}d^{bca}+\frac{i}{2}f^{bca})\\
&=&T^a(\partial_0\hat{\alpha}^{a}{\hat\psi})+gT^a\hat{\alpha}^{c}({\hat A}_{0}^b*{\hat\psi})(-\frac{i}{2}d^{bca}+\frac{1}{2}f^{bca}).
\label{ta0}
\end{eqnarray}
The space part is also obtained in a similar way
\begin{eqnarray}
\delta_{\hat{\alpha}}({\hat A}_{i}*{\hat\psi})=T^a(\partial_i\hat{\alpha}^{a}{\hat\psi})+gT^a\hat{\alpha}^{c}({\hat A}_{i}^b*{\hat\psi})(-\frac{i}{2}d^{bca}+\frac{1}{2}f^{bca}).
\label{tai}
\end{eqnarray}
Expressions (\ref{ta0}, \ref{tai}) are basically the time and space component of the equation (\ref{g.v.}). The gauge variations of the other composites are computed in the same way reproducing the results (\ref{gag}), (\ref{73}) obtained by using the twisted coproduct rule.
\section{Discussion}
We have studied a Hamiltonian formulation of gauge symmetry on canonical noncommutative space. The gauge generator, obtained in this formulation, is same for two different gauge transformations. It reproduced star deformed gauge transformations with a normal coproduct as well as undeformed gauge transformations with a twisted coproduct. This was based on an appropriate interpretation of computing the Poisson brackets that led to the gauge transformations. Some examples were worked out explicitly.

Our analysis provided a new interpretation of the twisted coproduct rule. It was found that the twisted coproduct was equivalent to the normal coproduct with the condition that the gauge parameter had to be taken outside the star operation at the end of the computations. This interpretation keeps the twisted gauge transformation identical to the corresponding commutative space gauge transformations even for the composite fields.

As mentioned in the previous chapter, recent study\cite{chai,zet,gaume} stressed that twisted symmetry is not a physical symmetry at all. It is quite different from the usual gauge transformations, in the sense that twisted transformations do not act only on fields. Nevertheless we were successful in suitably defining gauge generators and transformations. This was quite reassuring since for a genuine symmetry (twisted or otherwise), a generator must be appropriately defined from which transformation rules of the field variables can be obtained.


\chapter{\label{chap:ncgravity}Noncommutative gravity}

Field theories formulated on a noncommutative space include an intrinsic length scale which is supposed to be of the order of Planck length. Construction of a noncommutative general theory of relativity which is considered to be a necessity for quantizing gravity\cite{szabo1} remains an open subject. There are many approaches to this problem. In \cite{chams} a deformation of Einstein's gravity was studied using a construction based on gauging the noncommutative $SO(4,1)$ de Sitter group and applying the Seiberg -- Witten map \cite{SW} with subsequent contraction to $ISO(3,1)$. Construction of a noncommutative gravitational theory was also based on a twisted diffeomorphism algebra \cite{Aschierie,kur}. In these approaches physical symmetries such as general covariance and local Lorentz invariance are difficult to interpret. There is a formulation called minimal formulation of noncommutative gravity where physical symmetries are restored \cite{CK1} by taking a class of restricted coordinate transformations that preserve the canonical noncommutative algebra. This restriction corresponds to the theory of unimodular gravity \cite{Ein,Ein1,Ein2,Ein3,Ein4} where only volume preserving diffeomorphism is considered. Similar discussion for a covariantly constant $\theta$ is given in \cite{hari}.

      A remarkable feature is that there is no first order correction for various theories of noncommutative gravity for constant $\theta$\cite{chams,Aschierie,MS}. Nontrivial contribution starts from the second order term \cite{chams,Aschierie,CK2}. Since the discussions of noncommutative gravity are mostly performed on the canonical noncommutative spacetime (\ref{cano}) the question that naturally appears is whether the vanishing of the order $\theta$ correction is due to this restriction. Perhaps a more general noncommutative structure might lead to order $\theta$ effects.

In this chapter, instead of taking a constant $\theta^{\mu\nu}$ we take the noncommutative parameter in the Lie algebraic form 
\begin{eqnarray}
[\hat{x}^{\mu},\hat{x}^{\nu}]=i\theta^{\mu\nu}(\hat{x})=i\theta \ell^{\mu\nu}_{ \ \ \ \lambda}\hat{x}^{\lambda}
\label{lie}
\end{eqnarray}
where the structure constant $\ell^{\mu\nu}_{ \ \ \ \lambda}$ is antisymmetric in all the three indices and the constant $\theta$ is a measure of noncommutativity. We carry out our analysis in the framework of the tetrad formalism. Here general coordinate invariance is viewed as a local symmetry implemented by the tetrad as the gauge field along with the local Lorentz invariance $(SO(3,1))$ generated by the spin connection fields. Since this formalism is closely related to the usual gauge theory one can use the results of noncommutative gauge theory\cite{SW,W1,W2,W3} in the context of noncommutative gravity. The Seiberg -- Witten map \cite{SW} can be used to cast the theory of noncommutative gravity as a perturbative theory in the noncommutative parameter $\theta$. Such maps have been exhaustively available for the canonical structure \cite{W3,RK,RLee,Banerjee:2004rs}. Maps for the gauge field in a general noncommutative space have been given in \cite{Behr} without giving the map for the field strength tensor. In fact defining a field strength tensor in a general noncommutative space is not obvious. So we develop the appropriate maps for the gauge field and the field strength tensor in the Lie -- algebra valued noncommutative space.

In section -- 7.1 the class of the general coordinate transformations consistent with the algebra (\ref{lie}) is discussed. In the next section we construct the Seiberg--Witten maps for the gauge potentials and field strength tensors valid for the Lie algebraic noncommutativity. Using these maps noncommutative gravity is reduced to an equivalent commutative theory in section -- 7.3. We show that there is no first order correction in the action of noncommutative gravity, exactly as happens for canonical (constant) noncommutativity.
 
\section{General coordinate transformation in noncommutative space}
The formulation of gravity on noncommutative space time poses problems. This is seen by considering the general coordinate transformation,        
\begin{eqnarray}
\hat{x}^{\mu}\rightarrow \hat{x}'^{\mu}=\hat{x}^{\mu}+\hat{\xi}^{\mu}(\hat{x})
\label{xxi}
\end{eqnarray}
and realizing that, for arbitrary $\hat{\xi}^{\mu}(\hat{x})$, it is not compatible with the algebra (\ref{lie}). However it is possible to find a restricted class of coordinate transformations (\ref{xxi}) which preserves the algebra (\ref{lie}). Before demonstrating this point let us define the star product for the Lie algebra valued coordinate dependent noncommutative structure.

The structure of (\ref{WWeyl}) remains unchanged for the Lie algebraic noncommutativity. We write the product of exponentials in the following way,
\begin{eqnarray}
e^{ik_{\lambda}\hat{x}^{\lambda}}e^{ip_{\lambda}\hat{x}^{\lambda}}=e^{i\{k_{\lambda}+p_{\lambda}+\frac{1}{2}g_{\lambda}(k,p)\}\hat{x}^{\lambda}}.
\end{eqnarray}
Using the Baker-Campbell-Hausdorff formula (\ref{BCH}), an explicit form of $g_{\lambda}(k,p)$ is obtained for the coordinates ${\hat x}^{\mu}$ satisfying (\ref{lie})
\begin{eqnarray}
g_{\lambda}(k,p)&=&-\theta k_{\mu}p_{\nu}\ell^{\mu\nu}_{ \ \ \ \lambda}+\frac{1}{6}\theta^2k_{\mu}p_{\nu}(p_{\sigma}-k_{\sigma})\ell^{\mu\nu}_{ \ \ \ \delta}\ell^{\delta\sigma}_{ \ \ \ \lambda}\nonumber\\&&+\frac{1}{24}\theta^3(p_{\sigma}k_{\beta}+k_{\sigma}p_{\beta})k_{\mu}p_{\nu}\ell^{\mu\nu}_{ \ \ \ \delta}\ell^{\delta\sigma}_{ \ \ \ \alpha}\ell^{\alpha\beta}_{ \ \ \ \lambda}+...
\end{eqnarray}
Following the same method discussed in chapter 5, the star product between two functions is written as
\begin{eqnarray}
f(x)*g(x)=e^{\frac{i}{2}x^{\lambda}g_{\lambda}(i\frac{\partial}{\partial x},i\frac{\partial}{\partial y})}f(x)g(y)|_{y\rightarrow x}
\label{star}
\end{eqnarray}

Now we can replace the operator product between two noncommutative variables by the $*$ product (\ref{star}) between the corresponding commutative variables. Thus using (\ref{xxi}) in (\ref{lie}) we get
\begin{eqnarray}
[x'^{\mu},x'^{\nu}]_*=[x^{\mu},x^{\nu}]_*+[x^{\mu},\hat{\xi}^{\nu}(x)]_*+[\hat{\xi}^{\mu}(x),x^{\nu}]_*+\mathcal{O}(\hat{\xi}^2).
\end{eqnarray}

Using the formula (\ref{star}) one derives the following relation\cite{W1},
\begin{eqnarray}
[x^{\mu},f(x)]_*=i\theta \ell^{\mu\nu}_{ \ \ \ \lambda}x^{\lambda}\frac{\partial f}{\partial x^{\nu}}.
\label{formula}
\end{eqnarray}
It is then straightforward to find, using (\ref{formula}), that in order to preserve (\ref{lie}), $\xi^{\mu}$ must satisfy the condition,
\begin{eqnarray}
i\theta \ell^{\mu\sigma}_{ \ \ \ \lambda}x^{\lambda}\frac{\partial \hat{\xi}^{\nu}}{\partial x^{\sigma}}-i\theta \ell^{\nu\sigma}_{ \ \ \ \lambda}x^{\lambda}\frac{\partial \hat{\xi}^{\mu}}{\partial x^{\sigma}}=i\theta \ell^{\mu\nu}_{ \ \ \ \lambda}\hat{\xi}^{\lambda}(x).
\end{eqnarray}
A nontrivial solution of the above equation is given by,
\begin{eqnarray}
\hat{\xi}^{\mu}(x)=\ell^{\mu\alpha}_{ \ \ \ \beta}x^{\beta}\partial_{\alpha}g(x).
\label{restrtrans}
\end{eqnarray}
This can be checked by using the Jacobi identity following from (\ref{lie})
\begin{eqnarray}
\ell^{\mu\nu}_{ \ \ \ \sigma}\ell^{\sigma\lambda}_{ \ \ \ \delta}+\ell^{\nu\lambda}_{ \ \ \ \sigma}\ell^{\sigma\mu}_{ \ \ \ \delta}+\ell^{\lambda\mu}_{ \ \ \ \sigma}\ell^{\sigma\nu}_{ \ \ \ \delta} = 0.
\label{jacobi-lie}
\end{eqnarray}
Equation (\ref{restrtrans}) gives the restricted class of general coordinate transformations under which the noncommutative algebra (\ref{lie}) is preserved. From (\ref{restrtrans}) we find that 
\begin{eqnarray}
\partial_{\mu}\hat{\xi}^{\mu}(x) = 0.\nonumber
\end{eqnarray}
The Jacobian of the transformations (\ref{xxi}) is then unity which means the transformations are volume preserving. Thus the theory belongs to the noncommutative version of unimodular gravity.

\section{Seiberg-Witten map for Lie algebraic noncommutativity}
 The Seiberg -- Witten maps for the non-Abelian noncommutative gauge fields where the noncommutative coordinates satisfy the canonical algebra are elaborately worked out in the literature \cite{SW,W1,W2,W3,RK,RLee,Banerjee:2004rs}. But the corresponding results for Lie algebraic noncommutative structure are only sketched \cite{W1,W2,Behr}. Here we give a comprehensive analysis where the results are valid upto first order in $\theta$.

In the commutative space, for  non-Abelian gauge theory, the matter field $\psi(x)$ and the gauge potential $A_{\mu}(x)$ transform as,
\begin{eqnarray}
\delta_{\alpha}\psi(x)=i\alpha(x)\psi(x), \  \  \  \ \alpha=\alpha_aT^a
\end{eqnarray}
and
\begin{eqnarray}
\delta_{\alpha}A_{\mu}(x)=\partial_{\mu}\alpha(x)+i[\alpha(x),A_{\mu}(x)].
\label{mim09}
\end{eqnarray} 
The commutator of two gauge transformations is another gauge transformation 
\begin{eqnarray}
(\delta_{\alpha}\delta_{\beta}-\delta_{\beta}\delta_{\alpha})\psi(x)=\delta_{-i[\alpha,\beta]}\psi(x)
\label{closure}
\end{eqnarray}

In the noncommutative space, on the other hand, the closure (\ref{closure}) does not hold \cite{W1} within the Lie algebra but is satisfied in the enveloping algebra. Thus the noncommutative field $\hat{\psi}(x)$ transforms as\cite{W2}
\begin{eqnarray}
\delta_{{\alpha}}{\hat\psi}(\hat{x})=i\hat{\alpha}(\hat{x})\hat{\psi}(\hat{x})
\label{sihat}
\end{eqnarray}
which is written in * product formalism as
\begin{eqnarray}
\delta_{{\alpha}}\hat{\psi}(x)=i\hat{\alpha}(x)*\hat{\psi}(x)
\label{delsi}
\end{eqnarray}
Note that the structure of this equation is same as the star deformed gauge transformation ((\ref{si}) with $g=-1$) for canonical noncommutativity. We use the notation $\delta_{\alpha}$ instead of $\delta$ for later convenience. Throughout the chapter we shall not consider the other type of gauge transformation, namely the twisted gauge transformation.

Here the gauge parameter $\hat{\alpha}(x)$ is in the enveloping algebra \cite{W2} similar to (\ref{amu})
\begin{eqnarray}
\hat{\alpha}(x)=\hat{\alpha}_a(x):T^a:+\hat{\alpha}^1_{ab}(x):T^aT^b:+...+\hat{\alpha}^{n-1}_{a_1...a_n}(x):T^{a_1}...T^{a_n}:+...
\end{eqnarray}
All these infinitely many parameters $\hat{\alpha}^{n-1}_{a_1...a_n}(x)$ depend only on the commutative gauge parameter $\alpha(x)$, the gauge potential $A_{\mu}(x)$ and on their derivatives. We denote this as $\hat{\alpha}\equiv\hat{\alpha}(\alpha(x),A(x))$. Then it follows from (\ref{delsi}) that the variation of $\hat{\psi}$ is expressed as
\begin{eqnarray}
\delta_{\alpha}\hat{\psi}(x)=i\hat{\alpha}(\alpha(x),A(x))*\hat{\psi}(x).
\end{eqnarray}
Now we impose the requirement of closure,
\begin{eqnarray}
(\delta_{\alpha}\delta_{\beta}-\delta_{\beta}\delta_{\alpha})\hat{\psi}(x)=\delta_{-i[\alpha,\beta]}\hat{\psi}(x).
\end{eqnarray}
Then the above equation is written in the explicit form
\begin{eqnarray}
i\delta_{\alpha}\hat{\beta}(\beta,A)-i\delta_{\beta}\hat{\alpha}(\alpha,A)+\hat{\alpha}(\alpha,A)*\hat{\beta}(\beta,A)-\hat{\beta}(\beta,A)*\hat{\alpha}(\alpha,A)\nonumber\\
=i(\widehat{-i[\alpha,\beta]})(-i[\alpha,\beta],A).
\end{eqnarray} 
Expanding $\hat{\alpha}$ in $\theta$ 
\begin{eqnarray}
\hat{\alpha}(\alpha,A)=\alpha+\theta\alpha^1(\alpha,A)+\mathcal{O}(\theta^2)
\end{eqnarray}
we obtain up to first order,
\begin{eqnarray}
i\delta_{\alpha}\beta^1(\beta,A)-i\delta_{\beta}\alpha^1(\alpha,A)+[\alpha,\beta^1(\beta,A)]-[\beta,\alpha^1(\alpha,A)]\nonumber\\
-i(-i[\alpha,\beta])^1(-i[\alpha,\beta],A)=-\frac{i}{2}f^{\mu\nu}_{ \  \ \lambda}x^{\lambda}\{\partial_{\mu}\alpha,\partial_{\nu}\beta\}.
\end{eqnarray}
The solution is given by
\begin{eqnarray}
\theta\alpha^1(\alpha,A)=\frac{1}{4}\theta^{\mu\nu}\{\partial_{\mu}\alpha,A_{\nu}\}
\label{29}
\end{eqnarray}
where $\theta^{\mu\nu}$ was defined in (\ref{lie}).

The noncommutative gauge potential $\hat{A}_{\mu}$ is most naturally introduced by the covariant coordinate $\hat{X}^{\mu}$ \cite{W1}. It is defined in the following way
\begin{eqnarray}
\hat{X}^{\mu}(\hat{x})=\hat{x}^{\mu}+\hat{A}^{\mu}(\hat{x})
\label{XAA}
\end{eqnarray}
which, acting on $\hat{\psi}$, transforms covariantly\cite{W2}, i.e.
\begin{eqnarray}
\delta_{\alpha}\big(\hat{X}^{\mu}(\hat{x})\hat{\psi}(\hat{x})\big)=i\hat{\alpha}(\hat{x})\hat{X}^{\mu}(\hat{x})\hat{\psi}(\hat{x}).
\end{eqnarray}
This requirement together with (\ref{sihat}) fixes the transformation of $\hat{X}^{\mu}$
\begin{eqnarray}
\delta_{\alpha}\hat{X}^{\mu}=i[\hat\alpha,\hat{X}^{\mu}].
\label{dela}
\end{eqnarray}
Since gauge transformation does not act on the coordinates,
\begin{eqnarray}
\delta_{\alpha}\hat{x}^{\mu}=0
\end{eqnarray}
we get from (\ref{XAA}) and (\ref{dela})
\begin{eqnarray}
\delta_{\alpha}\hat{A}^{\mu}(\hat{x})=-i[\hat{x}^{\mu},\hat{\alpha}(\hat{x})]+i[\hat{\alpha}(\hat{x}),\hat{A}^{\mu}(\hat{x})].
\end{eqnarray}
In the star product formalism
\begin{eqnarray}
\delta_{\alpha}\hat{A}^{\mu}(x)=-i[x^{\mu},\hat{\alpha}(x)]_*+i[\hat{\alpha}(x),\hat{A}^{\mu}(x)]_*
=\theta^{\mu\rho}\frac{\partial}{\partial x^{\rho}}\hat{\alpha}+i[\hat{\alpha}(x),\hat{A}^{\mu}(x)]_*.
\end{eqnarray}
The gauge potential $\hat{A}_{\mu}$ is defined through $\hat{A}^{\mu}$ in the following way\cite{W3}
\begin{eqnarray}
\hat{A}^{\mu}=\theta^{\mu\rho}\hat{A}_{\rho}.
\end{eqnarray}
Since $\theta^{\mu\rho}$ is not a constant it is not possible to find the transformation of $\hat{A}_{\mu}$ in closed form. We give the result correct upto first order.
\begin{eqnarray}
\delta_{\alpha}\hat{A}_{\mu}=\partial_{\mu}\hat{\alpha}+i[\hat{\alpha},\hat{A}_{\mu}]-\frac{1}{2}\theta^{\lambda\sigma}\{\partial_{\lambda}\hat{\alpha},\partial_{\sigma}\hat{A}_{\mu}\}-\frac{1}{2}\theta_{\mu\alpha}\theta^{\lambda\sigma}\partial_{\sigma}\theta^{\alpha\beta}\{\partial_{\lambda}\hat{\alpha},\hat{A}_{\beta}\}
\label{deltaA}
\end{eqnarray}
where $\theta_{\mu\alpha}$ is the inverse of $\theta^{\mu\alpha}$ i. e.  $\theta_{\mu\alpha}\theta^{\alpha\sigma}=\delta^{\sigma}_{\mu}$.

  To get the Seiberg -- Witten map for the gauge potential
we expand it in a perturbative series in      $\theta$
\begin{eqnarray}
\hat{A}_{\mu}(A)=A_{\mu}+\theta A^1_{\mu}(A)+\mathcal{O}(\theta^2).
\label{SWA}
\end{eqnarray}
Computing the gauge transformation of $\hat{A}_{\mu}$ from the above equation (using (\ref{mim09})) and comparing with (\ref{deltaA}) we get
\begin{eqnarray}
\delta_{\alpha} A^1_{\mu}(A)=\partial_{\mu}\alpha^1(\alpha,A)+i[\alpha^1(\alpha,A),A_{\mu}]+i[\alpha,A^1_{\mu}(A)]-\frac{1}{2}f^{\nu\lambda}_{ \  \ \delta}x^{\delta}\{\partial_{\nu}\alpha,\partial_{\lambda}A_{\mu}\}.
\end{eqnarray}
The solution to the last equation is
\begin{eqnarray}
\theta A^1_{\mu}(A)=-\frac{1}{4}\theta^{\nu\lambda}\{A_{\nu},\partial_{\lambda}A_{\mu}+F_{\lambda\mu}\}-\frac{1}{4}\theta_{\mu\nu}\theta^{\lambda\sigma}\partial_{\sigma}\theta^{\nu\delta}\{A_{\lambda},A_{\delta}\}
\label{seibergA}
\end{eqnarray}
where,
\begin{eqnarray}
F_{\mu\nu}=\partial_{\mu}A_{\nu}-\partial_{\nu}A_{\mu}-i[A_{\mu},A_{\nu}].
\end{eqnarray}
Combining (\ref{SWA}) and (\ref{seibergA}) we get the map for the gauge potential\begin{equation}
\hat{A}_{\mu}(A)= A_{\mu} - \frac{1}{4}\theta^{\nu\lambda}\{A_{\nu},\partial_{\lambda}A_{\mu}+F_{\lambda\mu}\}-\frac{1}{4}\theta_{\mu\nu}\theta^{\lambda\sigma}\partial_{\sigma}\theta^{\nu\delta}\{A_{\lambda},A_{\delta}\}.
\label{A1}
\end{equation}
Note that in the case of constant $\theta$ the last term on the r.h.s. vanishes and one gets back the usual Seiberg--Witten map\cite{SW}.

   In order to construct the map for the Yang -- Mills field $\hat{F}_{\mu\nu}$ we first define a second rank tensor
\begin{eqnarray}
\hat{F}^{\mu\nu}(\hat{x})&=&-i\left([\hat{X}^{\mu}(\hat{x}),\hat{X}^{\nu}(\hat{x})]-i\theta \ell^{\mu\nu}_{ \ \ \ \lambda}\hat{X}^{\lambda}(\hat{x})\right).\nonumber
\end{eqnarray}
or equivalently
\begin{eqnarray}
\hat{F}^{\mu\nu}(x)=-i\left([x^{\mu},\hat{A}^{\nu}(x)]_*-[x^{\nu},\hat{A}^{\mu}(x)]_*+[\hat{A}^{\mu}(x),\hat{A}^{\nu}(x)]_*-i\theta \ell^{\mu\nu}_{ \ \ \ \lambda}\hat{A}^{\lambda}(x)\right)
\label{f107}
\end{eqnarray}
which transforms covariantly
\begin{eqnarray}
\delta_{\alpha}\hat{F}^{\mu\nu}=i[\hat{\alpha},\hat{F}^{\mu\nu}]_*.
\label{trans}
\end{eqnarray}
 Now we define the Yang -- Mills $\hat{F}_{\mu\nu}$ through
\begin{eqnarray}
\hat{F}^{\mu\nu}=\theta^{\mu\lambda}\theta^{\nu\sigma}\hat{F}_{\lambda\sigma}.
\label{45}
\end{eqnarray}
From (\ref{f107}), we get the following expression for $\hat{F}_{\mu\nu}$, 
\begin{eqnarray}
\hat{F}_{\mu\nu}&=&\partial_{\mu}\hat{A}_{\nu}-\partial_{\nu}\hat{A}_{\mu}-i[\hat{A}_{\mu},\hat{A}_{\nu}]+\frac{1}{2}\theta^{\lambda\sigma}\{\partial_{\lambda}A_{\mu},\partial_{\sigma}A_{\nu}\}\nonumber\\
&&+\frac{1}{2}\theta^{\lambda\sigma}\theta_{\mu\alpha}\theta_{\nu\beta}\partial_{\lambda}\theta^{\alpha\eta}\partial_{\sigma}\theta^{\beta\delta}\{A_{\eta},A_{\delta}\}+\frac{1}{2}\theta^{\lambda\sigma}\theta_{\mu\alpha}\partial_{\lambda}\theta^{\alpha\eta}\{A_{\eta},\partial_{\sigma}A_{\nu}\}\nonumber\\
&&+\frac{1}{2}\theta^{\lambda\sigma}\theta_{\nu\beta}\partial_{\sigma}\theta^{\beta\delta}\{\partial_{\lambda}A_{\mu},A_{\delta}\}+\mathcal{O}(\theta^2).
\label{F}
\end{eqnarray}
The gauge transformation of $\hat{F}_{\mu\nu}$ is obtained from (\ref{trans}) and (\ref{45}) as,
\begin{eqnarray}
\delta_{\alpha}\hat{F}_{\mu\nu}=i[\hat{\alpha},\hat{F}_{\mu\nu}]-\frac{1}{2}\theta^{\lambda\sigma}\{\partial_{\lambda}\hat{\alpha},\partial_{\sigma}\hat{F}_{\mu\nu}\}-\frac{1}{2}\theta^{\lambda\sigma}\theta_{\mu\alpha}\theta_{\nu\beta}\partial_{\sigma}(\theta^{\alpha\eta}\theta^{\beta\delta})\{\partial_{\lambda}\hat{\alpha},\hat{F}_{\eta\delta}\}.
\end{eqnarray}
The gauge transformation of $\hat{F}_{\mu\nu}$ can also be obtained from its definition (\ref{F}) and the gauge transformation (\ref{deltaA}) of $\hat{A}_{\mu}$. These are found to be identical.

 Using the map of $\hat{A}_{\mu}$ in (\ref{F}) we obtain the map for the field strength tensor $\hat{F}_{\mu\nu}$. It is given by
\begin{eqnarray}
\hat{F}_{\mu\nu}&=&F_{\mu\nu}+\frac{1}{2}\theta^{\lambda\sigma}\{F_{\mu\lambda},F_{\nu\sigma}\}-\frac{1}{4}\theta^{\lambda\sigma}\{A_{\lambda},(\partial_{\sigma}+\mathcal{D}_{\sigma})F_{\mu\nu}\}\nonumber\\
&&+\frac{1}{2}\theta_{\nu\lambda}\theta^{\beta\sigma}\partial_{\sigma}\theta^{\alpha\lambda}\{F_{\mu\alpha},A_{\beta}\}-\frac{1}{2}\theta_{\mu\lambda}\theta^{\beta\sigma}\partial_{\sigma}\theta^{\alpha\lambda}\{F_{\nu\alpha},A_{\beta}\}
\label{mapF}
\end{eqnarray}
where the commutative space covariant derivative was defined in the adjoint representation (\ref{YY1}). Here also for constant $\theta$ the map for $\hat{F}_{\mu\nu}$ (\ref{mapF}) reproduces the corresponding well known Seiberg--Witten map\cite{SW}.
\section{Noncommutative gravity}
In the framework of Poincar{$\acute{\textrm e}$} gauge gravity, the noncommutative gauge transformation is now decomposed into two generators- one is the generator of inhomogeneous translation $p_{\mu}$ and the other is the generator of the local Lorentz algebra $\Sigma_{ab}$
\begin{eqnarray}
\hat{\Lambda}(\hat{x})=\hat{\xi}^{\mu}(\hat{x})p_{\mu}+\frac{1}{2}\hat{\lambda}^{ab}(\hat{x})\Sigma_{ab}.
\label{gauge}
\end{eqnarray}
Here $\hat{\xi}^{\mu}$ is the local translation of the tetrad which must be restricted to the form given in equation (\ref{restrtrans}) in order to preserve the noncommutative algebra (\ref{lie}). The parameters $\hat{\lambda}^{ab}(\hat{x})$ characterize the local Lorentz transformations at $\hat{x}$. We consider the following vector representation of these generators for future calculation.
\begin{equation}
\left[\Sigma_{cd}\right]_{ab} = \eta_{ac}\eta_{bd} - \eta_{ad}\eta_{bc}.
\label{vector}
\end{equation}
where $\eta_{ab}$ is the Minkowski metric,
\begin{eqnarray}
\eta_{ab}={\textrm{diag}}(-,+,+,+)
\end{eqnarray}
 As is usual we denote the general coordinates by the Greek indices and components with respect to the tetrad by Latin indices.
 Corresponding to the noncommutative gauge transformations (\ref{gauge}) we introduce the gauge potential
\begin{eqnarray}
\hat{A}_{a}(\hat{x})=(\hat{D}_a)=i\hat{E}^{\mu}_a(\hat{x})p_{\mu}+\frac{i}{2}\hat{\omega}_a^{ \ bc}(\hat{x})\Sigma_{bc}
\end{eqnarray}
where $E^{\mu}_{a}(\hat{x})$ are the components of the noncommutative tetrad $\hat{E}_a$ which are also the gauge fields corresponding to general coordinate transformations and $\hat{\omega}_a^{ \ bc}(\hat{x})$ are the spin connection fields associated with local Lorentz invariance. Since $p_{\mu} = -i\partial_{\mu}$, the noncommutative tetrad maps trivially on the commutative one \cite{CK1}. Assuming the gauge transformations and the spin connection fields in the enveloping algebra we write
\begin{eqnarray}
\hat{\Lambda}=\Lambda(x)+\Lambda^{(1)}(x,\omega_a)+\mathcal{O}(\theta^2)
\end{eqnarray}
\begin{eqnarray}
\hat{\omega}_a=\omega_a(x)+\omega_a^{(1)}(x,\omega_a)+\mathcal{O}(\theta^2)
\end{eqnarray}
where
\begin{eqnarray}
\Lambda(x)&=&\xi(x)p_{\mu}+\frac{1}{2}\lambda^{ab}(x)\Sigma_{ab}\\
\omega_a(x)&=&\frac{1}{2}\omega_a^{ \ bc}\Sigma_{bc}
\end{eqnarray}
Invoking the results (\ref{29},\ref{seibergA}) obtained in the last section we write down the order $\theta$ corrections,
\begin{eqnarray}
\Lambda^{(1)}=\frac{1}{4}\theta^{ab}\{\partial_a\Lambda,\omega_b\}
\end{eqnarray}
\begin{eqnarray}
\omega_a^{(1)}=-\frac{1}{4}\theta^{bc}\{\omega_b,\partial_c\omega_a+F_{ca}\}-\frac{1}{4}\theta_{ab}\theta^{cd}\partial_d\theta^{be}\{\omega_c,\omega_e\}
\end{eqnarray}
The field strength tensor can also be expanded in a power series of $\theta$ and we obtain from (\ref{mapF})
\begin{eqnarray}
\hat{F}_{ab}=F_{ab}+F^{(1)}_{ab}+\mathcal{O}(\theta^2)
\end{eqnarray}
where,
\begin{eqnarray}
F^{(1)}_{ab}&=&\frac{1}{2}\theta^{cd}\{F_{ac},F_{bd}\}-\frac{1}{4}\theta^{cd}\{\omega_c,(\partial_d+\mathcal{D}_d)F_{ab}\}\nonumber\\&&
+\frac{1}{2}\theta_{bc}\theta^{de}\partial_e\theta^{fc}\{F_{af},\omega_d\}-\frac{1}{2}\theta_{ac}\theta^{de}\partial_e\theta^{fc}\{F_{bf},\omega_d\}.
\label{62}
\end{eqnarray}
The field strength $F_{ab}$ in general contains both Riemann tensor $R_{ab}^{ \  \ cd}$ and the torsion $T_{ab}^{\ \ c}$. Setting the classical torsion to be zero we get
\begin{eqnarray}
F_{ab}=\frac{1}{2}R_{ab}^{ \  \ cd}\Sigma_{cd}.
\end{eqnarray}
The noncommutative Riemann Tensor $\hat{R}_{ab}^{ \  \ cd}(\hat{x})$ is obtained from 
\begin{eqnarray}
\hat{R}_{ab}(\hat{x})=\frac{1}{2}\hat{R}_{ab}^{ \  \ cd}(\hat{x})\Sigma_{cd}
\end{eqnarray}
where $\hat{R}_{ab}$ is identified with $\hat{F}_{ab}$ under the condition of zero torsion. Explicitly
\begin{eqnarray}
\hat{R}_{ab}=R_{ab}+R^{(1)}_{ab}+\mathcal{O}(\theta^2)
\end{eqnarray}
where the $\mathcal{O}(\theta)$ correction term is obtained from (\ref{62}) as,
\begin{eqnarray}
R^{(1)}_{ab}&=&\frac{1}{2}\theta^{cd}\{R_{ac},R_{bd}\}-\frac{1}{4}\theta^{cd}\{\omega_c,(\partial_d+\mathcal{D}_d)R_{ab}\}\nonumber\\&&
+\frac{1}{2}\theta_{bc}\theta^{de}\partial_e\theta^{fc}\{R_{af},\omega_d\}-\frac{1}{2}\theta_{ac}\theta^{de}\partial_e\theta^{fc}\{R_{bf},\omega_d\}.
\label{67}
\end{eqnarray}
The Ricci tensor $\hat{R}_{a}^{ \ c}=\hat{R}_{ab}^{ \  \ bc}$ and the Ricci scalar $\hat{R}=\hat{R}_{ab}^{ \  \  ab}$ are formed to construct the action
\begin{eqnarray}
S&=&\int d^4x \ \frac{1}{2\kappa^2}\hat{R}(\hat{x})\\
&=&\int d^4x \ \frac{1}{2\kappa^2}\left(R(x)+R^{(1)}(x)\right)+\mathcal{O}(\theta^2).
\end{eqnarray}
The first order correction term to the Lagrangian is
\begin{eqnarray}
R^{(1)}(x)=R^{(1)ab}_{ab}=[R^{(1)}_{ab}]^{ab}.
\end{eqnarray}
It is convenient to arrange the correction as 
\begin{eqnarray}
[R^{(1)}_{ab}]^{ab}=\mathcal{R}_1+\mathcal{R}_2+\mathcal{R}_3+\mathcal{R}_4.
\label{correction}
\end{eqnarray}
where $\mathcal{R}_1,...,\mathcal{R}_4$ correspond to the contributions coming from the four pieces appearing on the r.h.s. of (\ref{67}) in the same order. It is now simple to get the first term,
\begin{eqnarray}
\mathcal{R}_1=2\theta^{cd}[R_{acg}^{ \  \  \ a}R_{bd}^{ \  \ bg}+R_{ac \ g}^{ \  \ b}R_{bd}^{ \  \ ga}].
\end{eqnarray}
For evaluating $\mathcal{R}_2$ we first compute the part containing the covariant derivative
\begin{eqnarray}
[(\partial_d+\mathcal{D}_d)R_{ab}]^e_{ \ f}=2\partial_dR_{ab \ f}^{ \  \ e}-i[\omega_d,R_{ab}]^e_{ \ f}
\label{covder}
\end{eqnarray}
where we have used the expression (\ref{YY1}) for the covariant derivative $\mathcal{D}_d$. Then the second correction term becomes
\begin{eqnarray}
\mathcal{R}_2&=&-\theta^{cd}\left[\frac{1}{2}(\omega_c^{ \ aj}\partial_dR_{abj}^{ \  \  \ b}-\omega_c^{ \ aj}\partial_dR_{ba \ j}^{ \  \ b})\right]\nonumber\\
&&+\frac{i}{4}\theta^{cd}\omega_c^{ \ ab}\left[\omega_{db}^{ \  \ g}R_{ajg}^{ \  \  \ j}+R_{bja}^{ \  \  \ g}\omega_{dg}^{ \  \ j}+\omega_d^{ \ jg}R_{jbga}+R_{ja}^{ \  \ jg}\omega_{dgb}\right].
\end{eqnarray}
Exploiting the various symmetries of the Riemann tensor, spin connection and the noncommutative structure $\theta^{ab}$ we can easily show that both $\mathcal{R}_1$ and $\mathcal{R}_2$ individually vanish. Now the last two terms on the r.h.s. of (\ref{correction}) are
\begin{eqnarray}
\mathcal{R}_3=\frac{1}{2}\theta_{jk}\theta^{nl}\partial_l\theta^{mk}[\omega_n^{ \ ab}R_{amb}^{ \  \  \  \ j}+\omega_n^{ \ aj}R_{im \ a}^{ \  \ i}]
\end{eqnarray}
and
\begin{eqnarray}
\mathcal{R}_4=-\frac{1}{2}\theta_{ik}\theta^{nl}\partial_l\theta^{mk}[\omega_n^{ \ ai}R_{jm \ a}^{ \  \ j}+\omega_n^{ \ ab}R_{amb}^{ \  \  \  \ i}].
\end{eqnarray}
Clearly these terms owe their existence to the Lie -- algebraic noncommutativity assumed in the present work. Most significantly
\begin{eqnarray}
\mathcal{R}_3+\mathcal{R}_4=0
\end{eqnarray}
identically, which can be demonstrated easily by changing dummy variables in any one of the terms on the l.h.s. We thus find that the first order correction of the Ricci scalar vanishes for the Lie algebraic noncommutativity.
\section{Discussion} 
We have constructed a noncommutative gravity theory where the spacetime satisfy a general Lie algebra. A set of general coordinate transformations has been found which keeps the noncommutative algebra covariant. This restricted transformation is volume preserving and hence the corresponding theory of gravity is a unimodular theory\cite{Ein,Ein1,Ein2,Ein3,Ein4}. Our formulation of noncommutative general relativity is based on Poincar$\acute{\textrm e}$ gauge gravity approach where each spacetime point is associated with a set of local inertial coordinates, mutually related by Lorentz transformation. Looking from the point of view of noncommutative field theories the problem reduces to solving a noncommutative Yang--Mills theory where the gauge group is $ISO(3,1)$. The Seiberg--Witten maps for the noncommutative gauge parameters, potential and field strengths have been worked out in detail. Using these results we have expanded the noncommutative Ricci scalar in the powers of the noncommutative parameter. The first order noncommutative correction is found to vanish. This result was previously known for the canonical noncommutative algebra \cite{chams,Aschierie,MS}. Here we see that the same result holds for Lie algebraic noncommutativity. From the present analysis it is clear that the nonexistence of the $\cal O(\theta)$ correction is due to various symmetries of the Riemann tensor and the spin connection of the zero order theory. Thus it appears that the vanishing of the first order correction is a general result which is perhaps due to the inherent symmetries of the spacetime itself.


\chapter{\label{chap:discussions}Conclusions}

The main purpose of this thesis was to study different noncommutative theories and the symmetries associated with them. In the second chapter we have shown how phase space noncommutativity emerges in a planar quantum mechanical problem, namely the generalized Landau problem. The noncommutativity in the coordinates or in the momenta was described as a dual aspect of the same phenomenon. We have adopted two different methods, the Batalin--Tyutin embedding technique and the doublet splitting approach to show this duality.

In the next chapter we had studied the gravitational quantum well problem in a constant noncommutative phase space setting. By making proper transformations we mapped the problem on the commutative space to find the energy spectrum. The results were then compared with experimental data to give an upper bound on the noncommutative parameter. Our analytical findings agreed with the results previously obtained by numerical method. 

The issues related to the space time symmetry for noncommutative spaces had been discussed in chapter 4. Both nonrelativistic and relativistic examples were considered. The deformed Schr\"odinger generators for the canonical noncommutativity and the deformed Poincar$\acute{\textrm{e}}$-conformal generators for the Snyder type noncommutativity were obtained. The deformed generators which satisfy the standard commutative space algebra were derived, in either case, both by an algebraic approach and by a dynamical approach. These two approaches were shown to be consistent.

In the next two chapters we had studied the gauge symmetries of noncommutative field theory. As a specific model noncommutative Yang--Mills theory was taken to analyze both the star deformed gauge transformation and the twisted gauge transformation. In the Lagrangian analysis the gauge generators were found to be different for two different gauge symmetries. On the other hand, in the Hamiltonian analysis, the generator was identical in either case. The gauge transformations of the fields were obtained from the computations of Poisson brackets which were different for two different gauge symmetries.

In chapter 7, Seiberg--Witten maps were used to write the action for Lie algebraic noncommutative gravity in terms of commutative variables. The noncommutative correction appeared as a perturbative expansion in the noncommutative parameter. By explicit computation we had shown that the leading order correction was zero.

Thus in this thesis we studied different aspects of noncommutativity in quantum mechanics, field theory and gravity. Symmetry analysis played an important role in this study. We analyzed deformations of usual symmetries like the external space time symmetry and the internal gauge symmetry in the noncommutative theories. Here the usual methods of analyzing commutative space theories were appropriately generalized to study the noncommutative theories. It was reassuring that in the limit of vanishing noncommutative parameter the theories reduced smoothly to their commutative versions. In this way we saw that the noncommutative theories had many novel properties and this subject can be approached as a consistent theory of physics.

\backmatter




\addcontentsline{toc}{chapter}{Bibliography}

\begin{thebibliography}{100}\raggedright \small \setlength{\itemsep}{0.0cm}
\bibitem{connes}
A. Connes,
 Noncommutative Geometry (Academic Press, New York, 1994).
\bibitem{mat1}                                                           ´
A. Connes,
\newblock ``Noncommutative Differential Geometry'',
\newblock {\em Inst. Hautes $\acute{\textrm{E}}$tudes Sci. Publ. Math.} {\bf 62} 257 (1985).
\bibitem{mat2}
S. L. Woronowicz,   
\newblock ``Twisted SU(2) Group: An Example of a Noncommutative Differ-
ential Calculus'',
\newblock {\em Publ. Res. Inst. Math. Sci.} {\bf 23} 117 (1987).
\bibitem{y1}
A. Connes and M. A. Rieffel,  
\newblock ``Yang-Mills for Noncommutative Two-Tori'',
\newblock {\em Contemp. Math.} {\bf 62} 237 (1987).
\bibitem{y2}
A. Connes and J. Lott,  
\newblock ``Particle Models and Noncommutative Geometry'',
\newblock {\em Nucl. Phys. (Proc. Suppl.)} {\bf 18 B} 29 (1991).
\bibitem{pauli1}
 See Wolfgang Pauli, Scientific Correspondence, Vol II, p.15, Ed. Karl von Meyenn, Springer-Verlag, 1985. 
\bibitem{pauli2}
 See Wolfgang Pauli, Scientific Correspondence, Vol III, p.380, Ed. Karl von Meyenn, Springer-Verlag, 1993.
\bibitem{sny}
H. S. Snyder,
\newblock ``Quantized spacetime,''
\newblock {\em Phys. Rev.} {\bf 71} 38 (1947).
\bibitem{yang}
C. N. Yang, 
\newblock ``On quantized spacetime",
\newblock {\em Phys. Rev.} {\bf 72} 874 (1947).
\bibitem{Weyl}
H. Weyl,
\newblock ``Quantenmechanik und Gruppentheorie,''
\newblock {\em Z. Physik} 46 1 (1927);
\newblock ``The theory of groups and quantum mechanics,''Dover, New-York (1931)
, translated from \newblock``Gruppentheorie und
Quantenmechanik",
Hirzel Verlag, Leipzig (1928).
\bibitem{neumann}
J. v. Neumann,
\newblock ``Die Eindeutigkeit der Schr\"odingerschen Operatoren'',
\newblock {\em Math. Ann.} {\bf 104} 570 (1931).
\bibitem{groenewold}
H. Groenewold,
\newblock ``On the principles of elementary quantum mechanics'',
\newblock {\em Physica} {\bf 12} 405 (1946).
\bibitem{moyal} 
J. E. Moyal,
\newblock``Quantum mechanics as a statistical theory'',
\newblock{\it Proc. Cambridge Phil. Soc.} {\bf 45}, 99 (1949).
\bibitem{mead}
C. A. Mead,
\newblock ``Possible connection between gravitation and fundamental length'',
\newblock {\em Phys. Rev.} {\bf 135} B849 (1964).
\bibitem{townsend}
P. K. Townsend,
\newblock ``Small scale structure of space time as the origin of the gravitational constant'',
\newblock {\em Phys. Rev.} {\bf D 15} 2795 (1977).
\bibitem{mag1}
M. Maggiore,
\newblock ``Quantum groups, gravity and the generalized uncertainty principle'',
\newblock {\em Phys. Rev.} {\bf D 49} 5182 (1994)
\newblock {\tt hep-th/9305163}.
\bibitem{mag2}
M. Maggiore,
\newblock ``The algebraic structure of the generalized uncertainty principle'',
\newblock {\em Phys. Lett.} {\bf B 319} 83 (1993)
\newblock {\tt hep-th/9309034}.
\bibitem{dop1}
S. Doplicher, K. Fredenhagen and J. E. Roberts,
\newblock ``Space-time quantization induced by classical gravity'',
\newblock {\it Phys. Lett.} {\bf B 331} 39 (1994).
\bibitem{dop2}
S. Doplicher, K. Fredenhagen and J.E. Roberts,
\newblock ``The Quantum structure of space-time at the Planck scale and quantum fields'',
\newblock {\em Commun. Math. Phys.} {\bf 172} 187 (1995)
\newblock{\tt hep-th/0303037}.
\bibitem{ven}
G. Veneziano,
\newblock ``A stringy nature needs just two constants'',
\newblock {\em Europhys. Lett.} {\bf 2} 199 (1986).
\bibitem{gross}
D. J. Gross and P. F. Mende,
\newblock ``String theory beyond the Planck scale'',
\newblock {\em Nucl. Phys.} {\bf B 303} 407 (1988).
\bibitem{amati}
D. Amati, M. Ciafaloni and G. Veneziano,
\newblock ``Can space time be probed below the string size'',
\newblock {\em Phys. Lett.} {\bf B 216} 41 (1989).
\bibitem{lizzi}
F. Lizzi and R. J. Szabo,
\newblock ``Duality symmetries and noncommutative geometry of string space time'',
\newblock {\em Commun. Math. Phys.} {\bf 197} 667 (1998)
\newblock{\tt hep-th/9707202}.
\bibitem{landi}
G. Landi, F. Lizzi and R. J. Szabo,
\newblock ``String geometry and the noncommutative torus'',
\newblock {\em Commun. Math. Phys.} {\bf 206} 603 (1999)
\newblock{\tt hep-th/9806099}.
\bibitem{aconnes}
A. Connes, M. R. Douglas and A. Schwarz,
\newblock `` Noncommutative Geometry and Matrix Theory: Compactification on Tori'',
\newblock {\em JHEP} {\bf 9802} 003 (1998)
\newblock{\tt hep-th/9711162}.
\bibitem{SW}
N. Seiberg and E. Witten,
\newblock ``String theory and noncommutative geometry,''
\newblock {\em JHEP} {\bf 9909} 032 (1999) 
\newblock{\tt hep-th/9908142}.
\bibitem{dou}
M. R. Douglas and N. A. Nekrasov,
\newblock ``Noncommutative field theory,''
\newblock {\em Rev. Mod. Phys.} {\bf 73} 977 (2001)
\newblock{\tt hep-th/0106048}.
\bibitem{boz}
R. J. Szabo,
\newblock ``Quantum field theory on noncommutative spaces'',
\newblock {\em Phys. Rept.} {\bf 378} 207 (2003) 
\newblock{\tt hep-th/0109162}.
\bibitem{NIK}
F. A. Schaposnik
\newblock `` Three lectures on noncommutative field theories'',
\newblock{\tt hep-th/0408132}.
\bibitem{chainew}
M. Chaichian, P. P. Kulish, K. Nishijima and A. Tureanu,
\newblock ``On a Lorentz-Invariant Interpretation of Noncommutative Space-Time and Its Implications on Noncommutative QFT,''
\newblock {\em Phys. Lett.} {\bf B 604} 98 (2004) 
\newblock{\tt hep-th/0408069}.
\bibitem{wessju}
J. Wess,
\newblock ``Deformed Coordinate Spaces: Derivatives,''
\newblock {\tt hep-th/0408080}.
\bibitem{113}
 A. P. Balachandran, A. Pinzul and B. A. Qureshi, 
\newblock ``UV-IR mixing in non-commutative plane'',
\newblock {\em Phys. Lett.} {\bf B 634} 434 (2006) 
\newblock{\tt hep-th/0508151}.
\bibitem{114}
A. P. Balachandran, G. Mangano, A. Pinzul and S. Vaidya,  
\newblock ``Spin and statistics on the Groenwald-Moyal plane: Pauli-forbidden levels and transitions'',
\newblock {\em Int. J. Mod. Phys.} {\bf A 21} 3111 (2006)
\newblock{\tt hep-th/0508002}.
\bibitem{M.M}
M. M. Sheikh-Jabbari,
\newblock ``Discrete symmetries (C, P, T) in noncommutative field theories'',
\newblock {\em Phys. Rev. Lett.} {\bf 84} 5265 (2000) 
\newblock{\tt hep-th/0001167}.
\bibitem{Hayakawa}
M. Hayakawa,
\newblock `` Perturbative analysis on infrared aspects of noncommutative QED on R**4'',
\newblock {\em Phys. Lett.} {\bf B 478} 394 (2000) 
\newblock{\tt hep-th/9912094}.
\bibitem{bala}
A. P. Balachandran, T. R. Govindarajan, C. Molina and P. Teotonio-Sobrinho,
\newblock ``Unitary Quantum Physics with Time-Space Noncommutativity,''
\newblock {\em JHEP} {\bf 0410} 072 (2004) 
\newblock{\tt hep-th/0406125}.
\bibitem{t1}
M. Chaichian, P. Presnajder, M.M. Sheikh-Jabbari and A. Tureanu,
\newblock `` Noncommutative standard model: Model building'',
\newblock {\em Eur. Phys. J.} {\bf C 29} 413 (2003) 
\newblock{\tt hep-th/0107055}.
\bibitem{t2}
M. Chaichian, A. Kobakhidze and A. Tureanu, 
\newblock `` Spontaneous reduction of noncommutative gauge symmetry and model building'',
\newblock {\em Eur. Phys. J.} {\bf C 47} 241 (2006) 
\newblock{\tt hep-th/0408065}.
\bibitem{schn}
L. Bonora, M. Schnabl, M. M. Sheikh-Jabbari and A. Tomasiello,
\newblock ``Noncommutative SO(n) and Sp(n) gauge theories'',
\newblock {\em Nucl. Phys.} {\bf B 589} 461 (2000) 
\newblock{\tt hep-th/0006091}.
\bibitem{W1}
J. Madore, S. Schraml, P. Schupp and J. Wess,
\newblock ``Gauge theory on noncommutative spaces,''
\newblock {\em Eur. Phys. J.} {\bf C} 16 161 (2000) 
\newblock{\tt hep-th/0001203}.
\bibitem{W2}
B. Jurco, S. Schraml, P. Schupp and J. Wess,
\newblock ``Enveloping algebra valued gauge transformations for nonAbelian gauge groups on noncommutative spaces,''
\newblock {\em Eur. Phys. J.} {\bf C 17} 521 (2000) 
\newblock{\tt hep-th/0006246}.
\bibitem{W3}
B. Jurco, L. Moller, S. Schraml, P. Schupp and J. Wess,
\newblock ``Construction of non -- Abelian gauge theories on noncommutative spaces,''
\newblock {\em Eur. Phys. J.} {\bf C 21} 383 (2001) 
\newblock{\tt hep-th/0104153}.
\bibitem{vas}
D. V. Vassilevich,
\newblock ``Twist to close,''
\newblock {\em Mod. Phys. Lett.} {\bf A 21} 1279 (2006) 
\newblock{\tt hep-th/0602185}.
\bibitem{wess}
P. Aschieri, M. Dimitrijevic, F. Meyer, S. Schraml and J. Wess,
\newblock ``Twisted Gauge Theories,''
\newblock {\em Lett. Math. Phys.} {\bf 78} 61 (2006)
\newblock {\tt hep-th/0603024}.
\bibitem{chai}
M. Chaichian and A. Tureanu,
\newblock ``Twist Symmetry and Gauge Invariance,''
\newblock {\em Phys. Lett.} {\bf B 637} 199 (2006) 
\newblock{\tt hep-th/0604025}.
\bibitem{zet}
M. Chaichian, A. Tureanu and G. Zet
\newblock `` Twist as a Symmetry Principle and the Noncommutative Gauge Theory Formulation''
\newblock {\em Phys. Lett.} {\bf B 651} 319 (2007) 
\newblock{\tt hep-th/0607179}.
\bibitem{chams}
A. H. Chamseddine,
\newblock ``Deforming Einstein's gravity,''
\newblock {\em Phys. Lett.} {\bf B 504} 33 (2001) 
\newblock{\tt hep-th/0009153}.
\bibitem{Aschierie}
P. Aschieri, C. Blohmann, M. Dimitrijevic, F. Meyer, P. Schupp and J. Wess,
\newblock ``A Gravity theory on noncommutative spaces,''
\newblock {\em Class. Quant. Grav.} {\bf 22} 3511 (2005) 
\newblock{\tt hep-th/0504183}.
\bibitem{gaume}
L. Alvarez-Gaume, F. Meyer and M. A. Vazquez-Mozo,
\newblock ``Comments on Noncommutative Gravity,''
\newblock {\em Nucl. Phys.} {\bf B 753} 92 (2006) 
\newblock{\tt hep-th/0605113}.
\bibitem{sa}
S. Samanta,
\newblock ``Noncommutativity from embedding techniques,''
\newblock {\em Modern Physics Letters A} {\bf 21} 675 (2006) 
\newblock{\tt hep-th/0510138}.
\bibitem{ssamanta}
R. Banerjee, B. Dutta Roy and S. Samanta,
\newblock ``Remarks on the noncommutative gravitational quantum well,''
\newblock {\em Phys. Rev.} {\bf D 74} 045015 (2006) 
\newblock{\tt hep-th/0605277}.
\bibitem{ssam}
R. Banerjee, S. Kulkarni and S. Samanta,
\newblock ``Deformed symmetry in Snyder space and relativistic particle dynamics'',
\newblock {\em JHEP} {\bf 0605} 077 (2006) 
\newblock{\tt hep-th/0602151}.
\bibitem{saurav}
R. Banerjee and S. Samanta,
\newblock ``Gauge generators, transformations and identities on a noncommutative space,''
\newblock {\em Eur. Phys. J.} {\bf C 51} 207 (2007)
\newblock {\tt hep-th/0608214}.
\bibitem{sau}
 R. Banerjee and S. Samanta, 
\newblock ``Gauge Symmetries on theta-Deformed Spaces'',
\newblock {\em JHEP} {\bf 0702} 046 (2007) 
\newblock {\tt hep-th/0611249}.
\bibitem{samanta}
R. Banerjee, P. Mukherjee and S. Samanta,
\newblock `` Lie algebraic noncommutative gravity'',
\newblock {\em Phys. Rev.} {\bf D 75} 125020 (2007) 
\newblock {\tt hep-th/0703128}.
\bibitem{sausa}
S. Samanta,
\newblock ``Diffeomorphism symmetry in the Lagrangian formulation of gravity'',
arXiv:0708.3300 [hep-th]
\bibitem{LUKE}
J. Lukierski, P. Stichel and W. Zakrzewski,
\newblock ``Galilean invariant (2+1)-dimensional models with a Chern-Simons-like term and D = 2 noncommutative geometry'',
\newblock {\em Ann. Phys.} 260, 224 (1997) 
\newblock {\tt hep-th/9612017}. 
\bibitem{x}V. P. Nair and A. P. Polychronakos,
\newblock ``Quantum mechanics on the noncommutative plane and sphere,''
\newblock {\em Phys. Lett.} {\bf B 505} 267 (2001) 
\newblock {\tt hep-th/0011172}.
\bibitem{x1} 
A. Hatzinikitas and I. Smyrnakis, 
\newblock `` The Noncommutative harmonic oscillator in more than one-dimensions'',
\newblock {\em J. Math. Phys.} {\bf 43} 113 (2002)
\newblock {\tt hep-th/0103074}. 
\bibitem{x2}
J. Gamboa, M. Loewe and J. Rojas, 
\newblock ``Noncommutative quantum mechanics'',
\newblock {\em Phys. Rev.} {\bf D 64} 067901 (2001)
\newblock {\tt hep-th/0010220}.
\bibitem{x3}
B. Muthukumar and P. Mitra,
\newblock ``Noncommutative oscillators and the commutative limit'', 
\newblock {\em Phys. Rev.} {\bf D 66} 027701 (2002), 
\newblock {\tt hep-th/0204149}.
\bibitem{x4}
 C. Duval and P. A. Horvathy,
\newblock `` The 'Peierls substitution' and the exotic Galilei group'', 
\newblock {\em Phys Lett.} {\bf B 479} 284 (2000) 
\newblock {\tt hep-th/0002233}.
\bibitem{y} M. Eliashvili and G. Tsitsishvili,
\newblock `` Chern-Simons theory and quantum fields in the lowest Landau level'', \newblock {\em Int. J. Mod. Phys.} {\bf B 14} 1429 (2000)
\newblock {\tt hep-th/0007033}.
\bibitem{y-1} 
L. Susskind, 
\newblock``The Quantum Hall fluid and noncommutative Chern-Simons theory'',
\newblock {\tt hep-th/0101029}.
\bibitem{y-2}
A. P. Polychronakos,
\newblock `` Quantum Hall states as matrix Chern-Simons theory'',
\newblock {\em JHEP} {\bf 0104} 011 (2001)
\newblock{\tt hep-th/0103013}.
\bibitem{y-3}
S. Hellerman and M. Van Raamsdonk, 
\newblock`` Quantum Hall physics equals noncommutative field theory'',
\newblock {\em JHEP} {\bf 0110} 039 (2001)
\newblock{\tt hep-th/0103179}.
\bibitem{m}
I. A. Batalin and I. V. Tyutin,
\newblock ``Existence theorem for the effective gauge algebra in the generalized canonical formalism with Abelian conversion of second class constraints'',
\newblock {\em Int. J. Mod. Phys.} {\bf A 6} 3255 (1991).
\bibitem{c}
R. Banerjee,
\newblock ``A novel approach to noncommutativity in planar quantum mechanics,''
\newblock {\em Modern Physics Letters} {\bf A 17} 631 (2002) 
\newblock{\tt hep-th/0106280}.
\bibitem{n}
P. A. M. Dirac,
\newblock {\em Lectures on Quantum Mechanics},
\newblock Yeshiva University Press, New York, 1964.
\bibitem{la}
L. Jonke and S. Meljanac,
\newblock `` Representations of noncommutative quantum mechanics and symmetries'',
\newblock {\em Eur. Phys. J} {\bf C 29} 433 (2003)
\newblock {\tt hep-th/0210042}
\bibitem{a}
G. 't Hooft,
\newblock ``Quantum gravity as a dissipative deterministic system'',
\newblock {\em Class. Quantum Grav.} {\bf 16} 3263 (1999)
\newblock {\tt gr-qc/9903084}.
\bibitem{aG}
G. 't Hooft,
\newblock ``Determinism and dissipation in quantum gravity''
\newblock{\tt hep-th/0003005}.
\bibitem{aGH}
G. 't Hooft,
\newblock ``Quantum mechanics and determinism''
\newblock{\tt hep-th/0105105}.
\bibitem{fa}
L. D. Faddeev and R. Jackiw,
\newblock ``Hamiltonian Reduction of Unconstrained and Constrained Systems'',
\newblock {\em Phys. Rev. Lett.} {\bf 60} 1692 (1988).
\bibitem{s}
S. Ghosh,
\newblock `` Extended space duality in the noncommutative plane'',
\newblock {\em Phys. Lett.} {\bf B 601} 93 (2004)
\newblock {\tt hep-th/0409138}.
\bibitem{new}
N. Banerjee, R. Banerjee and S. Ghosh,
\newblock`` Quantization of O(N) invariant nonlinear sigma model in the Batalin-Tyutin formalism''
\newblock {\em Nucl. Phys.} {B 417} 257 (1994) 
\newblock{\tt hep-th/9310044}.
\bibitem{newsubir}
S. Ghosh,
\newblock``Covariantly quantized spinning particle and its possible connection to noncommutative space-time'',
\newblock {\em Phys. Rev.} {\bf D 66} 045031 (2002)
\newblock {\tt hep-th/0203251}.
\bibitem{wot}
C. Wotzasek, 
\newblock ``Soldering formalism: Theory and applications'',
\newblock{\tt hep-th/9806005}.
\bibitem{wotrabin}
 R. Banerjee and C. Wotzasek,
\newblock `` Bosonization and duality symmetry in the soldering formalism'',
\newblock {\em Nucl. Phys.} {B 527} 402-432 (1998)
\newblock{\tt hep-th/9805109.}
\bibitem{ra}
R. Banerjee and S. Ghosh,
\newblock `` The Chiral oscillator and its applications in quantum theory'',
\newblock {\em J. Phys.} {\bf A 31} L603 (1998)
\newblock{\tt quant-ph/9805009}.
\bibitem{nes}
V. V. Nesvizhevsky {\it et al.}
\newblock ``Quantum states of neutrons in the Earth's gravitational field'',
\newblock {\em Nature} {\bf 415} 297 (2002)
\bibitem{vvnes}
V. V. Nesvizhevsky {\it et al.}
\newblock`` Measurement of quantum states of neutrons in the earth's gravitational field'',
 \newblock {\em Phys. Rev.} {\bf D 67} 102002 (2003)
\newblock{\tt hep-ph/0306198}.
\bibitem{nes1}
V. V. Nesvizhevsky {\it {et al.}},
\newblock ``Study of the neutron quantum states in the gravity field'',
\newblock {\em Eur. Phys. J.} {\bf C 40} 479 (2005) 
\newblock{\tt hep-ph/0502081}.
\bibitem{ba}
O. Bertolami, J. G. Rosa, C. M. L. de Aragao, P. Castorina and D. Zappala,
\newblock ``Noncommutative gravitational quantum well'',
\newblock {\em Phys. Rev.} {\bf D 72} 025010 (2005) 
\newblock{\tt hep-th/0505064}.
\bibitem{zhang}
Jian-zu Zhang,
\newblock ``Constraint on Quantum Gravitational Well in Noncommutative Space'',
\newblock {\tt hep-th/0508164}.
\bibitem{Ani}
A. Saha,\newblock ``Time-space noncommutativity in gravitational quantum well scenario,''\newblock {\em Eur. Phys. J.} {\bf C 51} 199 (2007)
\newblock{\tt hep-th/0609195}.
\bibitem{lan}
L. D. Landau and E. M. Lifshitz,
\newblock {\em Quantum Mechanics: Non-Relativistic Theory},
\newblock Pergamon London, 1965.
\bibitem{par}
Particle Data Group,
\newblock {\em Particle Physics Booklet}
\newblock (AIP, New York, 2002).
\bibitem{brau}
F. Brau and F. Buisseret,
\newblock ``Minimal Length Uncertainty Relation and gravitational quantum well,'',
\newblock{\em Phys. Rev.} {\bf D 74} 036002 (2006)
\newblock {\tt hep-th/0605183}.
\bibitem{bert}
O. Bertolami, J. G. Rosa, C. M. L. de Aragao, P.Castorina and D. Zappala,
\newblock ``Scaling of variables and the relation between noncommutative parameters in noncommutative quantum mechanics'',
\newblock {\em Mod. Phys. Lett.} {\bf A 21} 795 (2006) 
\newblock{\tt hep-th/0509207}.
\bibitem{car}
S. M. Carroll, J. A. Harvey, V. A. Kostelecky, C. D. Lane and T. Okamoto,
\newblock ``Noncommutative Field Theory and Lorentz Violation'',
\newblock {\em Phys. Rev. Lett.} {\bf 87} 141601 (2001) 
\newblock{\tt hep-th/0105082}.
\bibitem{ban}
R. Banerjee, B. Chakraborty and K. Kumar,
\newblock ``Noncommutative Gauge Theory and Lorentz Symmetry'',
\newblock {\em Phys. Rev.} {\bf D 70} 125004 (2004) 
\newblock{\tt hep-th/0408197}.
\bibitem{koch}
F. Koch and E. Tsouchnika,
\newblock ``Construction of Theta-Poincare Algebras and Their Invariants on Mu (Theta)'',
\newblock {\em Nucl. Phys.} {\bf B 717} 387 (2005) 
\newblock {\tt hep-th/0409012}.
\bibitem{rabin}
R. Banerjee,
\newblock ``Deformed Schrodinger symmetry on noncommutative space'',
\newblock {\it Eur. Phys. J.} {\bf C 47} 541 (2006)
\newblock {\tt hep-th/0508224}.
\bibitem{lee}
R. Banerjee, C. Lee and S. Siwach,
\newblock ``Deformed conformal and super-Poincar\'e symmetries in the non-(anti)commutative spaces'',
\newblock {\em Eur. Phys. J.} {\bf C 48} 305 (2006)
\newblock {\tt hep-th/0511205}.
\bibitem{cha}
M. Chaichian, P. Presnajder and A. Tureanu,
\newblock ``New Concept of Relativistic Invariance in NC Space-Time: Twisted Poincare Symmetry and Its Implications'',
\newblock {\em Phys. Rev. Lett.} {\bf 94} 151602 (2005) 
\newblock {\tt hep-th/0409096}.
\bibitem{luk}
J. Lukierski and M. Woronowicz,
\newblock ``New Lie-Algebraic and Quadratic Deformations of Minkowski Space from Twisted Poincare Symmetries'',
\newblock {\em Phys. Lett.} {\bf B 633} 116 (2006) 
\newblock {\tt hep-th/0508083}.
\bibitem{duvalhorvathy}
C. Duval, P. A. Horvathy,
\newblock ``Exotic Galilean symmetry in the noncommutative plane, and the Hall effect'',
\newblock {\em J. Phys.} {\bf A 34} 10097 (2001)
\newblock {\tt hep-th/0106089}.
\bibitem{deri}
A. A. Deriglazov,
\newblock ``Noncommutative Relativistic Particle'',
\newblock {\tt hep-th/0207274}.
\bibitem{wor}
J. Lukierski and M. Woronowicz,
\newblock ``Twisted Space-Time Symmetry, Noncommutativity and Particle Dynamics'',
\newblock {\tt hep-th/0512046}.
\bibitem{kk}
R. Banerjee and K. Kumar,
\newblock ``Deformed relativistic and nonrelativistic symmetries on canonical noncommutative spaces,'' {\tt Phys. Rev. } {\bf D 75} 045008 (2007)
\newblock {\tt hep-th/0604162}.
\bibitem{tooft}
G. 't Hooft,
\newblock ``Quantization of Point Particle in (2+1)-Dimensional Gravity and Space-Time Discreteness'',
\newblock {\em Class. Quant. Grav.} {\bf 13} 1023 (1996) 
\newblock{\tt gr-qc/9601014}.
\bibitem{glik}
J. Kowalski-Glikman and S. Nowak,
\newblock ``Noncommutative Space-Time of Doubly Special Relativity Theories'',
\newblock {\em Int. J. Mod. Phys} {\bf D 12} 299 (2003) 
\newblock {\tt hep-th/0204245}.
\bibitem{rom}
J. M. Romero and A. Zamora,
\newblock ``Snyder Noncommutative Space-Time from Two Time Physics'',
\newblock {\em Phys. Rev.} {\bf D 70} 105006 (2004) 
\newblock {\tt hep-th/0408193}.
\bibitem{gir}
F. Girelli, T. Konopka, J. Kowalski-Glikman and E. R. Livine,
\newblock ``The Free Particle in Deformed Special Relativity'',
\newblock {\em Phys. Rev.} {\bf D 73} 045009 (2006)
\newblock {\tt hep-th/0512107}.
\bibitem{ghosh}
S. Ghosh and P. Pal,
\newblock ``Kappa-Minkowski Space-Time through Exotic 'Oscillator' '',
\newblock {\em Phys. Lett.} {\bf B 618} 243 (2005) 
\newblock {\tt hep-th/0502192}.
\bibitem{rome}
J. M. Romero and D. Vergara,
\newblock ``The Parametrized Relativistic Particle and the Snyder space-time'',
\newblock {\tt hep-th/0602058}.
\bibitem{zachos}
C. Zachos,
\newblock ``A survey of star product geometry''
\newblock {\tt hep-th/0008010}.
\bibitem{jwess}
J. Wess,
\newblock ``Deformed gauge theories''
\newblock {\em J. Phys. Conf. Ser.} {\bf 53} 752 (2006)
\newblock {\tt hep-th/0608135}.
\bibitem{wes}
J. Wess,
\newblock ``Differential calculus and gauge transformations on a deformed space''
\newblock {\em Gen. Rel. Grav.} {\bf 39} 1121 (2007)
\newblock {\tt hep-th/0607251}.
\bibitem{gitman}
D. M. Gitman and I. V. Tyutin,
\newblock {\em Quantization of Fields with Constraints,}
\newblock Springer-Verlag Berlin, Heidelberg (1990).
\bibitem{shirzad}
A. Shirzad,
\newblock ``Gauge Symmetry in Lagrangian formulation and Schwinger Models'',
\newblock {\em J. Phys.} {\bf A 31} 2747 (1998).
\bibitem{rothe}
R. Banerjee, H. J. Rothe and K. D. Rothe,
\newblock ``Master equation for Lagrangian gauge symmetries'',
\newblock {\em Phys.Lett.} {\bf B 479} 429 (2000) 
\newblock{\tt hep-th/9907217}.
\bibitem{rothe1}
R. Banerjee, H. J. Rothe and K. D. Rothe,
\newblock ``Hamiltonian approach to Lagrangian gauge symmetries'',
\newblock {\em Phys.Lett.} {\bf B 463} 248 (1999) 
\newblock{\tt hep-th/9906072}.
\bibitem{wineberg} S. Weinberg {\it Gravitation and Cosmology} (John Wiley, 1972).
\bibitem{ortin} T. Ortin {\it Gravity and Strings} (Cambridge University, 2004).
\bibitem{Wigner}
E. P. Wigner,
\newblock ``On the quantum correction for
thermodynamic equilibrium'',
\newblock {\em Phys. Rev.} {\bf 40} 749 (1932). 
\bibitem{amorim}
R. Amorim and F. A. Farias,
\newblock ``Hamiltonian formulation of non-Abelian noncommutative gauge theories'',
\newblock {\em Phys. Rev.} {\bf D 65} 065009 (2002) 
\newblock{\tt hep-th/0109146}.
\bibitem{rabinb}
R. Banerje,
\newblock ``Noncommuting Electric Fields and Algebraic Consistency in Noncommutative Gauge Theories'',
\newblock {\em Phys. Rev.} {\bf D 67} 105002 (2003) 
\newblock{\tt hep-th/0210259}.
\bibitem{banora}
L. Bonora and M. Salizzoni,
\newblock ``Renormalization of noncommutative U(N) gauge theories'',
\newblock {\em Phys. Lett.} {\bf B 504} 80 (2001) 
\newblock{\tt hep-th/0011088}.
\bibitem{armoni}
 A. Armoni, \newblock`` Comments on perturbative dynamics of noncommutative Yang-Mills theory'',\newblock {\em Nucl. Phys.} {\bf B 593} 229 (2001) 
\newblock{\tt hep-th/0005208}.
\bibitem{henn} 
M. Hennaux and C. Teitelboim,
\newblock {\em Quantization of Gauge Systems},
\newblock Princeton University Press, Princeton (1992).
\bibitem{szabo1}
R. J. Szabo,
\newblock ``Symmetry, gravity and noncommutativity'',
\newblock {\em Class. Quant. Grav.} {\bf 23} R199 (2006)
\newblock{\tt hep-th/0606233}.
\bibitem{kur}
S. Kurkcuoglu and C. Saemann,
\newblock ``Drinfeld Twist and General Relativity with Fuzzy Spaces'',
\newblock {\em Class. Quant. Grav.} {\bf 24} 291 (2007) 
\newblock{\tt hep-th/0606197}.
\bibitem{CK1}
X. Calmet and A. Kobakhidze,
\newblock ``Noncommutative general relativity,''
\newblock {\em Phys. Rev.} {\bf D 72} 045010 (2005) 
\newblock{\tt hep-th/0506157}.
\bibitem{Ein}
J.~J.~van der Bij, H.~van Dam and Y. J. Ng,
\newblock ``The Exchange Of Massless Spin Two Particles'',
\newblock {\em Physica} {\bf 116 A} 307 (1982).
\bibitem{Ein1}
F.~Wilczek,
\newblock ``Foundations And Working Pictures In Microphysical Cosmology'',
\newblock {\em Phys. Rept.} {\bf 104} 143 (1984).
\bibitem{Ein2}
W.~Buchmuller and N.~Dragon,
\newblock ``Einstein Gravity From Restricted Coordinate Invariance'',
\newblock {\em Phys. Lett.} {\bf B 207} 292 (1988).
\bibitem{Ein3}
M.~Henneaux and C.~Teitelboim,
\newblock ``The Cosmological Constant And General Covariance'',
\newblock {\em Phys. Lett.}  {\bf B 222} 195 (1989).
\bibitem{Ein4}
W.~G.~Unruh,
\newblock ``A Unimodular Theory Of Canonical Quantum Gravity''.
\newblock {\em Phys. Rev.} {\bf D 40} 1048 (1989).
\bibitem{hari}
E. Harikumar and Victor O. Rivelles,
\newblock ``Noncommutative Gravity'',
\newblock {\em Class. Quant. Grav.} {\bf 23} 7551 (2006) 
\newblock{\tt hep-th/0607115}.
\bibitem{MS}
P. Mukherjee and A. Saha,
\newblock ``A Note on the noncommutative correction to gravity,''
\newblock {\em Phys. Rev.} {\bf D 74} 027702 (2006)
\newblock{\tt hep-th/0605287}.
\bibitem{CK2}
X. Calmet and A. Kobakhidze,
\newblock``Second order noncommutative corrections to gravity'',
\newblock {\em Phys. Rev.} {\bf D 74} 047702 (2006) 
\newblock{\tt hep-th/0605275}.
\bibitem{RK}
R. Banerjee and K. Kumar,
\newblock ``Maps for currents and anomalies in noncommutative gauge theories,''
\newblock {\em Phys. Rev.} {\bf D 71} 045013 (2005) 
\newblock{\tt hep-th/0404110}.
\bibitem{RLee}
R. Banerjee, C. Lee and H. S. Yang,
\newblock `` Seiberg-Witten-type maps for currents and energy momentum tensors in noncommutative gauge theories'',
\newblock {\em Phys. Rev.} {\bf D 70} 065015 (2004) 
\newblock{\tt hep-th/0312103}.
\bibitem{Banerjee:2004rs}
R. Banerjee and H. S. Yang,
\newblock ``Exact {S}eiberg--{W}itten map, induced gravity and topological invariants in noncommutative field theories'',
\newblock {\em Nucl.~Phys.} {\bf B 708}  434 (2005)
\newblock{\tt hep-th/0404064}.
\bibitem{Behr}
W. Behr and A. Sykora,
\newblock ``Construction of gauge theories on curved noncommutative space-time'',
\newblock {\em Nucl. Phys.} {\bf B 698} 473 (2004) 
\newblock{\tt hep-th/0309145}.
\end{thebibliography}
\end{document}